\documentclass[showpacs,onecolumn,aps,pra,floatfix,superscriptaddress,10pt,secnumarabic]{revtex4-1}
\pdfoutput=1
\usepackage[T1]{fontenc}
\usepackage{amsmath}
\usepackage{amsfonts}
\usepackage{amssymb}
\usepackage{graphicx} 
\usepackage{bm}
\usepackage{hyperref}
\usepackage{bookmark}
\hypersetup{colorlinks, allcolors={blue}}
\makeatletter
\renewcommand{\p@subsection}{}
\renewcommand{\p@subsubsection}{}
\makeatother
\newcommand{\HE}     {Hall effect}
\newcommand{\QHE}    {quantum Hall effect}

\newcommand{\SHE}    {spin Hall effect}
\newcommand{\QSHE}   {quantum spin Hall effect}
\newcommand{\VHE}    {valley Hall effect}
\newcommand{\QVHE}   {quantum valley Hall effect}
\newcommand{\ZSHE}   {Zeeman spin Hall effect}
\newcommand{\soc}    {spin-orbit coupling}

\newcommand{\maglen} {\ell_{\rm B}}

\newcommand{\TR}     {\text{Tr}}

\newcommand{\Frac}   {\displaystyle\frac} 
\renewcommand{\vec}[1]{\boldsymbol{#1}}
\newcommand{\fref} [1]{Fig.~\ref{#1}}
\newcommand{\Fref} [1]{Figure~\ref{#1}}
\newcommand{\ffref}[1]{Figs.~\ref{#1}}
\newcommand{\FFref}[1]{Figures~\ref{#1}}
\newcommand{\sref} [1]{Sec.~\ref{#1}}
\newcommand{\Sref} [1]{Section~\ref{#1}}
\newcommand{\ssref}[1]{Sects.~\ref{#1}}

\newcommand{\eref} [1]{Eq.~\eqref{#1}}

\newcommand{\eeref}[1]{Eqs.~\eqref{#1}}

\newcommand{\cref} [1]{ref.~[\onlinecite{#1}]}
\newcommand{\Cref} [1]{Reference~[\onlinecite{#1}]}
\newcommand{\ccref}[1]{refs.~[\onlinecite{#1}]}

\newcommand{\aref} [1]{Appendix~\ref{#1}}

\newcommand{\aaref}[1]{Appendices~\ref{#1}}

\begin{document}

%%%%% TITLE %%%%%%%%%%%%%%%%%%%%%%%%%%%%%%%%%%%%%%%%%%%%%%%%%%%%%%%%%%%%%%%%%%%%%%%%%%%%%

\title{Charge, Spin and Valley Hall Effects in Disordered Graphene}

%%%%% AUTHORS %%%%%%%%%%%%%%%%%%%%%%%%%%%%%%%%%%%%%%%%%%%%%%%%%%%%%%%%%%%%%%%%%%%%%%%%%%%

\author{Alessandro Cresti}
\affiliation{Univ. Grenoble Alpes, IMEP-LAHC, F-38000 Grenoble, France}
\affiliation{CNRS, IMEP-LAHC, F-38000 Grenoble, France}
\author{Branislav K. Nikoli\'{c}}
\affiliation{Department of Physics and Astronomy, University of Delaware, Newark, DE 19716, U.S.A.}
\author{Jose Hugo Garc\'ia}
\affiliation{Catalan Institute of Nanoscience and Nanotechnology (ICN2), CSIC and The Barcelona Institute of Science and Technology, Campus UAB, Bellaterra, 08193 Barcelona, Spain}
\author{Stephan Roche}
\affiliation{Catalan Institute of Nanoscience and Nanotechnology (ICN2), CSIC and The Barcelona Institute of Science and Technology, Campus UAB, Bellaterra, 08193 Barcelona, Spain}
\affiliation{ICREA - Institucio Catalana de Recerca i Estudis Avan\c{c}ats, 08010 Barcelona, Spain}

%%%%% ABSTRACT %%%%%%%%%%%%%%%%%%%%%%%%%%%%%%%%%%%%%%%%%%%%%%%%%%%%%%%%%%%%%%%%%%%%%%%%%%
\begin{abstract}
The discovery of the integer quantum Hall effect in the early eighties of the last century, with highly precise quantization values for the Hall conductance in multiples of $e^2/h$, has been the first fascinating manifestation of the topological state of matter driven by magnetic field and disorder, and related to the formation of non-dissipative current flow. 
In 2005, several new phenomena such as the \SHE\ and the \QSHE\ were predicted in the presence of strong spin-orbit coupling and vanishing external magnetic field. 
More recently, the Zeeman spin Hall effect and the formation of valley Hall topological currents have been introduced for graphene-based systems, under time-reversal or inversion symmetry-breaking conditions, respectively. 
This review presents a comprehensive coverage of all these Hall effects in disordered graphene from the perspective of numerical simulations of quantum transport in two-dimensional bulk systems (by means of the Kubo formalism) and multiterminal nanostructures (by means of the Landauer-B\"{u}ttiker scattering and nonequilibrium Green function approaches).
In contrast to usual two-dimensional electron gases, the presence of defects in graphene generates more complex electronic features such as electron-hole asymmetry, defect resonances or percolation effect between localized impurity states, which, together with extra degrees of freedom (sublattice pseudospin, valley isospin), bring a higher degree of complexity and enlarge the transport phase diagram.

%%%%% COVER FIGURE %%%%%%%%%%%%%%%%%%%%%%%%%%%%%%%%%%%%%%%%%%%%%%%%%%%%%%%%%%%%%%%%%%%%%%

\vspace*{0.75cm}
\begin{center}
\resizebox{10cm}{!}{\includegraphics{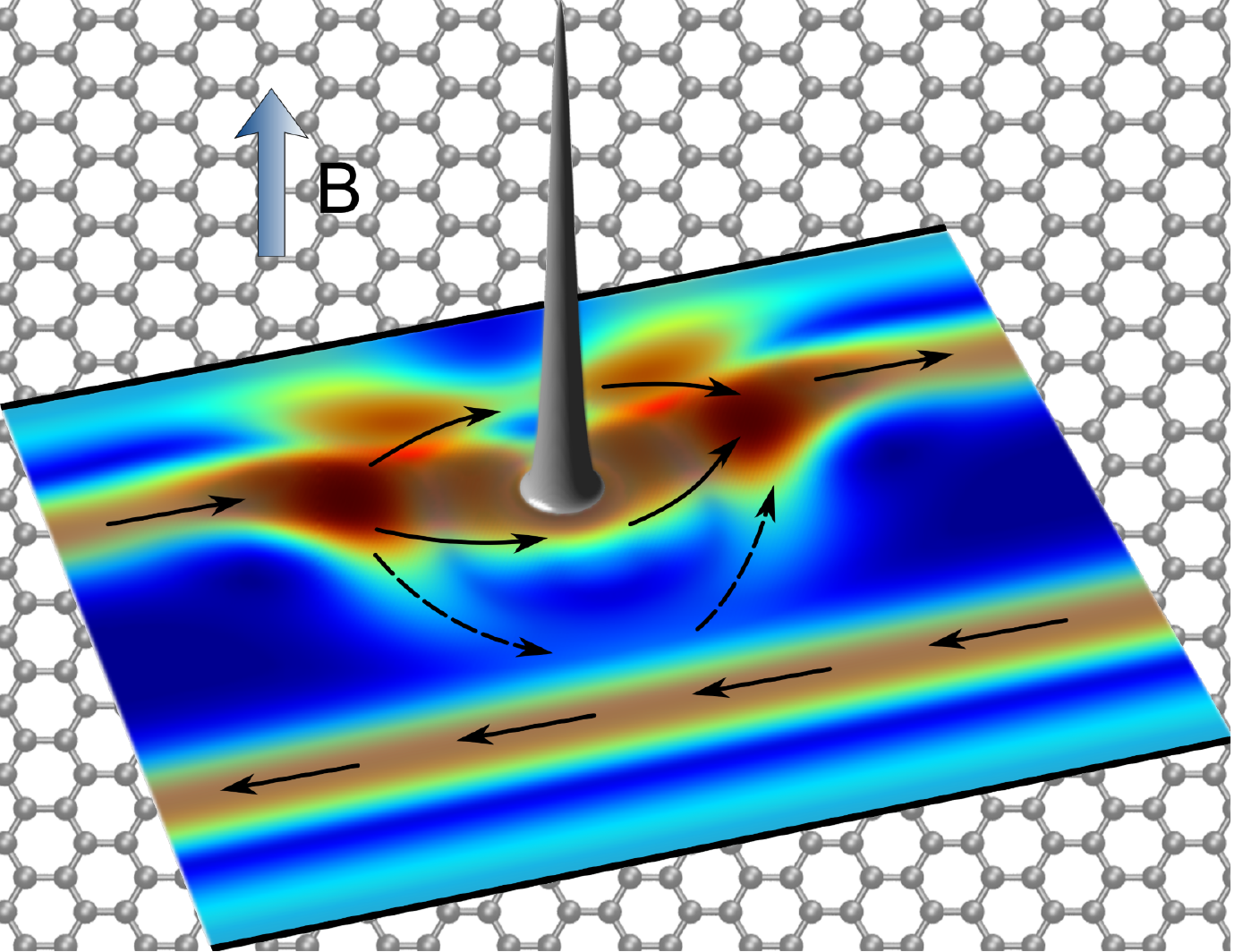}}
\end{center}

\vspace*{0.25cm}

{\footnotesize \noindent  Equilibrium spectral current distribution in a 100 nm wide graphene ribbon at an energy between the first and the second Landau levels in the presence of a Gaussian potential impurity. The current flows along edge states in opposite directions and it circumvents the impurity. Some small current reaches the bottom edge thus breaking spatial chirality of currents and being backscattered.}

\end{abstract}

\maketitle

%%%%% TABLE OF CONTENTS %%%%%%%%%%%%%%%%%%%%%%%%%%%%%%%%%%%%%%%%%%%%%%%%%%%%%%%%%%%%%%%%%

\newpage
\tableofcontents

%%%%% LIST OF ACRONYMS %%%%%%%%%%%%%%%%%%%%%%%%%%%%%%%%%%%%%%%%%%%%%%%%%%%%%%%%%%%%%%%%%%

\section*{Acronyms and main variables}
\vspace*{1.5cm}
\noindent
\hspace*{2.5cm}
\begin{minipage}{2in}
		\begin{tabular}{@{}p{4cm}@{}l}
			2D        \dotfill & two-dimensional \\
			2DEG      \dotfill & two-dimensional electron gas \\
			AGNR      \dotfill & graphene nanoribbon with armchair edges  \\
			BTE       \dotfill & Boltzmann transport equation  \\
			BZ        \dotfill & Brillouin zone \\
			CNP       \dotfill & charge neutrality point \\
			CVD       \dotfill & chemical vapor deposition  \\
			DP        \dotfill & Dyakonov-Perel  \\
			DOS       \dotfill & density of states \\	
			EY        \dotfill & Elliot-Yafet \\
			GB        \dotfill & grain boundary \\
			GF        \dotfill & Green's function \\
			GNR       \dotfill & graphene nanoribbon \\
			h-BN      \dotfill & hexagonal boron-nitride \\
			HE        \dotfill & Hall effect \\
			IQHE      \dotfill & integer quantum Hall effect \\
			ISHE      \dotfill & inverse spin Hall effect \\
			KPM       \dotfill & kernel polynomial method \\
			LB        \dotfill & Landauer-B\"{u}ttiker \\
			LL        \dotfill & Landau level \\
			MBZ       \dotfill & magnetic Brillouin zone \\
			NEGF      \dotfill & nonequilibrium Green function \\
			PIA       \dotfill & pseudospin inversion asymmetry \\
			QHE       \dotfill & quantum Hall effect \\
			QSHE      \dotfill & quantum spin Hall effect \\
			QVHE      \dotfill & quantum valley Hall effect \\
			SH        \dotfill & spin Hall \\
			SHE       \dotfill & spin Hall effect \\
			SJ        \dotfill & side jump \\
			SO        \dotfill & spin-orbit \\
			SOC       \dotfill & spin-orbit coupling \\
			SS        \dotfill & skew-scattering \\
			TB        \dotfill & tight-binding \\
			TI        \dotfill & topological insulator \\
			VHE       \dotfill & valley Hall effect \\
			ZGNR      \dotfill & graphene nanoribbon with zigzag edges \\
			ZSHE      \dotfill & Zeeman spin Hall effect \\
			$E_{\rm F}$                                       \dotfill & Fermi energy \\
			$\maglen$                                         \dotfill & magnetic length \\
			$R_{\rm NL}$                                      \dotfill & nonlocal resistance \\
			$\vec{s}\ = \ [s_x,s_y,s_z]$                      \dotfill & Pauli matrices operating on the spin degree of freedom \\
			$\vec{\sigma}\ = \ [\sigma_x,\sigma_y,\sigma_z]$  \dotfill & Pauli matrices operating on the sublattice degree of freedom \\
			$\sigma_{xx}, \ \sigma_{xy}$                      \dotfill & longitudinal and transverse (Hall) conductivity \\
			$\sigma_{xy}^z$                                   \dotfill & spin Hall conductivity \\
			$\sigma^v_{xx}$                                   \dotfill & valley Hall conductivity \\
			$\tau_z$                                          \dotfill & $z$-Pauli matrix operating on the valley degree of freedom \\
			$\theta_{\rm sH}$                                 \dotfill & spin Hall angle
		\end{tabular}
  \end{minipage}
\newpage

%%%%% INTRODUCTION %%%%%%%%%%%%%%%%%%%%%%%%%%%%%%%%%%%%%%%%%%%%%%%%%%%%%%%%%%%%%%%%%%%%%%

\section{Introduction} \label{sec:intro}

In 1878, Edwin Hall designed an experiment aiming at measuring the change of electrical resistance in a thin gold leaf, in the presence of a steady magnetic field. He found that the magnetic field permanently altered the charge distribution, with a transverse potential difference \cite{HAL_AJM2}. The Hall conductance was defined as the longitudinal current divided by the transverse voltage. The magnitude and even the sign of the Hall voltage was found to be material-dependent, making the \HE\ (HE) a useful characterization tool for inspecting the transport properties (including the nature of charge carriers, electron versus holes) in a given solid. The classical Hall resistivity follows a typical law with charge density as $R_{xy}=-1/ne$.

The \QHE\ (QHE) was further discovered in 1980 at the High Magnetic Field Laboratory in Grenoble (France) by Klaus von Klitzing, who was measuring the Hall conductance of two-dimensional electron gas (2DEG) in the ultralow temperature regime, more precisely in Si(100) MOS inversion layers at $B=19$ T and $T=1.5$ K \cite{KLI_PRL45}. Von Klitzing found that the Hall conductance exhibited a staircase sequence of wide plateaus as a function of the strength of an applied magnetic field perpendicular to the 2DEG, that is $R_{xy}=R_{K90}/\nu$ with $R_{K90}=h/e^2=25812.807572\ \Omega$ the universal von Klitzing resistance constant and $\nu=1,2,3,4....$. This quantization is of incredible precision (1 part in $10^{10}$), vanishingly sensitive to measurement geometry and material degree of imperfection. Since 1990, the {\it von Klitzing resistance constant} stands as the international standard for resistance calibrations. In 1985, von Klitzing was awarded the Physics Nobel prize for the discovery of the integer QHE \cite{RevModPhys.58.519}.

In the quantum Hall regime, the Hall conductivity $\sigma_{xy}$ is thus also quantized $\sigma_{xy}= \nu e^2/h$, while the longitudinal conductivity becomes vanishingly small  $\sigma_{xx}\sim 0$. The prefactor $\nu$ in $\sigma_{xy}$ is the filling factor, and is either an integer number ($\nu = 1, 2, 3, \ldots$ ) or a fractional number ($\nu = 1/3, 2/5, 3/7, 2/3, 3/5, \ldots)$.  The integer QHE is explained in terms of single-particle orbitals of an electron in a magnetic field and is related to the Landau quantization. Differently, the fractional QHE fundamentally relies on strong electron-electron interactions, and the existence of so-called charge-flux composites known as composite fermions \cite{Tsui1982,Laughlin1983}.

The \SHE\ (SHE) was predicted theoretically by Dyakonov and Perel in 1971 \cite{dp2,dp3}, as the formation of spin accumulation on the lateral surfaces of an electric current-carrying sample, the signs of the spin directions being reversed at opposite boundaries. Differently from the case of classical HE, where opposite charges accumulate at the boundaries as a result of the Lorentz force generated by an external magnetic field, the formation of SHE takes place in the absence of magnetic field and is driven by \soc\ (SOC) either through scattering off impurities (extrinsic SHE) or spin-split band structure (intrinsic SHE) \cite{SHE2,VIG_JSNM23}. The SHE belongs to the same family as the anomalous HE, known for a long time in ferromagnets, which originates from the combined effect of SOC and magnetization (in fact, SHE can be viewed as the zero magnetization limit of anomalous HE) \cite{Sinitsyn}. 

The experimental confirmation of SHE detection has been first achieved by optical spectroscopy, in both the extrinsic regime \cite{KAT_SCI306} and the intrinsic regime \cite{WUN_PRL94}. Then the electrical detection was accomplished using the inverse SHE (ISHE) by Saitoh and coworkers \cite{Saitoh2006}, Valenzuela and Tinkham \cite{Valenzuela2006}, and Zhao and coworkers \cite{Zhao2006}. The ISHE, which is Onsager reciprocal of direct SHE, measures a charge imbalance at the sample edges resulting from injection of pure spin current or spin-polarized charge current into a spin-orbit-coupled sample. Despite being observed only a decade ago, these effects are already ubiquitous within spintronics, as standard spin-current generators and detectors \cite{SHE2}.

In 2005, Kane and Mele predicted the possibility of the \QSHE\ (QSHE) in graphene due to intrinsic SOC \cite{KAN_PRL95,KAN_PRL95b}.
Within the QSHE, the presence of SOC results in the formation of chiral edge channels for spin-up and spin-down electrons. The observation of the QSHE has been however jeopardized in clean graphene owing to vanishingly small intrinsic SOC on order of $\mu$eV \cite{YAO_PRB75}, but demonstrated in strong SOC materials (such as CdTe/HgTe/CdTe quantum wells or bismuth selenide and telluride alloys), giving rise to the new exciting field of {\it topological insulators} (TIs) \cite{BER_PRL96,BER_SCI314,HAS_RMP82,QI_RMP83}.
Recent theoretical proposals to induce a topological phase in graphene include the functionalization with heavy adatoms \cite{WEE_PRX1,JIA_PRL109}, covalent functionalization of the edges \cite{AUT_PRB87}, proximity effect with TIs \cite{JIN_PRB87,LIU_PRB87,KOU_NL13}, or intercalation and functionalization with $5d$ transition metals \cite{HU_PRL109,LI_PRB87}.

Finally, well-separated in momentum space degenerate valleys of energy bands (in graphene and other two-dimensional materials) constitute a discrete degree of freedom for low-energy carriers with long intervalley relaxation time. The valley degree of freedom may be used as a non-volatile information carrier, provided that it can be coupled to external probes. In the presence of inversion symmetry breaking in graphene---for instance having graphene onto a substrate such as hexagonal boron-nitride (h-BN)---the valley index plays a similar role as spin in conventional semiconductors, driving towards quantum phenomena such as Hall transport, magnetization, optical transition selection rules, and chiral edge modes \cite{LU_PRB81,YAO_PRL102}. These make possible the control of valley dynamics by magnetic, electric, and optical means, which form the basis of valley based information processing.

Beyond graphene, valley Hall physics also occurs in two-dimensional semiconductors such as ${\rm MoS}_{2}$ monolayers and other group VI transition metal dichalcogenides, which are direct bandgap semiconductors with band edges located at the $K$ points. The low energy electrons and holes are well described by massive Dirac fermions with strong spin-valley coupling. In analogy to the classical Hall and the SHE, the \VHE\ (VHE) is theoretically determined by different valley currents moving in opposite directions perpendicular to the drift current. Moreover, the large SOC in the valence band causes valley-spin locking, whereby optically excited valley populations are also spin polarized, and valley accumulation at sample edges is accompanied by spin accumulation \cite{XIA_PRL108,PhysRevB.86.165108}. Such a coexistence of VHE and SHE could make possible valley and spin controls for potential integrated spintronics and valleytronics applications on this platform. To date, VHE has only been directly observed in highly disordered ${\rm SiO}_{2}$-supported (low-quality) MoS$_2$ samples \cite{Mak1489}.

%%%%% FUNDAMENTAL OF HALL EFFECTS AND KEY CONCEPTS IN GRAPHENE %%%%%%%%%%%%%%%%%%

\section{Fundamental aspects of Hall effects and key concepts in graphene} \label{sec:intro_hall}

\subsection{Quantum Hall effect and the geometrical nature of the quantization of the Hall conductivity} \label{sec:intro_hall_top1}

\begin{figure} [b!]
\begin{center}
  \resizebox{12cm}{!}{\includegraphics{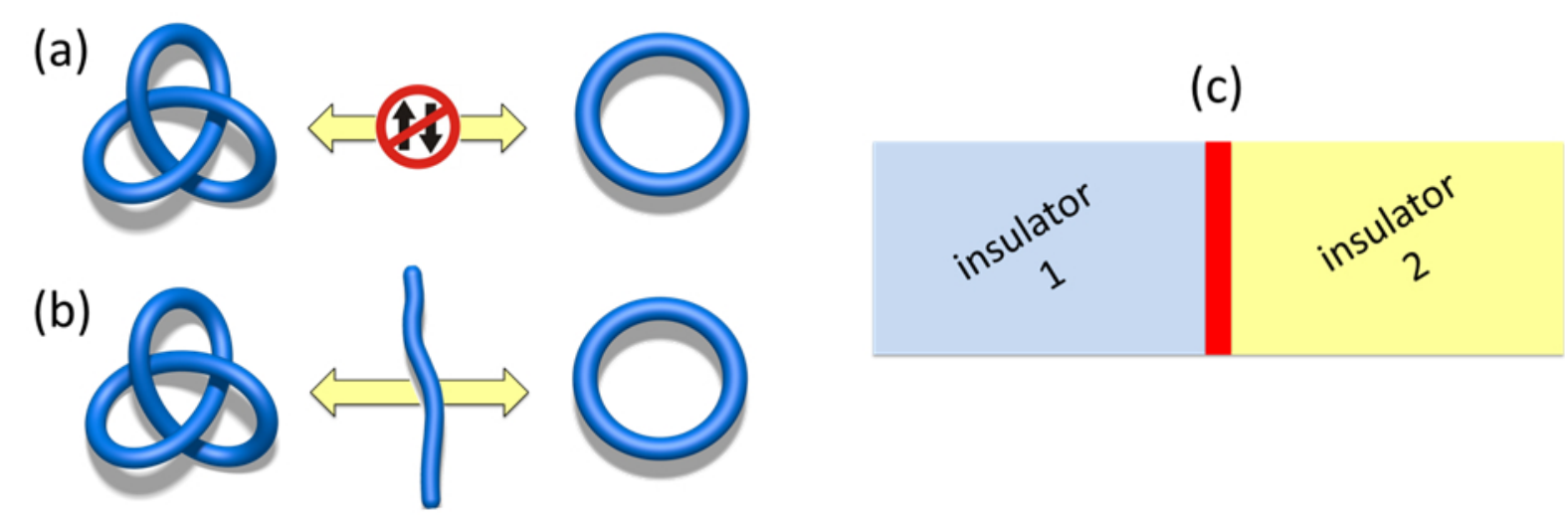}}
  \caption{(a) Knots with different topological index cannot be transformed between them in a continuous way. (b) To do this, the rope must be cut. (c) This is analog to the formation of metallic states at the interface between two insulators with different topological index.}
  \label{fig:knots}
	\end{center}
\end{figure}

The exact quantization of the (charge and spin) Hall conductivities and their robustness against disorder are direct consequence of the specific topological nature of the system band structure. 
The discovery of the Berry curvature \cite{BER_PRSA392} revealed the deep connection between certain aspects of condensed matter and differential geometry, whose important effects are partly illustrated in the rest of this section.
In particular, we will show that the transport properties of quantum Hall systems can be obtained in terms of topological invariant quantities, which characterize the wave functions and do not vary under continuous deformations of the system. 
A popular analogy to illustrate this concept is offered by knot theory.
The number of crossings in a knot is a topological invariant: knots with different number of crossings, as those in \fref{fig:knots}(a), cannot be transformed between them in a continuous way and their index cannot change under continuous deformations. This defines classes of equivalence. To transform one knot to another belonging to a different class, we need to cut the rope as illustrated in \fref{fig:knots}(b). This means a transition to a geometry that is not a knot. Analogously, at the interface between insulators with different topological index, a metallic phase must appear, see \fref{fig:knots}(c). For example, in the \QHE\ edge states are formed at interface with the vacuum.

\subsection{Topological aspect of the integer \QHE} \label{sec:intro_hall_top2}

The main manifestation of the QHE in high magnetic fields is the occurrence of a non-dissipative transport regime defined by a series of quantized plateaus $\sigma_{xy}=1/\rho_{xy}=n e^2/h$ and $\sigma_{xx}=\rho_{xx}=0$. For clean enough materials, Hall conductance plateaus develop at the Fermi energies ($E_{\rm F}$) where there is no dissipation in the bulk, whereas chiral edge states are formed at opposite edges of the sample and are identified through a transverse Hall voltage.  Historically, the topological origin of the Hall quantization has been demonstrated by using bulk Kubo conductivity calculations, assuming weak enough disorder to maintain the integrity of the electronic structure of the underlying electronic system \cite{Aoki87}.  

In their seminal paper, Thouless, Kohmoto, Nightingale, and den Nijs \cite{THO_PRL49} computed the Kubo Hall conductivity of an electron gas in 2D lattice and related it to the topological property of the ground wave function, which turns out to be a topological invariant, known as Chern number. In the basis of eigenstates, the Kubo formula \cite{KUB_JPSJ6} for the current density reads:
\begin{equation}
  j_{y}=\sigma_{xy}E_{x}=-\frac{ie^{2}\hbar}{L^{2}}\sum_{kq} \frac{\langle k|v_{x}|q\rangle \langle q|v_{y}|k\rangle} {(\varepsilon_{k}-\varepsilon_{q})^{2}}E_{x}+{ \rm c.c.} \ , 
\end{equation}
which is a bulk result derived for a clean system (translational invariance), but weak disorder and interaction do not destroy this invariance property of the wave function \cite{KOH_AP160,NIU_PRB31}. 

To obtain the quantization of the conductivity, one has first to write the Hamiltonian for noninteracting electrons in a uniform magnetic field perpendicular to the 2D surface as \cite{KOH_AP160}
\begin{equation}
   H\Psi({\bf r}) \ = \ \left[\frac{1}{2m}({\bf p}+e{\bf A})^{2}+U({\bf r}) \right]\Psi({\bf r})=E\Psi({\bf r}) \ ,
\end{equation}
where ${\bf B}={\bf \nabla}\times {\bf A}$ is the magnetic field in terms of the vector potential ${\bf A}({\bf r})$, and $U({\bf r})$ is an energy potential periodic over the Bravais lattice with translation vectors ${\bf a}$ and ${\bf b}$.
The magnetic field is translationally invariant, but $H$ is not invariant under the discrete translation operator $T_{\bf R}=e^{\frac{i}{\hbar}{\bf R}\cdot{\bf p}}$, because ${\bf A}({\bf r})$ is not constant.  The Hamiltonian can be made invariant under the combined action of a translation and a gauge transformation, which defines the ``magnetic translation operators'' $T_{\bf R}=e^{\frac{i}{\hbar}{\bf R}\cdot[{\bf p}+e({\bf r} \times{\bf B})/2]}$, with ${\bf R}=n{\bf a}+m{\bf b}$ and $(n,m)\in\mathbb{Z}^{2}$. It can be verified by inspection that $[T_{\bf R},H]=0$.
 However the two magnetic translation operators do not generally commute since $T_{\bf a}T_{\bf b}=e^{2i\pi\varphi}T_{\bf b} T_{\bf a}$,  with $\varphi=(eB/\hbar)ab$ the magnetic flux {\it per} unit cell in units of the elementary magnetic flux.  
To solve this problem, we can define an enlarged magnetic unit cell defined by the transformation ${\bf R}'=nq{\bf a}+m{\bf b}$, with $q\in\mathbb{Z}$, in such a way that an integer number of flux quanta thread the cell. This leads to $[T_{q{\bf a}},T_{\bf b}]=0$.
The eigenstates of $H$ can now be labelled with good quantum numbers in the Bloch form $\Psi_{\bf k}(x,y)=e^{i{\bf k}\cdot{\bf r}}u_{\bf k}(x,y) $, with $T_{q{\bf a}}\Psi=e^{i k_{x}qa}\Psi$ and $T_{{\bf b}}\Psi=e^{i k_{y}b}\Psi$, thus defining the generalized crystal momenta restricted to the magnetic Brillouin zone MBZ ($0\leq k_{x}\leq 2\pi/(qa)$, $0\leq k_{y}\leq 2\pi/b$).
By exploiting the relation $\langle k|v_{x}|q\rangle  = \langle k|\frac{\partial H({\bf k})}{\partial k_{x}}|q\rangle / \hbar$, one can write
\begin{equation}
   \sigma_{xy} = -\frac{ie^{2}\hbar}{L^{2}}\sum_{kq} \frac{\langle k|v_{x}|q\rangle \langle q|v_{y}|k\rangle} {(\varepsilon_{k}-\varepsilon_{q})^{2}}=
	              \frac{e^{2}}{2i\pi h}\int_{\rm MBZ}d^{2}k\int d^{2}r \Biggl(\frac{\partial u_{\bf k}^{*}}{\partial k_{y}}
                 \frac{\partial u_{\bf k}}{\partial k_{x}}-\frac{\partial u_{\bf k}^{*}}{\partial k_{x}} \frac{\partial u_{\bf k}}{\partial k_{y}}\Biggr) \ .
\end{equation}
Finally, the Hall conductance turns out to be connected to topological properties of the ground state-wave function through \cite{KOH_AP160}
\begin{equation}
    \sigma_{xy}=\frac{e^{2}}{h}\frac{1}{2\pi}\int_{\rm MBZ}d^{2}k[{\bf \nabla}_{\bf k}\times\vec{\omega}({\bf k})]_{z}
		 \ \ \ \ \ {\rm with} \ \ \ \ \ 
		\vec{\omega}({\bf k})=-i\int d^{2}r u_{\bf k}^{*}{\bf \nabla}_{\bf k} u_{\bf k}=-i\langle u_{\bf k}|{\bf \nabla_{\bf k}}|u_{\bf k}\rangle \ .
\end{equation}
Note that $\vec{\omega}({\bf k})$ is the Berry connection and $\vec{\Omega}={\bf \nabla}_{\bf k}\times\vec{\omega}({\bf k})$ is the Berry curvature, whose integral over the closed Brillouin zone (BZ) surface is a topological invariant. Indeed, by means of the Stokes theorem, we can write 
\begin{equation}
  \sigma_{xy} \ = \ \frac{e^{2}}{h}\frac{1}{2\pi}\int_{\rm MBZ}d^{2}k \vec{\Omega}_z({\bf k})
	            \ = \ \frac{e^{2}}{h}\frac{1}{2\pi} \oint_{c}\vec{\omega}({\bf k}) dk = \frac{e^{2}}{h} n  
\end{equation}
with $n$ the integer Chern number, which entails the conductivity quantization \cite{KOH_AP160}. One notes that the concept of the Berry phase, as an accumulated phase of the wave function undergoing adiabatic evolution, was introduced in 1981 by Michael Berry, and has proven to be ubiquitous in quantum transport phenomena \cite{BER_PRSA392}. The existence of a topological invariant (Chern number) is a direct consequence of the known {\it failure of parallel transport around a closed loop}, which is measured by the Berry phase. The local adiabatic curvature of the bundle of ground states in the parameter space is then defined as the limit of Berry phase mismatch divided by the loop area (see \cref{Avron} for pedagogical introduction).

The Chern number is a topological invariant in the sense that its value is unchanged under small changes in the Hamiltonian (due for instance to disorder perturbation). However, when large deformations of the Hamiltonian are taking place, the wave function ground state can cross over other eigenstates, and such a level crossing triggers transitions between Chern numbers and Hall conductance plateaus. This topological interpretation of the quantized Hall conductivity stands as a major milestone for the understanding of QHE, but has also generated many inspired subsequent developments of topological physics in condensed matter.

\subsection{Fundamentals on Dirac materials} \label{sec:intro_hall_dm}
The electronic transport properties of graphene are known to be very peculiar \cite{CAS_RMP81, Goerbig2011, DAS_RMP83, AVO_NL10, GIA_RNC35, WEH_AP63} with unprecedented manifestations of quantum phenomena as Klein tunneling, weak antilocalization, or anomalous QHE, all driven by the additional degree of freedom (pseudospin) and related $\pi$-Berry phase endowed by the bipartite nature of graphene and its sublattice degeneracy \cite{Roche_Book}. These fascinating properties are robust as long as disorder preserves a long range character and valley mixing is minimized.

The simplest tight-binding (TB) model Hamiltonian for pristine graphene in the absence of SOC considers a single $2p_z$ orbital per atom and first nearest neighbor coupling
\begin{equation}
		H \ = \ -\gamma_0 \ \sum_{\langle i,j \rangle} c_i^\dagger c_j \ ,
\end{equation}
where $\gamma_0\approx 2.7$ eV is the coupling energy, $\langle i,j \rangle$ indicates the couples of neighbor carbon atoms with indices $i$ and $j$, $c^\dagger_i$ is the creation operator for an electron in the $2p_z$ orbital of the atom with index $i$, and $c_j$ is the annihilation operator for an electron in the $2p_z$ orbital of the atom with index $j$.
The carbon atoms are spatially distributed on a triangular lattice and over the two sublattices $A$ and $B$. The corresponding BZ is hexagonal, with two nonequivalent highly symmetric points at its corners, which are called $K$ and $K'$ points.  
The Bloch theorem allows the definition of a $\mathbf{k}$-dependent Hamiltonian, which operates on a vector containing the coefficient for the two Bloch sums corresponding to the $A$ and $B$ sublattices
\begin{equation}
  H(\mathbf{k}) \ = \ \left(\begin{array}{cc} 0 & f(\mathbf{k}) \\[3mm] f(\mathbf{k})^* & 0 \end{array}\right)
	\ \ \ \ \ {\rm with} \ \ \ \ \ 
	f(\mathbf{k})  \ = \   1 \ + \ 2 \ e^{i3ak_y/2}\ \cos(\sqrt{3}ak_x/2)
	\ ,
\end{equation} 
where $a=1.42$ \AA\ is the interatomic distance. The eigenvalues for conduction and valence bands are then
\begin{equation}
  E_\pm (\mathbf{k}) \ = \ \pm \gamma_0 \ \sqrt{3+2\cos(\sqrt{3}ak_x)+4\cos(3ak_y/2)\cos(\sqrt{3}ak_x/2)} \ . 
\end{equation} 
Note that the two bands touch each other at the $K$ and $K'$ points, where the charge neutrality point (CNP) is located. There, the energy dispersion is approximately linear and forms two Dirac cones.
If we develop the Hamiltonian around the $K$ and $K'$ points by considering $\mathbf{k}$ as the displacement from those points and by replacing it with the momentum operator $\mathbf{p}/\hbar=-i\mathbf{\nabla}$, we obtain two effective Hamiltonians
\begin{equation} \label{eq:Hkk}
    H_{K} \ = \ v_{\rm F} \ ( \sigma_x p_x + \sigma_y p_y ) \ \ \ \ \ {\rm and} \ \ \ \ \ H_{K'} \ = \ v_{\rm F} ( \sigma_x p_x - \sigma_y p_y ) \ ,
\end{equation} 
which have the same form of a couple of massless 2D Dirac equations, where the light speed is replaced by the Fermi velocity $v_{\rm F}=\sqrt{3} a \gamma_0 /(2\hbar)$, and the Pauli matrices $\sigma_x$ and $\sigma_y$ act on the sublattice degree of freedom (pseudospin) similarly to the mathematical structure defining the spin degree of freedom. For this reason, the $K$ and $K'$ points are also names Dirac points. The time-reversal invariance of the Hamiltonian is guaranteed by $H_K=H_{K'}^*$.  
The eigenvalues and eigenvectors of \eref{eq:Hkk} are
\begin{equation}
  E_\pm(\mathbf{k}) = \pm \hbar v_{\rm F} |\mathbf{k}|
	\ \ \ \ \ 
	\Psi_{\pm,K}  (\mathbf{r}) = \Frac{e^{i\mathbf{k}\cdot\mathbf{r}}}{\sqrt{2}} \left(\begin{array}{c} 1\\[1mm] \pm e^{i\theta_\mathbf{k}} \end{array}\right)
	\ \ \ \ \ 
	\Psi_{\pm,K'} (\mathbf{r}) = \Frac{e^{i\mathbf{k}\cdot\mathbf{r}}}{\sqrt{2}} \left(\begin{array}{c} 1\\[1mm] \pm e^{-i\theta_\mathbf{k}} \end{array}\right)
\end{equation}
where $\theta_\mathbf{k}$ is the angle of the momentum with respect to the $x$-axis, i.e. $\tan\theta_\mathbf{k}=k_{y}/k_{x}$.
It is worth noting that the pseudo-spinors are also eigenstates of the helicity operator $\vec{h}= \vec{\sigma}\cdot\vec{ p}/(2 |\vec{p}|)$, which defines the chirality of the electrons. This means that the pseudospin polarization is locked to the momentum.
By introducing the valley degree of freedom (also called isospin) and the corresponding $z$-Pauli matrix $\tau_z$, we can recast \eref{eq:Hkk} in a single equation
\begin{equation} \label{eq:Hkktau}
    H \ = \ v_{\rm F} \ \tau_z \vec{\sigma}\cdot\vec{p}  \ ,
\end{equation}
which acts on a 4-component wave functions describing the two valley isospin and two sublattice pseudospin degrees of freedom $[\psi_{A}^{K_{+}} , \ \psi_{B}^{K_{+}} , \ \psi_{B}^{K_{-}} , \ \psi_{A}^{K_{-}} ]$. Note the A and B components are inverted for the $K'$ valley.

\subsection{Spin lifetimes in Rashba spin-orbit-coupled materials} \label{sec:intro_hall_rashba}

In quantum physics, the SOC (also called spin-orbit effect or spin-orbit interaction) is an interaction of a particle spin with the magnetic field induced by particle motion relative to the surrounding electric fields (for an alternative interpretation see \aref{sec:soc}). This is detectable as a splitting of spectral lines, which can be thought of as a Zeeman effect due to the magnetic field in the particle rest frame. In the field of spintronics, spin-orbit (SO) effects for electrons in semiconductors and other materials are explored for technological applications. 

In solids, the atomic SOC splits bands that would be otherwise degenerate. The particular form of this SO splitting (typically of the order of few to few hundred meV) depends on the particular system and broken symmetries. The bands of interest can be then described by various effective models, usually based on some perturbative approach. A 2DEG in an asymmetric quantum well (or heterostructures) will feel the Rashba SOC, which is a momentum-dependent splitting of spin bands similar to the splitting of particles and anti-particles in the Dirac Hamiltonian \cite{RashbaSOC}. The splitting is a combined effect of atomic SOC and asymmetry of the potential in the direction perpendicular to the two-dimensional plane. The understanding of band structure and spin dynamics in the presence of Rashba SOC has been essential for the proposal of spintronic devices such as the Datta-Das spin transistor \cite{DattaDas}, and for the prediction of fundamental physical phenomena such as the intrinsic SHE \cite{SHE2,SHE1}. 

From a practical point of view, understanding the relaxation mechanisms and spin lifetimes in clean materials is a prerequisite to realizing spintronic devices, since they determine the upper time and length scales on which spin devices can operate.
In Rashba SO-coupled materials, the spin lifetime is commonly dictated by the Dyakonov-Perel (DP) mechanism \cite{dp1}, where SOC triggers the spin precession of charge carriers. 
The DP mechanism is an efficient mechanism of spin relaxation due to SOC in systems lacking inversion symmetry. Examples of materials without inversion symmetry include semiconductors from groups III-V (e.g. GaAs) or II-VI (e.g. ZnSe), where inversion symmetry is broken by the presence of two different atoms in the Bravais lattice. Electron spins process along an effective magnetic field that depends on the momentum. At each scattering event, the direction and frequency of the precession changes randomly. After many scattering events, the randomization of the precession leads to dephasing and a loss of the spin signal, thus resulting in a spin relaxation time $\tau_s^{\rm DP} \propto 1/\tau_p$ that is inversely proportional to the momentum scattering time $\tau_p$. 
This scaling behavior contrasts with the Elliot-Yafet (EY) mechanism \cite{ELL_PR96}. 
The EY mechanism has been derived for spin relaxation in metals, and relates the spin dynamics with electron scattering off impurities or phonons. Each scattering event changes the momentum, with a finite spin-flip probability, which is derived by a perturbation theory (assuming weak spin-orbit scattering). This gives rise to weak antilocalization phenomena in the low temperature regime, and to a typical scaling behavior of the spin relaxation time with momentum relaxation as $\tau_s^{\rm EY}\sim  \tau_p$.
The properties of Rashba SOC allow the manipulation of spin states by electrostatic means, thus making it possible to perform elementary operations and paving the way towards non-charge-based computing and information processing technologies \cite{MAN_NM14}. Beyond traditional III-V semiconductor quantum wells -- such as InAs, InGaAs, or InSb -- 2D graphene and monolayers of ${\rm MoS_2}$ and other group-VI dichalcogenides have recently raised a lot of interest. In addition to their predicted long spin lifetimes \cite{ERT_PRB80,HUE_PRL103,ZHO_PRB82,OCH_PRB87,WAN_PRB89}, the possibility to harness proximity effects or to couple the spin and valley degrees of freedom makes these materials very interesting from both a fundamental and a technological perspective \cite{XIA_PRL108, OCH_PRB87, graphene_prox1}.

The nature of spin relaxation in graphene has been fiercely debated since the first paper by van Wees and coworkers \cite{Tombros2007}, and initial theoretical predictions of millisecond spin lifetimes (see \cref{ROC_JPD47} for a discussion). Additionally, following what was known for metals and semiconductors, the two different EY type and the DP mechanisms were first considered in graphene \cite{ZUT_RMP76,FAB_APS57}. Some theoretical derivation in monolayer graphene (taking into account the Dirac cone physics) proposed a revision of the scaling behavior of the spin lifetime as $\tau_s\sim \epsilon_F^2\tau_p/ \lambda_R^2$, which would remain of the EY-type \cite{OCH_PRL108}, but such a prediction has failed to explain experimental data. Besides, this result was derived assuming the absence of intervalley scattering and perturbative effect of the SOC. 
All such approximations are incapable to explain experimentally observed smallness ($\sim 100$ ps) of spin relaxation time in graphene. 
The effect of magnetic impurities was also proposed to jeopardize long spin-diffusion lengths, but the predicted EY mechanism does not consistently account for all experimental observations \cite{Kochan_2014,Soriano_2015}.

\Cref{gmitrahydogenatedgraphene} has proposed that small spin relaxation time in graphene could be explained by resonant scattering off local magnetic moments. A completely different mechanism---spin-pseudospin entanglement \cite{DIN_NP10}---which is specific to graphene was unveiled by quantum simulations of spin dynamics in disordered graphene with Rashba SOC due to substrate-induced electric field or small density of metallic adatoms (such as gold or nickel).
Such a mechanism results from the entangled dynamics of spin and pseudospin degrees of freedom, which is particularly predominant close to the CNP \cite{DIN_NP10, VanTuan2016sr,Cummings2016}. This is actually first evidenced in the peculiar form of low-energy eigenstates, which write $\Psi \sim (1,0)^{\rm T}\times|\!\downarrow\rangle \pm i (0,1)^{\rm T} \times |\!\uparrow\rangle$, where $|\!\!\downarrow\rangle$ and $|\!\!\uparrow\rangle$ denote the spin state, whereas (1,0) and (0,1) the pseudospin state \cite{RAS_PRB79}. Such a spin-pseudospin locking effect results in a fast spin dephasing, even when approaching the ballistic limit \cite{DIN_NP10}, with increasing spin lifetimes away from the CNP, as observed experimentally \cite{GUI_PRL113}. 
Depending on the type of substrate (silicon oxide or h-BN), the impact of disorder (through electron-hole puddles) was found to yield either a DP or an EY-type of spin relaxation. 
Finally, this unique phenomenon offers opportunities to create spin manipulation by controlling the pseudospin degree of freedom, which could be useful in the development of spin logics \cite{DER_TED59,KAV_2DM3,ROC_2DM3,VanTuan2016}.

%%%%% QUANTUM HALL EFFECT IN CLEAN AND DISORDERED GRAPHENE %%%%%%%%%%%%%%%%%%%%%%%%%%%%%%

\section{Quantum Hall effect in clean and disordered graphene} \label{sec:qhe}

\begin{figure} [t!]
  \begin{center}
  \resizebox{9cm}{!}{\includegraphics{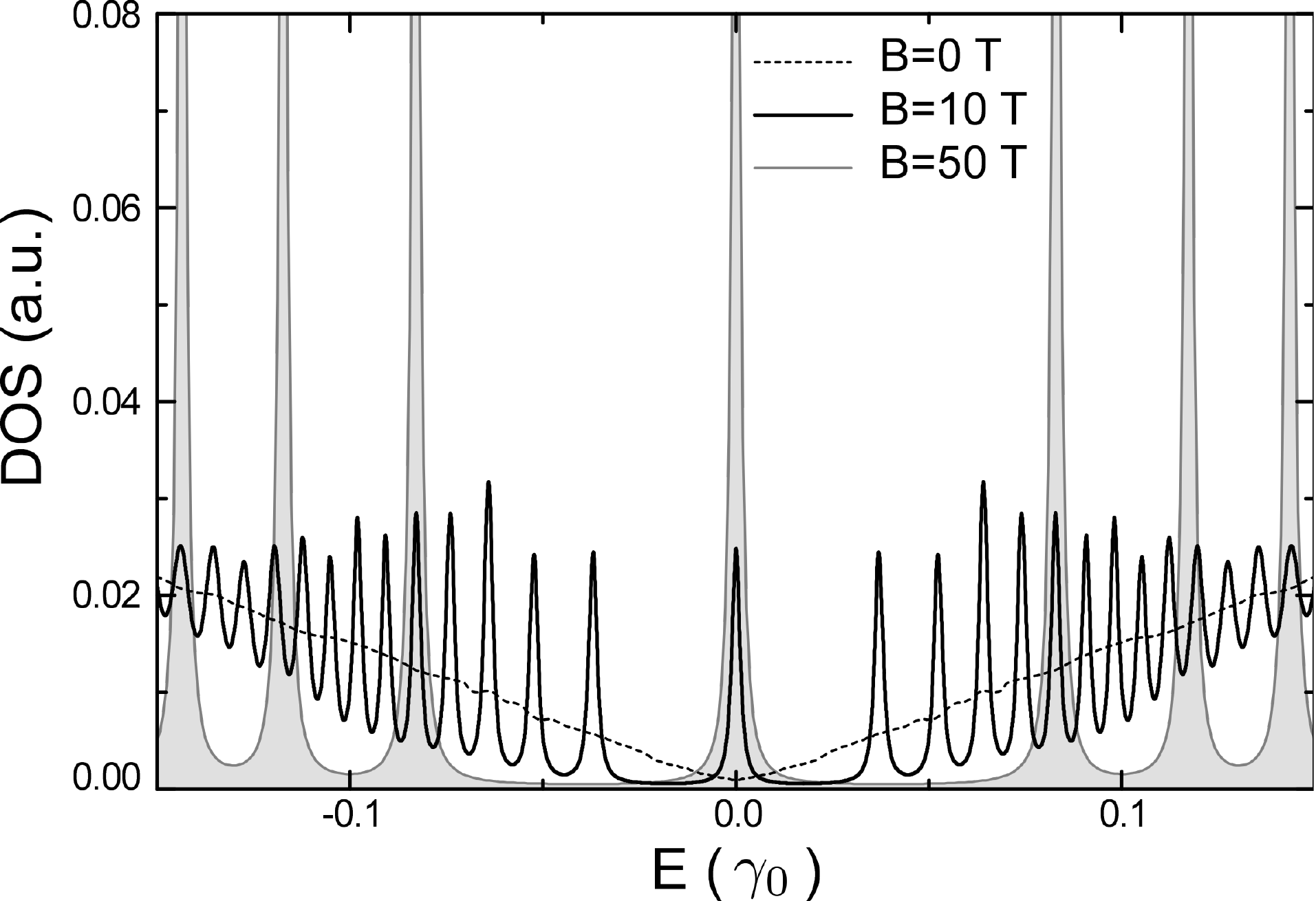}}
  \caption{Density of states showing the LLs in pristine graphene in the absence of magnetic field (dashed lines), for $B=10$ T (black solid line) and $B=50$ T (grey solid line). The DOS is given in arbitrary units.}
  \label{FigQHE00}
\end{center}
\end{figure}

The massless Dirac fermion nature of electronic excitations in monolayer graphene is manifested through the formation of non-equidistant Landau levels (LLs) in the high magnetic field regime \cite{Goerbig2011,McClure1956}. 
The eigenvalues for graphene under magnetic field are given by $E_{\pm n} = \pm (\hbar v_F/\maglen) \sqrt{2n}$ with $\pm$ denotes the levels with positive and negative energy, and $\maglen=\sqrt{\hbar/eB}$ is the magnetic length \cite{Goerbig2011}. 
As a matter of illustration, the LL spectrum for magnetic fields of 10 and 50 T are shown in \fref{FigQHE00}. 
Note that the term \emph{relativistic} is used to distinguish the $\sim\sqrt{Bn}$ dispersion of the levels from that of the conventional (non-relativistic) LLs, which disperse linearly in $Bn$. 
A remarkable difference with respect to non-relativistic LLs in metals (with parabolic bands) is the presence of a zero-energy LL with $n=0$. 

Owing to the peculiar nature of the LL spectrum, the well-known integer \QHE\ (IQHE) \cite{KLI_PRL45} observed in conventional two-dimensional electron systems transforms into a relativistic half-integer (anomalous) QHE in graphene, whose quantized Hall conductivity becomes $\sigma_{xy}= 4e^{2}/h\times(n+1/2)$ \cite{Goerbig2011,Novoselov2005a,Zhang2005}.

\begin{figure}[t!]
\begin{center}
\resizebox{9cm}{!}{\includegraphics{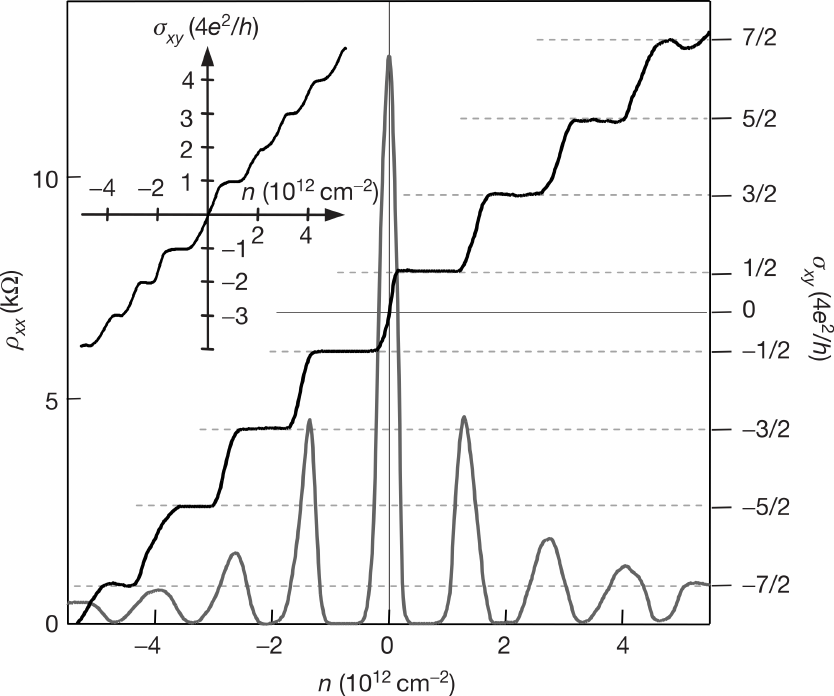}}
\caption{Longitudinal resistivity $\rho_{xx}$ and Hall conductivity $\sigma_{xy}$ as a function of charge density in monolayer graphene at 14 Tesla and 4 K. The inset shows the case of bilayer graphene. Reprinted with permission from \cref{Novoselov2005a}, Macmillan Publishers Ltd.}
\label{FigQHE0} 
\end{center}
\end{figure}

Such an anomalous QHE was simultaneously reported in the groups of Manchester University \cite{Novoselov2005a} and Columbia University \cite{Zhang2005}. \Fref{FigQHE0} shows both the charge density dependence of the longitudinal resistivity ($\rho_{xx}$) and Hall conductivity ($\sigma_{xy}$) at 14 Tesla and 4K \cite{Novoselov2005a}. Quantized plateaus of the Hall conductivity have been also reported at room temperature and low magnetic fields \cite{Novoselov2007}.

In what follows, we will illustrate some aspects of the QHE in clean and disordered graphene from the complimentary points of view of two-terminal and 2D systems.

\subsection{Two-terminal magnetoconductance and edge currents}  \label{sec:qhe_2t}

\begin{figure}[b!]
  \begin{center}
  \resizebox{9cm}{!}{\includegraphics{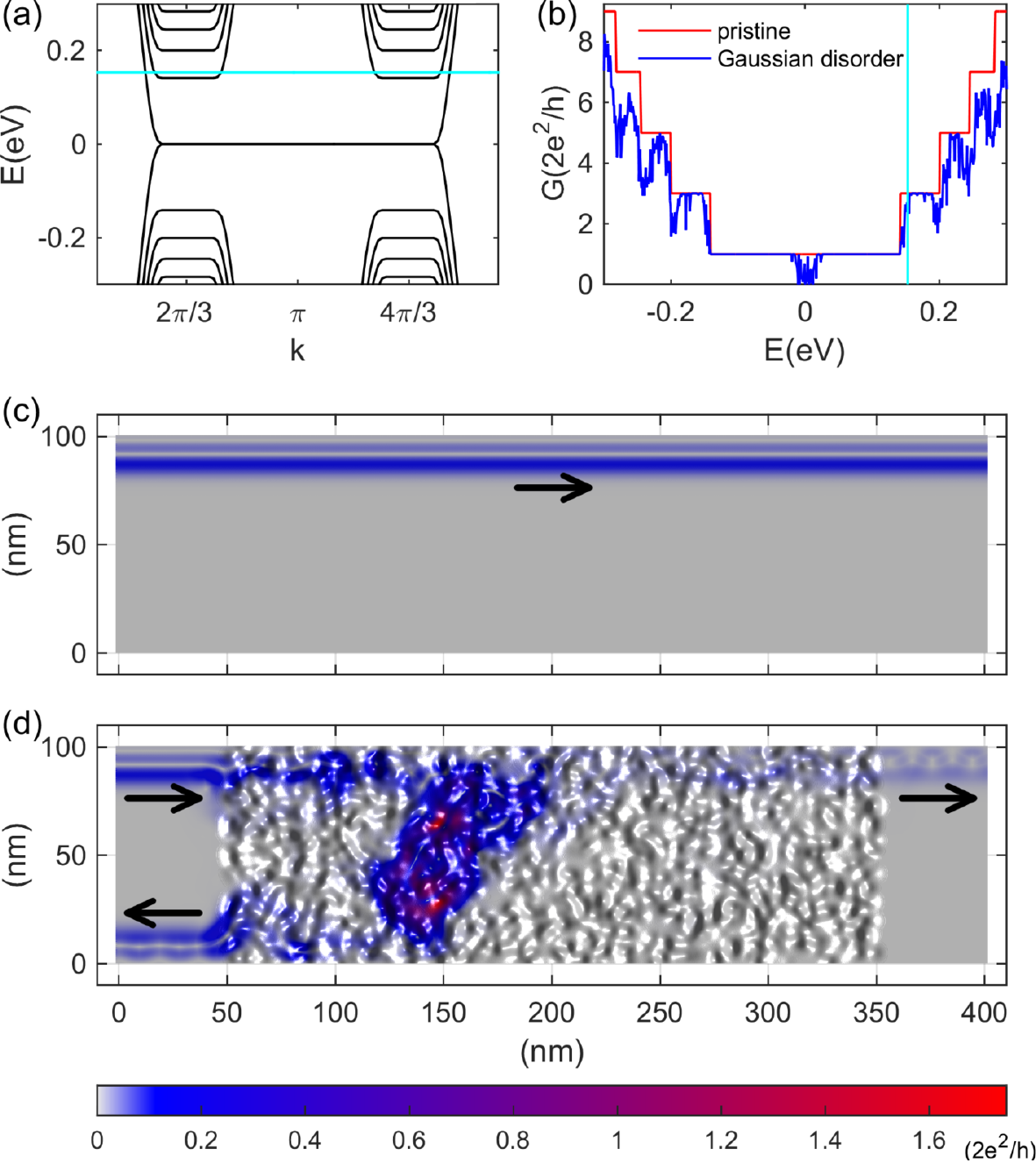}}
  \caption{
	             (a) Band structure of a 100 nm wide ZGNR under a perpendicular homogeneous magnetic field of 20 T. We observe the sequence of LLs and the valley degeneracy structure. 
							 (b) Zero temperature differential conductances of the pristine ribbon and of the defected ribbon. Gaussian disorder with concentration $n=5\times 10^{12}$ cm$^{-2}$, maximum strength 100 meV and range 2 nm, over a 300 nm long section. 
							 (c) Spatial distribution of the spectral current for the pristine ribbon at $E=0.153$ eV, as indicated by a cyan line in panels (a) and (b). 
							 (d) Spatial distribution of the spectral current at $E=0.153$ eV for the disordered ribbon. The correspondent conductivity is about 1.4 $(2e^2/h)$. The low relief in the disordered region indicates the local potential profile generated by Gaussian impurities.}
  \label{fig:fig_iqhe_zigzag}
  \end{center}
\end{figure}

We start by scrutinizing the regime of strong magnetic fields in ultraclean graphene nanoribbons (GNRs). 
In the absence of any disorder, the localization/delocalization transition is provoked by the edges of the samples, which break the symmetry of the system and allow for delocalized edge states (skipping orbits), in contrast with the magnetic-field-induced localized states of the bulk.
 
Before showing the results, we briefly recall the TB description of graphene under magnetic field.  
We consider a first-nearest neighbor TB Hamiltonian with a single $2p_z$ orbital {\it per} carbon atom
\begin{equation}\label{eq:TBH}
			H \ = \ -\sum_{\langle ij \rangle} \gamma_{ij} \ c^\dag_i c_j 
			   \ \ \ \ {\rm with} \ \ \ \ 
			\gamma_{ij} \ = \ \gamma_0 \ \exp\left[\Frac{e}{\hbar c} \ \int_{\vec{r}_i}^{\vec{r}_j} d\vec{r}\cdot \vec{A}(\vec{r})\right] \ ,
\end{equation} 
where $c_i^\dag$ and $c_i$ are the creation and annihilation operators for electrons on the $2p_z$ orbital of the carbon atom with index $i$, $\langle ij \rangle$ indicates couple of indices corresponding to first neighbor atoms, $\gamma_{ij}$ is the coupling parameter, which is proportional to the coupling $\gamma_0=2.7$ eV and to the Peierls phase factor, whose argument is given by the integral of the vector potential $\vec{A}$ along the straight line connecting the positions of the two atoms. 
Such a description is applicable as long as the magnetic length remains much larger than the atomic spacing $a$. This is actually the case for experimentally accessible magnetic fields, since $a/\maglen\approx 0.005 \ \sqrt{B{\rm [T]}}$. 
Note that the Peierls phase depends on the chosen gauge, however the product of phases along a closed circuit is invariant and its argument is given by the magnetic flux through the encircled area over the elementary magnetic flux $\hbar c/e$.  

In our two-terminal simulations, we consider graphene ribbons that are infinitely extended along the transport direction $\hat{x}$. 
In the presence of a uniform magnetic field $\vec{B}=B\hat{z}$, we can conveniently choose the vector potential in the first Landau gauge $\vec{A}=-By\hat{x}$. 
With such a gauge, and if the ribbon is pristine, the translation invariance along $\hat{x}$ is preserved, thus allowing the band structure calculation. 
As an example, \fref{fig:fig_iqhe_zigzag}(a) shows the bands of a 100 nm-wide zigzag graphene nanoribbon (ZGNR) under a magnetic field $B=20$ T. 
We can observe the two-valley structure, which is preserved by the zigzag edge geometry, as well as a series of flat LLs with four-fold degeneracy due to valley and spin. 
The dispersive bands correspond to edge states with opposite group velocity, whose direction and intensity is proportional to the energy band derivative with respect to the wave number. 
This means that, for energies in between the LLs, electrons move in one direction along the top edge of the ribbon and in the opposite direction along the bottom edge, i.e. they are \emph{spatially chiral} and no state is available in the bulk. 
As shown later on, this phenomenon is at the origin of the IQHE robustness.

To simulate two-terminal electron transport, we consider a standard configuration where the system under investigation is connected to two periodic contacts (source and drain) at different chemical potentials. 
All the results presented in this review refer to quasi-equilibrium conditions, where the source-drain potential difference is very small, as it usually occurs in magnetotransport experiments. The details of the nonequilibrium Green's function (NEGF) approach adapted to calculate the Landauer-B\"uttiker (LB) conductance, the local density of states (DOS) and density of occupied states, and the local spectral currents are presented in \aref{sec:qtalgorithms}.
 
Let us consider again our 100 nm-wide pristine ZGNR. 
Its zero-temperature differential conductance is reported in \fref{fig:fig_iqhe_zigzag}(b). 
As expected, the conductance is quantized in units of $2e^2/h$, and the height of each plateau depends on the number of active bands at the considered energy. 
For example, at $E\approx 150$ meV, indicated by the cyan lines in \ffref{fig:fig_iqhe_zigzag}(a,b), there are 3 spin-degenerate active bands and then the conductance is $3\times2e^2/h$. 
The corresponding local distribution of spectral currents, obtained from \eref{eq:lcur}, is shown in \fref{fig:fig_iqhe_zigzag}(c). 
We observe that the current, injected from the left source contact, flows toward the right drain contact along the top edge of the ribbon. 
The width of the edge channel is proportional to the magnetic length $\ell_{\rm B}\approx 25.7/\sqrt{B}$ nm.
A conductive channel where electrons flow in the opposite direction is present at the lower edge. 
However, it is empty, since electrons are injected from the left and thus only occupy the channel that allows them to move from left to right. 
This is why the bottom edge channel is not visible in the figure. 
As long as these two channels do not come into contact (i.e. they keep the spatial chirality), the backscattering is suppressed and the conductance is quantized. 
The large spatial separation between them thus explains the robustness of the \QHE.
To better clarify this point, let us introduce disorder in the system. We consider a random distribution of impurities with Gaussian potential profile over a 300 nm-long section. 
The concentration of impurities is $n=5\times10^{12}$ cm$^{-2}$, their maximum strength is 100 meV and they have a spatial range of 2 nm.
The resulting conductance is reported in \fref{fig:fig_iqhe_zigzag}(b). 
We can see that, especially at low energies, the conductance remains quantized over large energy ranges. 
This means that the spatial deviation of the current due to disorder is not large enough to break spatial chirality. 
On the contrary, at the energies where the conductance is not quantized, electrons are (partially) deviated from the top edge to the bottom one. 
Such a picture is confirmed by the local spectral currents shown in \fref{fig:fig_iqhe_zigzag}(d), where we observe part of injected electrons penetrating into the bulk and reaching the bottom edge. As a consequence, only part of the current is transmitted to the drain along the top edge, while the rest is backscattered to the source along the bottom edge.

\begin{figure}[t!]
  \begin{center}
  \resizebox{9cm}{!}{\includegraphics{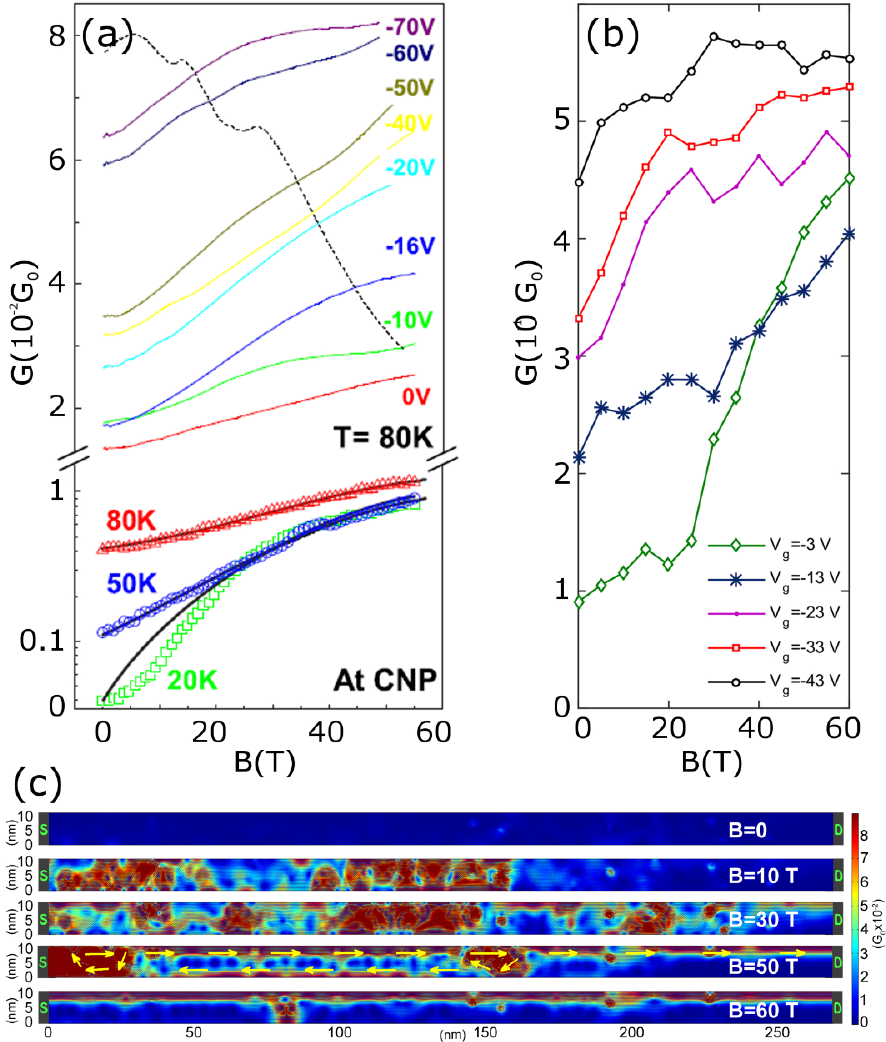}}
  \caption{(a) Experimental magnetoconductance at 80 K for the 10 nm wide GNR for various gate potential values $V_{\rm g}$ (top); and at 80, 50, and 20 K for $V_{\rm g}$=0 (curves with symbols: measurements; solid lines: simulated data). The magnetoconductance for a 90-nm-wide GNR is also shown, varying from 0.6$G_0$ to 0.4$G_0$ (dashed curve). 
	(b) Simulated magnetoconductance for a 10 nm wide AGNR at different $V_{\rm g}$ with Gaussian impurities. 
	 (c) Corresponding spectral current distribution at the energy $E$=200 meV in the presence of Gaussian at with $B$ varying from 0 to 60 T. Charge flows from left to right.
	Adapted from \cref{POU_PRB82}.}
  \label{fig:narrow_ribbon_prb82} 
  \end{center}
\end{figure}

When the graphene ribbon is narrow, also the spatial separation between counter propagating conductive channels is small. 
Therefore, a minimum amount of disorder suffices to break spatial chirality and induce backscattering. 
Such a situation is investigated by Poumirol and coworkers in \cref{POU_PRB82}, where the magnetoconductance of a chemically derived 10 nm-wide graphene ribbon is measured under fields up to 60 T and for different back gate potentials. 
The experimental results, shown in \fref{fig:narrow_ribbon_prb82}(a), clearly indicate a positive magnetoconductance, which, however, is far from being quantized. 
When the magnetic field increases, the magnetic length decreases, thus tending toward a progressive separation of the edge channels. 
Even though spatial chirality is never reached due to the narrow ribbon section, this phenomenon explains the positive magnetoconductance. 
\Fref{fig:narrow_ribbon_prb82}(b) shows the simulation of magnetoconductance for a 10 nm-wide armchair graphene nanoribbon (AGNR) in the presence of Gaussian potential impurities and for different back gate potentials, i.e. for different chemical potentials as calculated from the capacitive coupling between back gate and graphene. 
The Gaussian potential impurities mimic charged impurities trapped in the substrate. 
As observable in the figure, such a model fairly reproduces the experimental trend. 
Disorder of different nature (edge roughness, for example), entails a different behavior, thus indicating that the transport properties of the sample are dominated by charged impurities. 
The spatial distribution of the spectral current, at a representative energy $E=200$ meV, clarifies the mechanisms at work. 
\Fref{fig:narrow_ribbon_prb82}(c) shows the evolution of the current when increasing the magnetic field from 0 to 60 T. 
Up to $B\approx 30$ T, backscattering is maximum and the current spreads almost homogeneously over the ribbon. 
At higher fields the formation of edge channels is more visible, as indicated by the arrows in the case $B=50$ T.
This recalls what already observed in \fref{fig:fig_iqhe_zigzag}(d) for the larger ribbon. 
Finally, at $B=60$ T, the formation of the top edge channel is evident, though backscattering is still present and spatial chirality is not achieved. 

The results illustrated above clarify the origin of the integer quantum Hall regime and emphasize the role of the spatial chirality of the edge currents. 
This concept will often come up in the rest of the paper.

\subsection{Quantum Hall effect in disordered graphene and Kubo Hall conductivity} \label{sec:qhe_kubo}

The effect of disorder on the localization phenomena in the quantum Hall regime is central to understand the integer QHE from a bulk perspective, where $\sigma_{xy}$ plateaus develop by varying the charge density in a region coinciding with a vanishing $\sigma_{xx}$ \cite{Aoki87}. Historically, the phase diagram of $(\sigma_{xy},\sigma_{xx})$ was explained in terms of a localization/delocalization transition of the wave functions driven by the competition between magnetic and disorder length scales. Given that LLs need to be formed to observe QHE, weak disorder can be treated as a perturbation on top of the LL spectrum, and it was found that the states are much more robust at the center of LLs and remain mostly delocalized in the whole system, in contrast to states in between LL centers, which are typically localized. Actually, the onset of $\sigma_{xy}$ quantization is connected to the formation of mobility edges separating extended from localized states. Inside the plateau, the state localization is then consistent with $\sigma_{xx}\sim 0$ (but also the resistivity $\rho_{xx}\sim 0$), in the absence of bulk dissipative transport \cite{Aoki87}.

The most standard method to treat bulk Hall conductivities is the linear-response theory developed by Ryogo Kubo \cite{KUB_JPSJ6}, and applied for the first time to QHE by Aoki and Ando \cite{AokiSSC81}. Bulk quantities can actually be connected to measurements in the Corbino geometry for which electrodes (coaxial contacts) are attached to the inner and outer perimeters of a disk-shaped sample in which the current flows radially from an inner contact to an outer ring contact. This geometry not only eliminates any unknown edge effects that might interfere with determining the Hall conductance, but is also insensitive to the formation of the known quantized edge conductance of other filling factors, thus enabling us to directly probe the bulk conduction \cite{Corbino}.

Understanding the anomalous features of QHE in disordered graphene is a very challenging issue. Indeed, surprisingly, QHE in graphene seems very robust to very large disorder, as for instance that introduced by hydrogenation in \cref{Guillemette:PRL:2013}. Furthermore, QHE in graphene presents additional peculiarities depending on the symmetry breaking aspects conveyed by defects. In weakly oxidized graphene or in the presence of structural vacancies, the formation of disorder-induced resonant critical states in the zero-energy LL is observed, as evidenced by lower bound values of the dissipative conductivity $\sigma_{xx}$ and a zero-energy $\sigma_{xy}$ quantized plateau in between resonances \cite{LEC_2DM1,LEC_PRB93}, see \sref{sec:qhe_vac}.

Exploring Hall transport in disordered systems of large size and for moderate magnetic fields demands for real space order-$N$ computational methods. Recently, two approaches where developed to calculate $\sigma_{xy}$ conductivity, one by Garcia and coworkers based on the kernel polynomial method (KPM) \cite{Garcia2015, Garcia2016} and a second one due to Ortmann and coworkers based on time-dependent propagation methods \cite{Ortmann2013,Ortmann2015,Roche1999}. The advantage of these methods relies on the ability to compute the conductivity tensor without having to know all the eigenstates of the system, which would be prohibitive for the large systems required to simulate disordered systems under moderate magnetic fields \cite{AokiSSC81}.

The general framework is provided by Kubo linear response theory \cite{KUB_JPSJ6} in which the dc conductivity is given as 
\begin{equation}\label{KubosConductivity}
\sigma_{\alpha\beta}(\omega=0)=\lim_{}\frac{1}{V}\int_{0}^{\infty}dt  e^{-s t}\int_0^\beta d\lambda
\TR\left[\rho_0 j_\beta(0) j_\alpha(t+i\hbar\lambda) \right]
\end{equation}
where $V$ is the volume of the sample, $\rho_0$ is the equilibrium density matrix and $j_\alpha(\tau)$ is the current operator in the $\alpha$ direction, evaluated at $\tau=0$ and $\tau=t+i\hbar\lambda$ in the formula. A simple derivation of the Kubo formula, together with the different representations used in this review, is presented in  \aref{sec:kubo-formula}. 

\begin{figure}[t!]
  \begin{center}
  \resizebox{12cm}{!}{\includegraphics{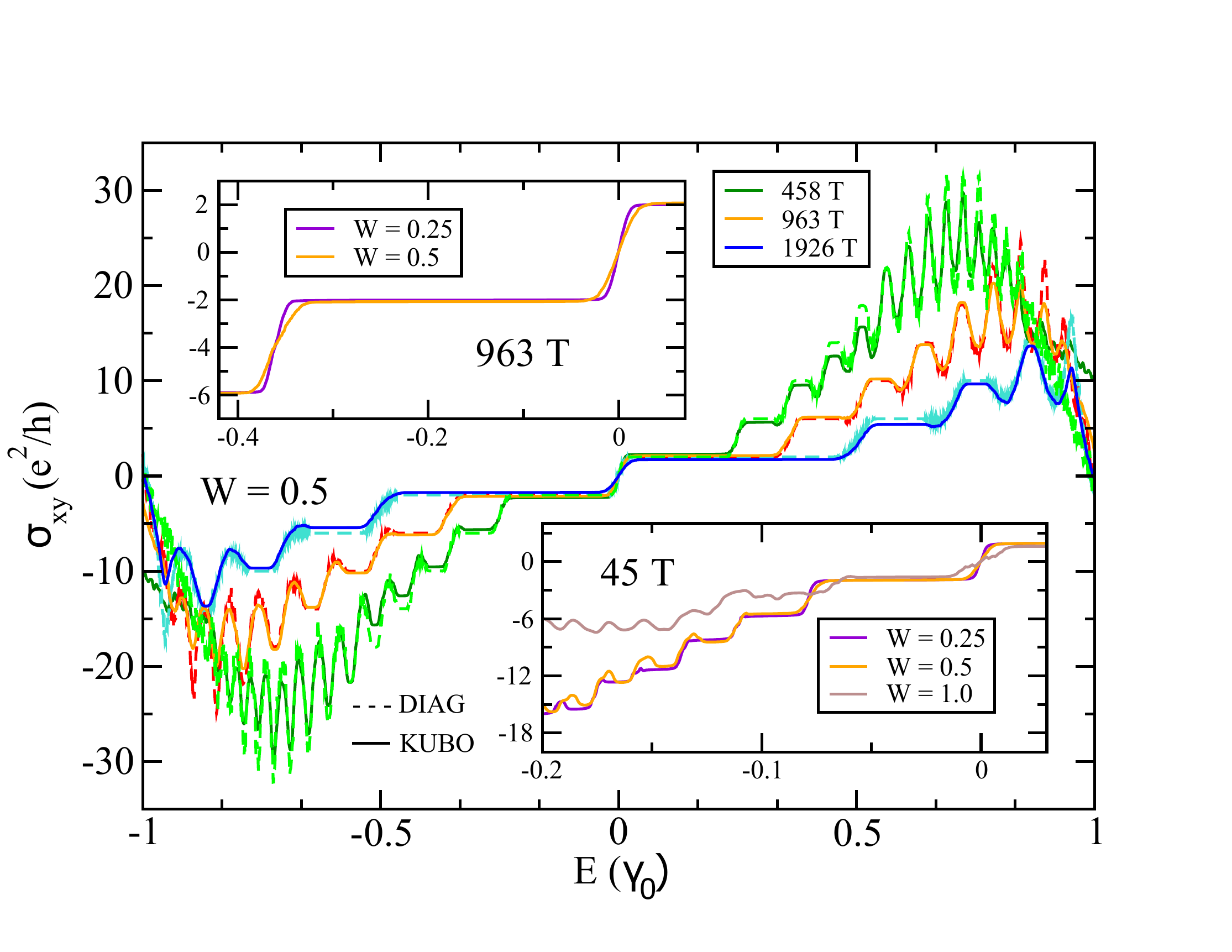}}
  \caption{Hall conductivity for Anderson disorder (dashed curves: exact diagonalization; solid curves: real space $\sigma_{xy}$ at different magnetic fields (main frame). Same information for larger disorder for high field 963 T (upper inset) and intermediate fields of 45 T (lower inset). Adapted from \cref{Ortmann2015}.}
  \label{FigQHEr} 
  \end{center}
\end{figure}

For disordered systems, where the correlations are small compared to the disorder scale, one can focus only on the non-interacting regime, where the density matrix $\rho_0$ can be fully described by two parameters: the temperature $T$ and the chemical potential $\mu$. One representation that can be exploited using time-evolution method is
\begin{equation} \label{sigma_ab}
\sigma_{\alpha,\beta}=\frac{1}{V}  \int_{0}^{\infty}dt  \int_{-\infty}^{\infty}d\varepsilon f(\varepsilon)\text{Tr}\left[ \delta(\varepsilon-H) j_\beta  \frac{1}{\varepsilon - H+ i\eta }j_\alpha(t) + h.c \right] \ ,
\end{equation}
where $\eta$ is a small convergence parameter, which is starting point  of the real space implementation developed in  \cite{Ortmann2013,Ortmann2015} . This formula was used to calculate some results for graphene with Anderson disorder, which are presented in \fref{FigQHEr}, where the robustness of quantized Hall plateaus is studied as a function of magnetic and disorder strengths.

The study of Anderson disorder is indeed a traditional reference for metallic systems,  allowing the simple investigation of the transition from the QHE regime to the conventional Anderson insulating state by varying a single parameter $W$, while keeping the initial LL spectrum unchanged by disorder (beyond broadening). \Fref{FigQHEr} indicates an excellent agreement (within few percent) between the real space Kubo algorithm and exact diagonalization techniques at low energy for high magnetic fields \cite{Ortmann2015}. 

To illustrate the effect of increasing disorder, a zoom on the first two Hall steps is provided for B=$963$ T in the upper inset of \fref{FigQHEr}. The LLs are broadened with $W$ around the critical states, while the first and second plateaus at $\pm 2 e^2/h$ and $\pm 6 e^2/h$ remain unaffected.  It is clear that by increasing disorder broadening, one progressively suppresses the LL integrity, which results in the disappearance of the Hall plateaus, starting from high energy levels, which are closer in energy.

One sees that the physics is identical (lower inset) for a more realistic magnetic field ($45$ T). This illustrates the performance of the algorithm to probe low magnetic fields and the possibility to destroy quantization at higher energy for $W=\gamma_0$. For such a disorder, only the zero-energy LL fully develops, while all the other states become localized.

\subsection{Effect of disorder grain boundary in polycrystalline graphene} \label{sec:qhe_gb}

As mentioned in the introduction, the QHE in graphene is characterized by a large energy separation between LLs. 
As a consequence, in graphene Hall bars, the transverse resistance quantization can be observed even at relatively high temperature and low magnetic fields. 
Graphene is thus an ideal material to replace the GaAs-based metrological standard of resistivity. 
Thanks to the development of the chemical vapor deposition (CVD) technique, it is now possible to fabricate very large Hall bars made of monolayer graphene, thus incontrovertibly opening the doors to metrological applications, as recently verified experimentally \cite{RIB_NN10}. 
However, despite its general high quality, CVD graphene is intrinsically polycrystalline. 
As shown below, this may induce dissipative transport \cite{LAF_PRB90}, thus putting at risk the quality of the Hall resistance quantization.

\begin{figure}[b!]
  \begin{center}
  \resizebox{9cm}{!}{\includegraphics{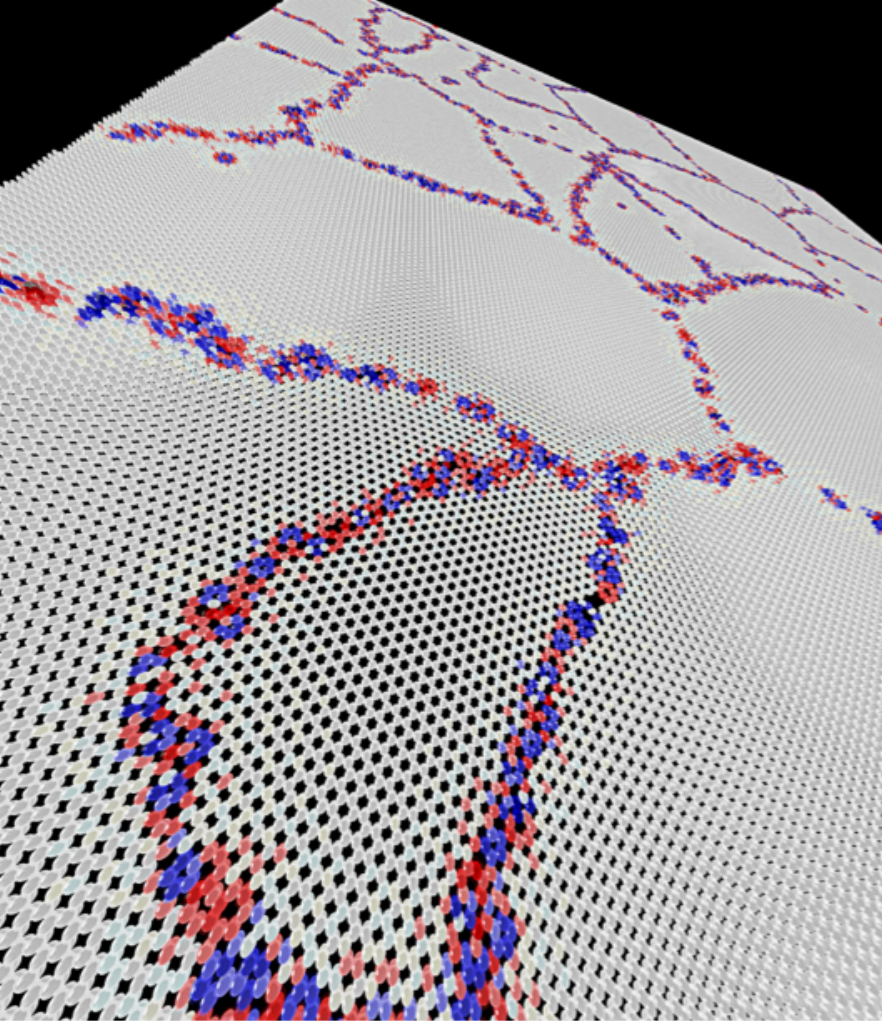}}
  \caption{Ball-and-stick model of a polycrystalline sample, with GBs manifested by red/blue colored atoms, which pictured local electron/hole local doping. Courtesy of Jani Kotakoski (University of Vienna, Austria).}
  \label{fig:puddles} 
  \end{center}
\end{figure}

The particularly complex and tunable morphology of polycrystalline graphene, made of a distribution of grains with varying size and orientation, interconnected by irregular grain boundaries (GBs), which appear as one dimensional dislocations full of odd-membered rings defects, opens challenging questions concerning the formation of LLs and the conditions for the QHE observation. \Fref{fig:puddles} shows a typical structural model of a polycrystalline sample with various grains of different orientations connected {\it via} GBs, which mainly contain pentagonal and heptagonal rings \cite{DinhPolyX}. In \cref{DinhPolyX}, the mean free path was found to upscale with the average GB sizes ($\ell_{e}\sim d_{\rm grain~size}$), with mobility in the order of $\mu \sim 3\times 10^5 \ {\rm cm}^{2}/(Vs)$ for grain size of about 1$\mu$m and charge density $n=3\times10^{11}{\rm cm}^{-2}$.

QHE has been investigated experimentally in low mobility polycrystalline graphene irregularly decorated with disordered multilayer graphene patches \cite{Nam2013}, and in CVD grown monolayer samples \cite{LAF_PRB90}. In \cref{LAF_PRB90}, the temperature and magnetic field dependence of the longitudinal conductance are found to follow smooth power-law scaling, which is incompatible with variable range hopping or thermal activation, and indicates the existence of extended or poorly localized states at energies between LLs \cite{LAF_PRB90}. Such a phenomenon has been theoretically understood in \cref{CUM_PRB90}.

\begin{figure}[t!]
  \begin{center}
  \resizebox{9cm}{!}{\includegraphics{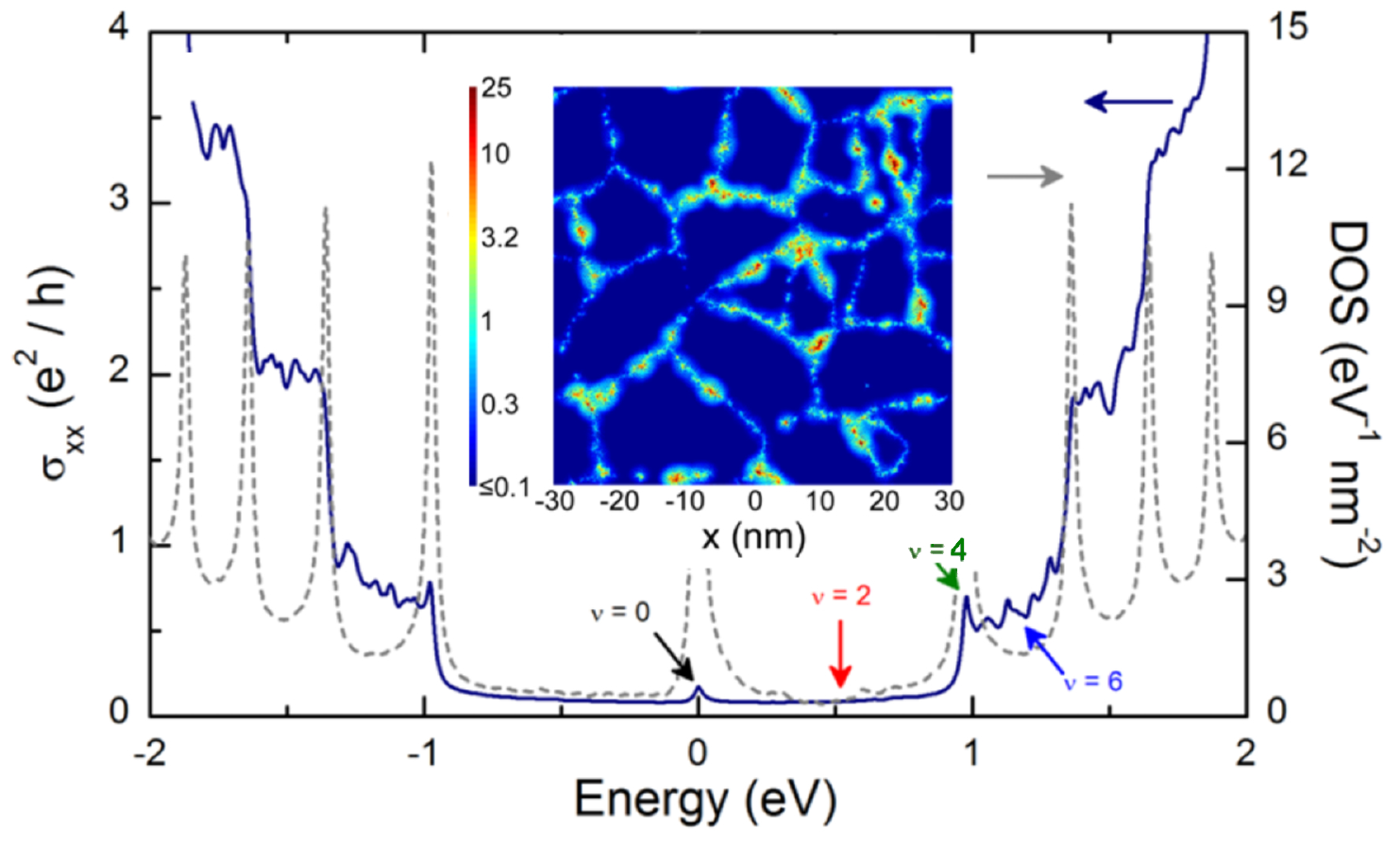}}
  \caption{Kubo conductivity (left axis) and superimposed DOS (right axis) of a polycrystalline sample with 15 nm average grain diameter and $k$ = 9, where $k$ is the ratio between the average grain diameter and the magnetic length. The conductivity has been calculated at a simulation time of 10 ps. Inset: Local DOS for an energy of 0.5 eV located between the zero-energy LL and the first LL. In both panels, $k$ = 9. Adapted from \cref{CUM_PRB90}.}
  \label{fig:polyX_prb90a} 
  \end{center}
\end{figure}

To simulate the electronic properties of large-area model of disordered polycrystalline graphene samples, containing hundreds of thousands atoms and morphology defined by varying grain misorientation angles, realistic carbon ring statistics, and unrestricted GB structures, a real space method is here again compulsory \cite{DinhPolyX}. A TB Hamiltonian and an efficient quantum transport method particularly well-suited for large samples of disordered low-dimensional systems can be used.  High-magnetic field quantum Kubo conductivity $\sigma_{xx}(E,B)$ is investigated with an order-$N$, real space approach \cite{Roche1999}. The scaling properties of $\sigma_{xx}$ are computationally followed through wave packets dynamics from $\sigma(E,t)=e^{2}\rho(E)\Delta X^{2}(E,t)/t$, where $\rho(E)$ is the DOS and $\Delta X^{2}(E,t) = Tr\left[\delta(E-\hat{H})\left|\hat{X}(t)-\hat{X}(0)\right|^2\right]/Tr\left[\delta(E-\hat{H})\right]$ (with $\hat{X}(t)$ the position operator in Heisenberg representation) is the energy- and time-dependent mean quadratic displacement of the wave packet. Calculations are performed on systems containing up to ten millions carbon atoms, which corresponds to sizes larger than $500\times 500~{\rm nm}^{2}$. In all of our simulations, the energy smearing factor is 13 meV. 

The main frame of \fref{fig:polyX_prb90a} shows the longitudinal conductivity $\sigma_{xx}$ (solid line) of a polycrystalline sample, superimposed with the total DOS (dashed line).  In contrast to other forms of disordered graphene \cite{Sheng2006}, the energy dependence of $\sigma_{xx}$ does not reflect that of the LL spectrum. In particular, the conductivity is suppressed at the LLs, while it remains finite and constant between LLs. This situation is opposite that of the conventional QHE, for which states at the center of LLs are robust against localization, while bulk states beyond the mobility edges all become localized, thus enabling both a quantized Hall conductivity and a longitudinal conductivity that qualitatively resembles the DOS.

The nature of the states at and between LLs is revealed by their time-dependent behavior. By scrutinizing the diffusion coefficient $D(t)$ for energies at the center of two LLs (marked by arrows at $\nu = 0$ and $4$) and energies between LLs (at $\nu = 2$ and $6$), the localized nature of the states at the center of the LLs is clear from the fast decay of $D(t)$ \cite{CUM_PRB90}. On the contrary, $D(t)$ exhibits a weak time-dependent decay between LLs, which is typical of extended states in the weak localization regime. This behavior connects to the finite value of $\sigma_{xx}$ between LLs, where the current is conveyed by states that propagate through the GB network as pictured in the inset of \fref{fig:polyX_prb90a}. 

The dissipative conductivity at the CNP has actually been the subject of intense study. In the absence of magnetic fields, experiments indicate the possible existence of some quantum critical states at the CNP with a finite conductivity that is insensitive to localization effects down to zero temperature \cite{Bolotin2008}. Under strong external transverse magnetic fields, finite values of the conductivity at the CNP are further predicted and suggested the existence of critical states \cite{Ortmann2013,Gattenlohner2014} and robustness of the QHE \cite{Sheng2006,Evers2008}. Recently, the observation of a quantized Hall conductance in highly resistive hydrogenated graphene, with mobility less than 10 cm$^2$V$^{-1}$s$^{-1}$  and mean free path far beyond the Ioffe-Regel limit \cite{Guillemette:PRL:2013}, or in low mobility polycrystalline graphene irregularly decorated with disordered multilayer graphene patches \cite{Nam2013}, further confirmed some amazing robustness of QHE in strongly disordered graphene. Here, we have shown that polycrystalline geometries, at the origin of zero-energy states, strongly jeopardize the formation of the QHE (for grains smaller than the magnetic length), and also suppress the robustness of the zero-energy conductivity, in total contrast with smooth disorder potentials \cite{Ortmann2013}.

Even for polycrystalline graphene with large grains, we expect that electrons can flow along the GB network thus entailing dissipative transport. However, the presence of disorder can localize the states at the GBs, thus {\it de facto} interrupting the network and partially restoring dissipationless transport. Indeed, experimentally \cite{LAF_PRB90}, despite the presence of GBs, the quality of the Hall quantization is found to be much better than expected from the simulation results above. This means that the conductive network along the GBs is indeed partially neutralized by additional disorder.   

\begin{figure}[t!]
  \begin{center}
  \resizebox{9cm}{!}{\includegraphics{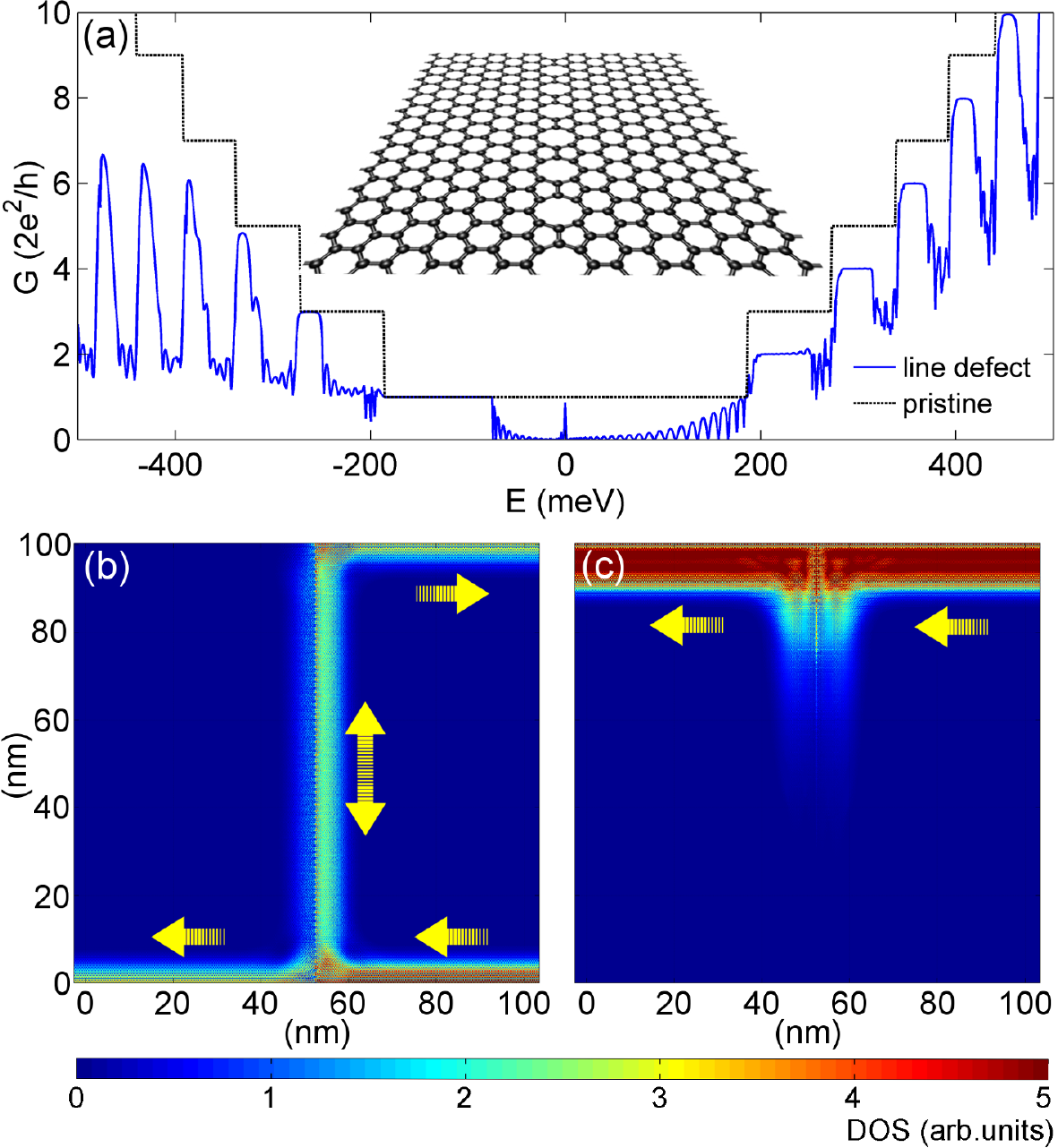}}
  \caption{(a) Transmission coefficient vs energy $E$ of a 100 nm wide AGNR with a transverse GB consisting of pentagonal and octagonal rings, shown in the inset, at B = 40 T. (b) Spatial distribution of the right-injected local DOS at $E$ = 140 meV. (c) Same as (b) at $E$ =-262 meV.
Adapted from \cref{CUM_PRB90}.}
  \label{fig:polyX_prb90b} 
  \end{center}
\end{figure}

Electron transport in the two-terminal configuration confirms the 2D picture and proves a further insight in the mechanism that breaks the spatial chirality of the edge currents.
\Fref{fig:polyX_prb90b}(a) shows the conductance of a 100 nm wide AGNR as a function of the electron energy in the presence and in the absence of a transverse line defect constituted of pentagonal and octagonal rings, and in the presence of a 40T magnetic field. 
Over a large region of the spectrum, the quantization observed for the pristine ribbon is affected by the line defect. 
Note that the graphene sublattice symmetry is broken by the presence of odd numbered (pentagonal) rings, which entails the electron-hole asymmetry observed in the figure. 
For a large energy range within the first conductance plateau, a series of periodic oscillations appear. Their frequency is inversely proportional to the width of the ribbon (not shown here), i.e. the length of the line defect. This suggests that they are generated by electrons travelling back and forth along the line defect itself.

\begin{figure}[t!]
  \begin{center}
  \resizebox{9cm}{!}{\includegraphics{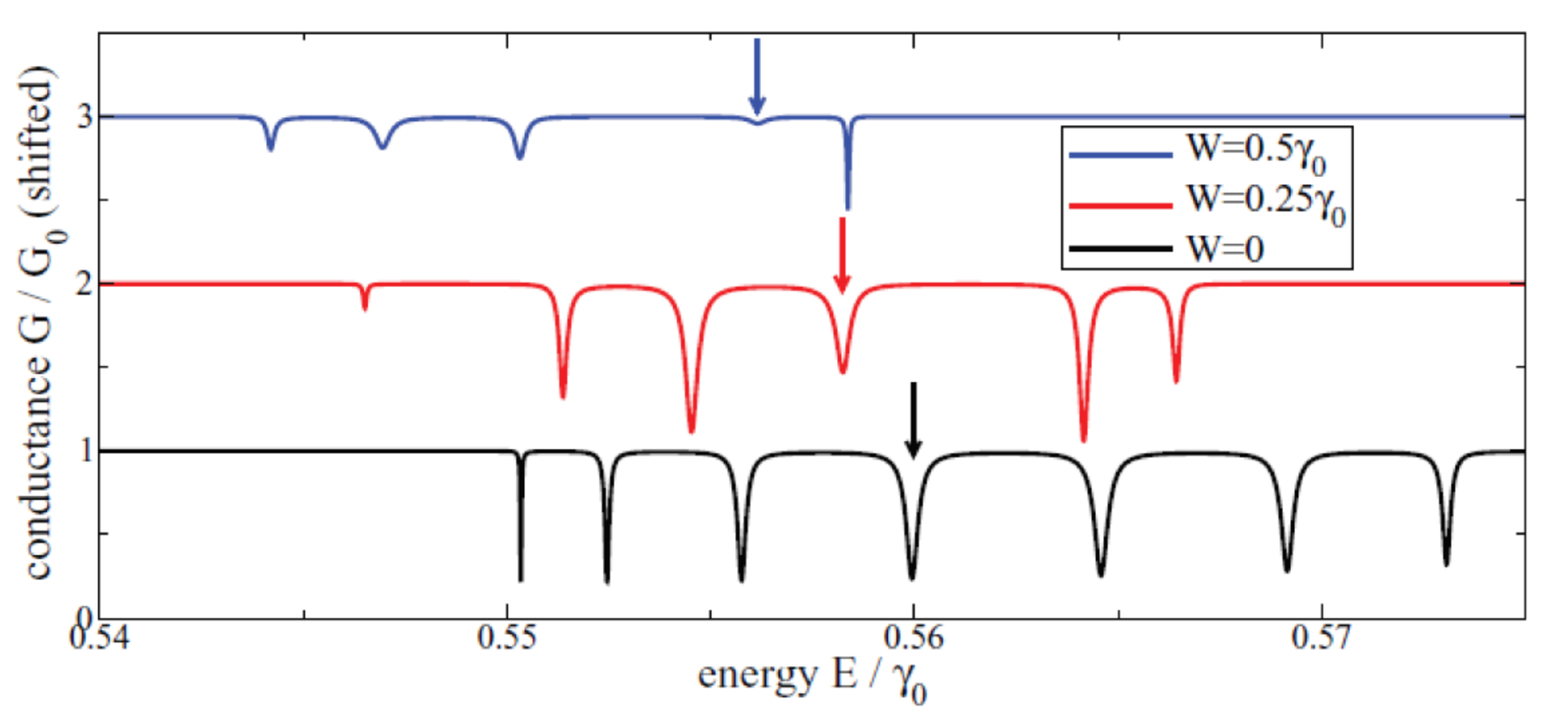}}
  \caption{Evolution of the conductance in of a ribbon with a GB and increasing Anderson disorder with strength $W$. Adapted with permission from \cref{BER_PRB91}, American Physical Society.}
  \label{fig:ber_prb91} 
  \end{center}
\end{figure}

By looking at the spatial distribution of the occupied density of states, it is possible to visualize the mechanism underlying the breakdown of the Hall regime. 
\Fref{fig:polyX_prb90b}(b) shows the distribution at $E=140$ meV. The electrons are injected from the right and flow along the bottom edge. When reaching the line defect, they are partially transmitted to the left still along the bottom edge, and partially deviated along the line defect before reaching the top edge and being backscattered. This ``short-circuit'' mechanism clearly shows how GBs can entail dissipative transport, in agreement with the 2D picture. 
The ribbon with the line defect can be seen as the union of two semi-infinite ribbons along whose edges the chiral currents flow clockwise. At the junction between the semi-infinite ribbons, i.e. along the line defect, two chiral channels with opposite group velocity are present. Therefore, the spatial chirality of current is largely lost along the line defect, where electrons can flow in both directions. This explains the periodic oscillations observed in \Fref{fig:polyX_prb90b}(a) and suggests that the inter-edge channel can be easily localized in the presence of additional disorder. 
This picture is confirmed in the literature \cite{LAF_PRB90,BER_PRB91}. In \fref{fig:ber_prb91}, the conductance of a graphene ribbon in the presence of Anderson disorder along the line defect is shown for a small energy interval along the first Hall plateau \cite{BER_PRB91}. Disorder progressively localizes the states along the defect, thus reducing backscattering and partially restoring quantization.

As seen in \fref{fig:polyX_prb90b}(a), for some energy ranges the conductance is not affected by the presence of the line defect. In this case, the penetration of electrons along the nonchiral channels is limited, and the current does not reach the opposite ribbon edge. This is seen in \fref{fig:polyX_prb90b}(c) for $E=$-262 meV. Note that, compared to \fref{fig:polyX_prb90b}(b), the chirality is here opposite, since we are in the hole region of the spectrum.

An analogous mechanism of spatial chirality breaking has been predicted \cite{L_fwander_2013} and observed \cite{Yager_2013} in epitaxial SiC graphene, where bilayer regions behave as metallic patches that can connect opposite edges.

\subsection{Anomalous quantum Hall effect} \label{sec:qhe_vac}

\begin{figure}[t!]
  \begin{center}
  \resizebox{14cm}{!}{\includegraphics{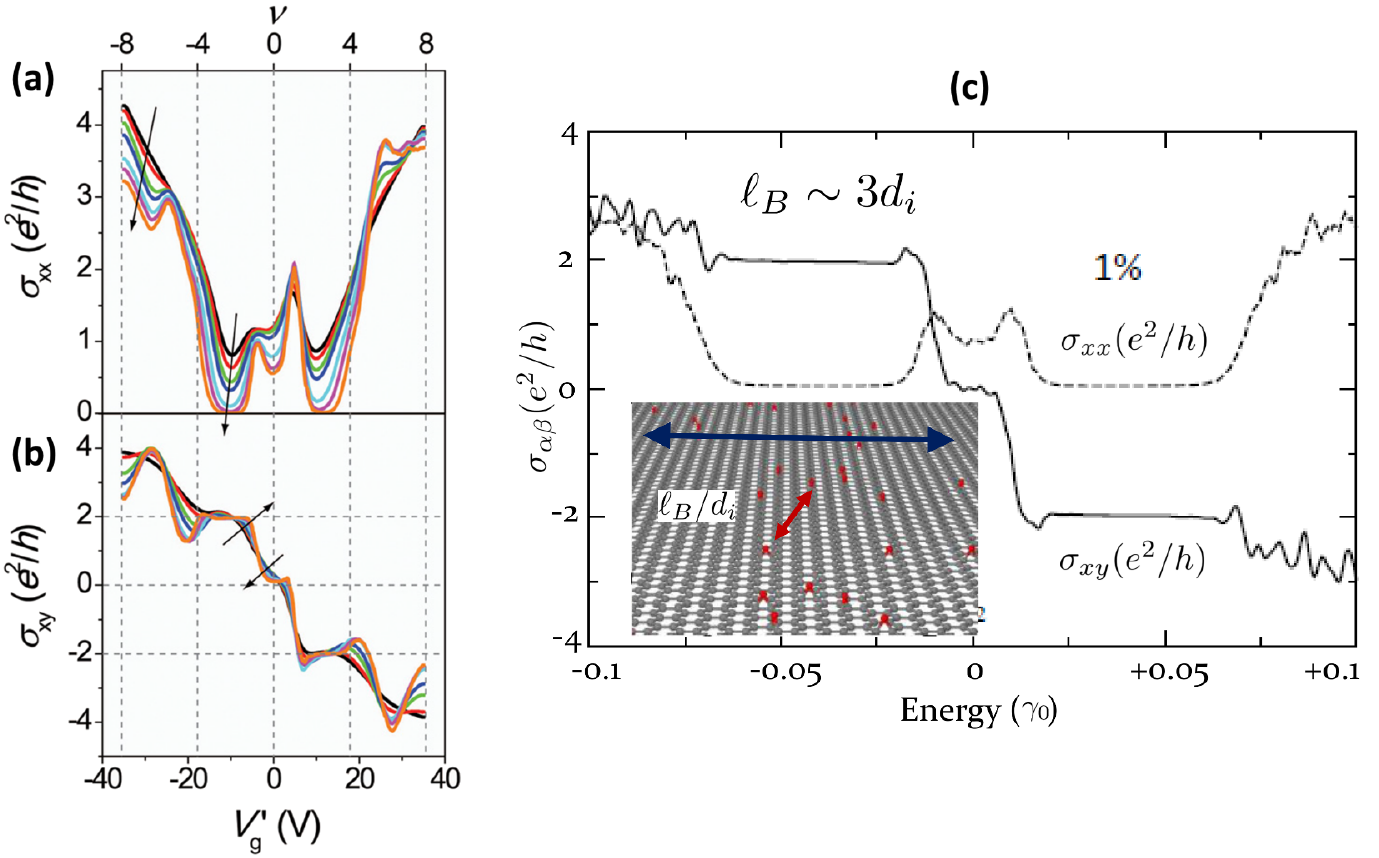}}
  \caption{(a) The longitudinal conductivity $\sigma_{xx}$ and (b) Hall conductivity $\sigma_{xy}$ as functions of the shifted gate voltage $V_{\rm g}'= (V_{\rm g} -V_{\rm CNP})$ for $T$=150, 100, 50, 30, 10, 5, and 2 K in the magnetic field $B=13$ T. The arrows indicate the changes corresponding to lowering temperature. The vertical dashed lines represent filling factors corresponding to $\nu=0, \pm 4, \pm 8$. Reprinted with permission from \cref{NAM_APL104}, AIP Publishing. (c): longitudinal conductivity $\sigma_{xx}$ (dashed lines) and Hall conductivity $\sigma_{xy}$ (solid lines) for $1\%$ of epoxy defect density. Courtesy of Nicolas Leconte (University of Seoul, Korea). 
}
  \label{fig:vacancies_qhe_1A} 
  \end{center}
\end{figure}

Experiments on disordered graphene with low mobility have reported surprising features such as a double peak in the dissipative conductivity and some onset of a zero-energy quantized Hall conductance, see \ffref{fig:vacancies_qhe_1A}(a,b), which cannot be easily interpreted by LL splitting since disorder is too strong and mixes valleys \cite{NAM_APL104}. 

\begin{figure}[h!]
  \begin{center}
  \resizebox{9cm}{!}{\includegraphics{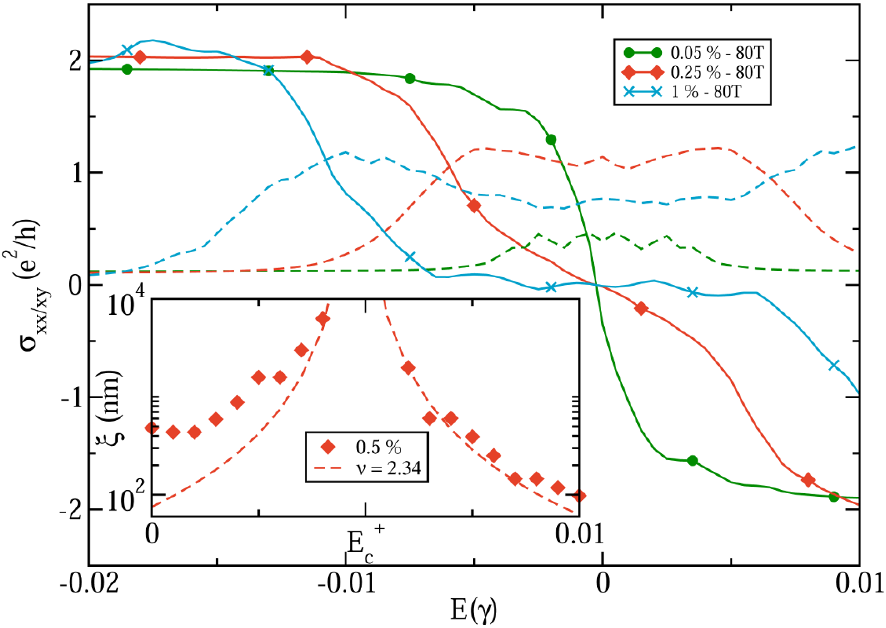}}
  \caption{$\sigma_{xx}$ (dashed lines) and $\sigma_{xy}$ (solid lines) for double vacancies at 80 T. inset: Estimated localization lengths for 0.5\% at 80 T with theoretical critical exponential ($\nu =2.34$) decay around $E^{+}_c$. Adapted from \cref{LEC_PRB93}, by courtesy of Nicolas Leconte (University of Seoul, Korea).}
  \label{fig:vacancies_qhe_1B} 
  \end{center}
\end{figure}

Such an occurrence of a zero-energy plateau in the presence of strong disorder is truly puzzling since it calls for revisiting the understanding of conductance quantization in terms of topological invariants in disordered materials.  Here, we show that similar unconventional magnetotransport fingerprints in the quantum Hall regime (with applied magnetic fields from one to several tens of Tesla) can occur in chemically or structurally disordered graphene. \Fref{fig:vacancies_qhe_1A}(c) shows a low-energy double-peaked conductivity for $1\%$ of adsorbed monoatomic oxygen on graphene, which results from the formation of critical states conveyed by the random network of defects-induced impurity states.  In conjunction with the double peak $\sigma_{xx} (E)$, the onset of a quantized Hall conductivity plateau $\sigma_{xy} (E)=0$ emerges, similarly to the experimental measurements of \ffref{fig:vacancies_qhe_1A}(a,b). Such features are observed when the magnetic length is larger than the typical distance ($d_{i}$) between impurities (here $\ell_{B}/d_{i}=3$), and have been extensively discussed in \cite{LEC_PRB93}.  When $\ell_{B}/d_{i}\ll 1$, these features disappear, as seen in \fref{fig:vacancies_qhe_1B}. 

The conductivity $\sigma_{xx} (E)$ is further shown in \fref{fig:vacancies_qhe_1B} for impurity densities of $0.05\%$, $0.25\%$, $1\%$ at B = 80 T \cite{LEC_PRB93}. In the inset, we observe how the localization length of electronic states diverges in the vicinity of some resonant energy ($E_{c}$), which correspond to one peak value of $\sigma_{xx} (E)$. It turns out that $\sigma_{xx} (E=E_{c})$ is scale-independent (does not vary with system length), thus suggesting that critical states are formed at $E_{c}$. In \cite{LEC_PRB93}, we discussed the origin of such a {\it critical delocalization} as the result of a percolation between local impurity states in real space \cite{LEC_PRB93}. At other energies (above or below $E_{c}$), $\sigma_{xx} (E)$ is found to decay with the system length, thus pointing towards localized states. Accordingly, the computed double-peaked conductivity together with the onset of a zero-energy Hall conductance plateau, very similar to the experimental features reported in \ffref{fig:vacancies_qhe_1A}(a,b), indicate a genuine metal-insulator transition, which is unique to disordered graphene.  

\begin{figure}[t!]
  \begin{center}
  \resizebox{9cm}{!}{\includegraphics{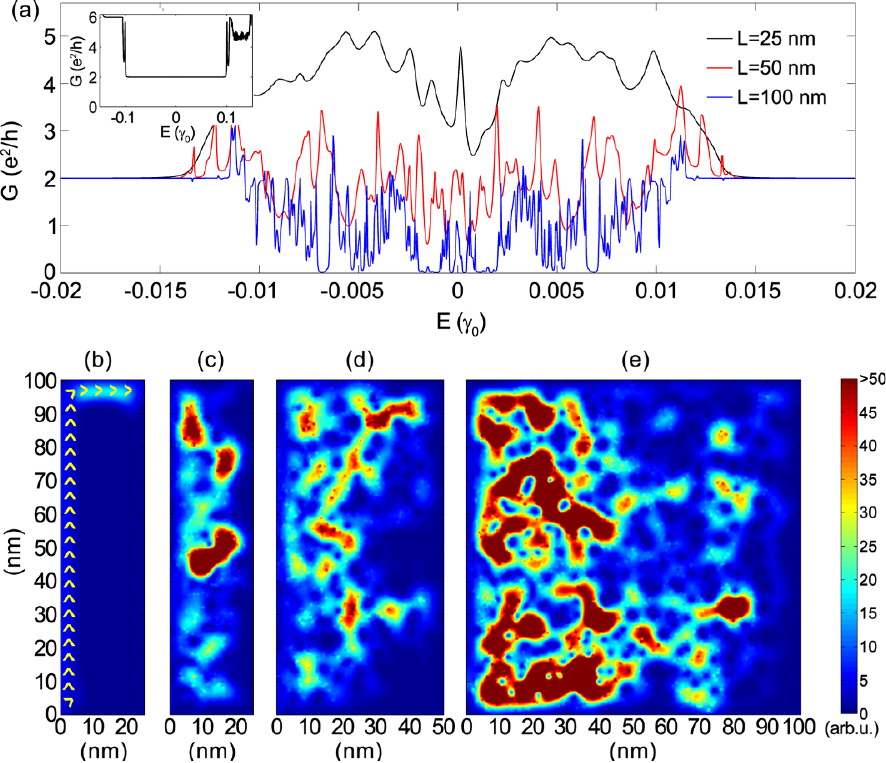}}
  \caption{(a) Conductance of the two-terminal system with a density $n$=0.5\% of double vacancies over a ribbon of length $L$ for 25 nm to 100 nm. Inset: pristine case for $L=25$ nm. (b) Spatial distribution of the spectral current at $E=0.008$ $\gamma_0$ for the pristine ribbon with $L=25$ nm. The arrows indicate the current direction. (c-e) Same as (b) for double vacancy density of 0.5\% and $L=25$ nm, 50 nm and 100 nm. All simulations at 80 T. Adapted from \cref{LEC_PRB93}.}
  \label{fig:vacancies_qhe_2}
  \end{center}
\end{figure}

The two-terminal simulations of large graphene ribbons with a random distribution of double vacancies and under high magnetic field, as reported in \cref{LEC_PRB93}, better illustrate the percolative behavior of electrons.
\Fref{fig:vacancies_qhe_2}(a) shows the conductance of a 100 nm wide graphene ribbon with a density 0.5\% of double vacancies over a section of the ribbon with length $L$ varying between 25 and 100 nm. The source and drain contacts, at the side of the disordered region, are doped so to inject many electrons, and a magnetic field of 80 T is considered.

For $L=25$ nm and in the absence of disorder, the conductance is quantized to $2e^2/h$ around $E=0$, see the inset of \fref{fig:vacancies_qhe_2}(a). 
As anticipated, in this case the current entering from the source is deviated to the top edge, where it flows without being scattered to the drain, as shown in \fref{fig:vacancies_qhe_2}(b). 
When including disorder, states are expected to form around the double vacancies with an extension comparable to the magnetic length, as observed in the 2D simulations and in the literature \cite{PER_PRB78}. 
For $L=25$ nm, the conductance increases above $2e^2/h$. Such a behavior, which may appear counterintuitive, is due to the formation of bulk conductive channels created by the coupling of states localized around the vacancies. The current distribution of \fref{fig:vacancies_qhe_2}(c) clearly shows this phenomenon. When increasing the length of the disordered section to $L=50$ nm and $L=100$ nm, backscattering increases due to the lengthening of the percolative path that electrons have to travel before reaching the drain contact, see \ffref{fig:vacancies_qhe_2}(d,e). This determines a progressive decrease of the conductance with $L$. In qualitative agreement with the 2D simulations, for $L=100$ nm, transport is strongly suppressed around $E=0$ and two energy windows of more extended states are present at the sides of this point.

%%%%% QUANTUM SPIN HALL EFFECT IN GRAPHENE WITH ENHANCED SPIN-ORBIT COUPLING %%%%%%%%%%%%

\section{Quantum spin Hall effect in graphene with enhanced spin-orbit coupling} \label{sec:qshe}

In 2005, Kane and Mele predicted the existence of the QSHE in graphene due to intrinsic SOC \cite{KAN_PRL95,KAN_PRL95b}.
Within the QSHE, the presence of SOC, which can be understood as a momentum-dependent magnetic field coupling to the spin of the electron, results in the formation of chiral edge channels for up-spin up and down-spin electron population. The observation of the QSHE is however prohibited in clean graphene owing to vanishingly small intrinsic SOC of the order of $\mu$eV \cite{YAO_PRB75} (see \aref{sec:soc}), but demonstrated in strong SO-coupled materials (such as CdTe/HgTe/CdTe quantum wells or bismuth selenide and telluride alloys), giving rise to the new exciting field of TIs \cite{BER_PRL96}.
Recent proposals to induce a topological phase in graphene include functionalization with heavy adatoms \cite{WEE_PRX1,JIA_PRL109}, proximity effect with TIs \cite{JIN_PRB87}, or intercalation and functionalization with 5$d$ transition metals \cite{HU_PRL109}.

In particular, the seminal theoretical study \cite{WEE_PRX1} by Weeks and co-workers has revealed that graphene endowed with modest coverage of heavy adatoms (such as indium and thallium) could exhibit a substantial band gap and QSH fingerprints (detectable in transport or spectroscopic measurements). For instance, one signature of such a topological state would be a robust quantized two-terminal conductance ($2e^{2}/h$), with an adatom density-dependent conductance plateau extending inside the bulk gap induced by SOC \cite{WEE_PRX1,QIA_PRB82}. To date, such a prediction lacks experimental confirmation, despite some recent results on indium-functionalized graphene that have shown a surprising reduction of the CNP resistance with increasing indium density \cite{COR_ACR46}, although other attempts have been unsuccessful in revealing anomalous transport phenomena \cite{JIA_PRB91}. On the other hand, it is known that adatoms deposited on graphene inevitably segregate, forming islands rather than a homogeneous distribution \cite{SUT_SS17}, which significantly affects most transport features \cite{MCCREARY2010}.
A detailed description of the origin of the different types of SOC in graphene and the corresponding TB Hamiltonians are reported in \aref{sec:soc}. 

\subsection{Homogeneous distribution of heavy adatoms} \label{sec:qshe_adatoms}

In this section, we illustrate the model Hamiltonian proposed by Weeks and coworkers \cite{WEE_PRX1} to describe graphene in the presence of heavy adatoms.
Heavy atoms, such as indium and thallium, preferably sit on the hollow site of graphene.  
Due their high atomic number, they show a strong intrinsic SOC, which makes them an ideal mean to induce an effective SOC in graphene.

The Hamiltonian terms corresponding to the adatom orbitals can be removed by including an energy-dependent self-energy and thus obtaining an all-graphene renormalized Hamiltonian, where only the carbon $2p_z$ orbitals appear. 
In the low-energy limit, such a Hamiltonian can be approximated by \eref{eq:kanemele}, with $V_\mathrm{R} =V_\mathrm{PIA} =0$.

This is equivalent to having two independent Hamiltonians for spin-up and spin-down electrons, which allows us to unambiguously define spin currents. Indeed, it can be shown that a moderate Rashba SOC, with strength weaker than the topological gap, does not sensibly affect the phenomena illustrated below.

\begin{figure}[t!]
  \begin{center}
  \resizebox{13cm}{!}{\includegraphics{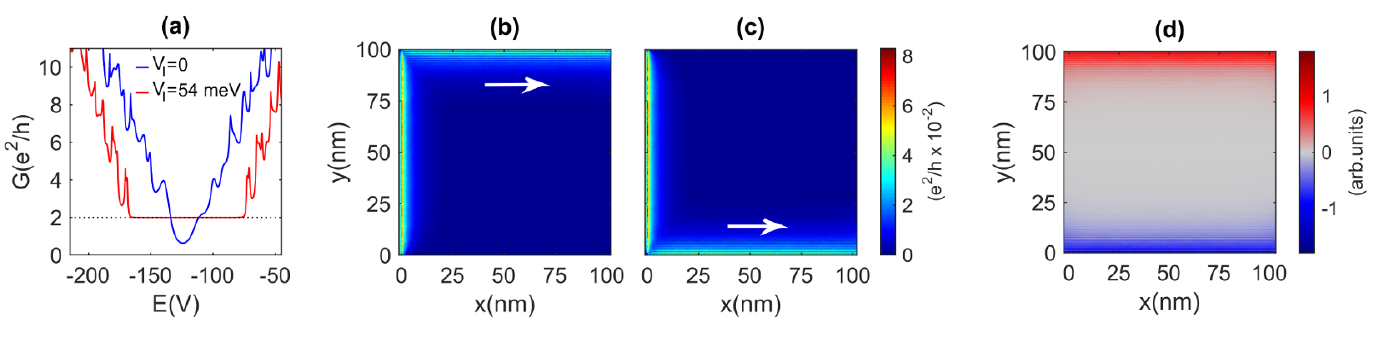}}
  \caption{(a) Differential conductance of a 100 nm wide ribbon with a homogeneous distribution of thallium adatoms with concentration $n_\mathrm{ad}=15\%$ over a 100 nm long section, in the presence and in the absence of SOC. (b) Spectral spin current distribution at $E=-100$ meV for spin up electrons injected from the left contact. (c) Same as (b) for spin down electrons. (d) Polarized density of occupied states at $E=-100$ meV. The color is blue for spin down and red for spin up electrons.}
  \label{fig:weeks_hom}
  \end{center}
\end{figure}

When adatoms are scattered over graphene with a high enough concentration \cite{WEE_PRX1}, we expect that the resulting effective SOC induces a quantum spin Hall phase in graphene, with the raise of counter propagating spin-polarized edge channels and quantized conductance \cite{WEE_PRX1,QIA_PRB82,QIA_PRL107}.
Let us consider a 100 nm wide graphene ribbon with a homogeneous distribution of thallium adatoms ($V_{\rm I}=54$ meV and $\mu$=-270 meV) with concentration $n_\mathrm{ad}=15\%$ over a length of 100 nm. The leads are electrically doped by adding an energy potential of -2.5 eV, thus mimicking source and drain contacts with a high DOS. 
\Fref{fig:weeks_hom} (a) shows the differential conductance as a function of the electron energy in such a system when $V_{\rm I}=0$, i.e. SOC is switched off, and $V_{\rm I} =54$ meV.  
First, we observe that, in both cases, the CNP is shifted to $E\approx -120$ meV. This is due to the doping effect of the adatoms, which is given by the concentration of atoms on the involved carbon rings ($3n_\mathrm{ad}$) times the energy shift ($\mu$). 
Then, and most importantly, \fref{fig:weeks_hom}(a) shows that in the presence of SOC, the transport gap observed for $V_{\rm I}=0$ closes, and a $2e^2/h$ plateau appears over an energy window of about 80 meV. Such a width corresponds to the topological gap $6\sqrt{3}V_{\rm I}^{\rm eff}$ expected for an effective SOC $V_{\rm I}^{\rm eff}\approx 7.9$ meV $\approx n_\mathrm{ad} \ V_{\rm I}$.
We can thus conclude that the effect of a homogeneous distribution of heavy adatoms is to induce an effective SOC with strength proportional to the adatom concentration.
To further verify the topological nature of this phenomenon, \fref{fig:weeks_hom}(b) reports the local distribution of spin spectral currents at $E=-100$ meV. We observe that electrons, injected from the left contact, are transmitted along the top or bottom edge channel according to their spin polarization. No current flows in the bulk, where a topological gap is present.

\subsection{Clustering of adatoms and transition to the \SHE} \label{sec:qshe_cluster}

As shown in the previous section, a homogeneous distribution of heavy adatoms (with concentration 15\% in the example) is expected to induce a spin Hall phase. 
However, at present no topological phase has been observed experimentally in this kind of systems \cite{JIA_PRB91}.  
Here, we illustrate the case of thallium adatoms and show how their clustering might be responsible for such a failure \cite{CRE_PRL113}. 
Clustering and segregation are typical phenomena for adatoms on 2D materials \cite{SUT_SS17}. 
This is already known to have a significant impact on several material features concerning doping \cite{PI_RPB80,SAN_PRB84}, transport \cite{MCC_PRB81,ALE_PRB86,EEL_PRL110,KAT_PRB79} and optical properties \cite{YUA_PRB84,YUA_PRB90}.

\begin{figure}[t!]
  \begin{center}
  \resizebox{14cm}{!}{\includegraphics{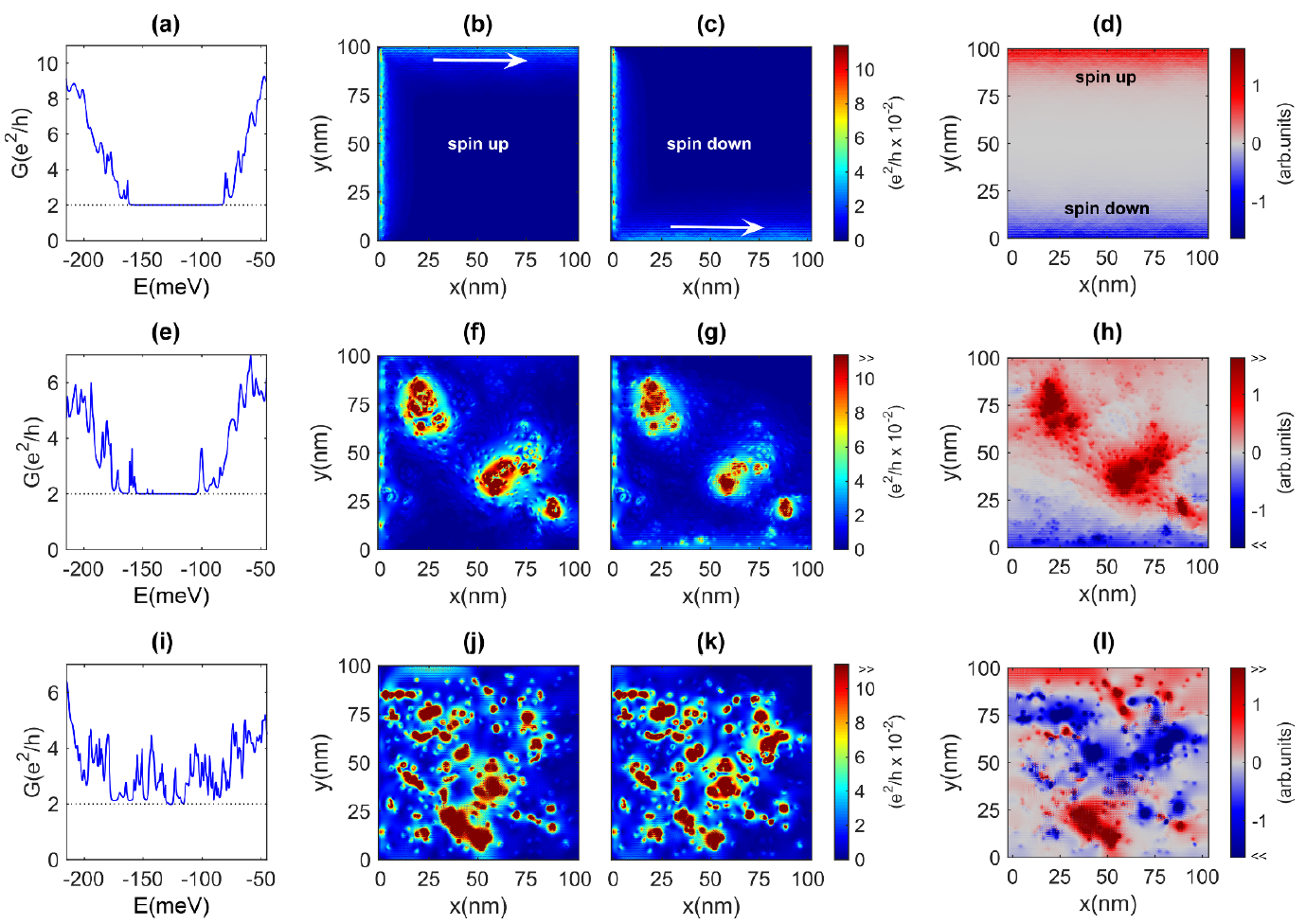}}
  \caption{(a) Differential conductance of a 100 nm wide ribbon with a distribution of thallium adatoms with concentration $n_\mathrm{ad}=15\%$ in islands with radius $r=0.5$ nm, over a 100 nm long section, in the presence and in the absence of SOC. (b) Spectral spin current distribution at $E=-100$ meV for spin up electrons injected from the left contact. (c) Same as (b) for spin down electrons. (d) Polarized density of occupied states at $E=-100$ meV. The color is blue for spin down and red for spin up electrons. (e-h) Same as (a-d) for islands with radius $r=1$ nm. Note that we cut the colorbar edges to have the same scale as in (b,c,d). (i-l) Same as (e-h) for islands with radius $r=1.5$ nm.}
  \label{fig:thallium_prl113} 
  \end{center}
\end{figure}
\begin{figure}[h!]
  \begin{center}
  \resizebox{11cm}{!}{\includegraphics{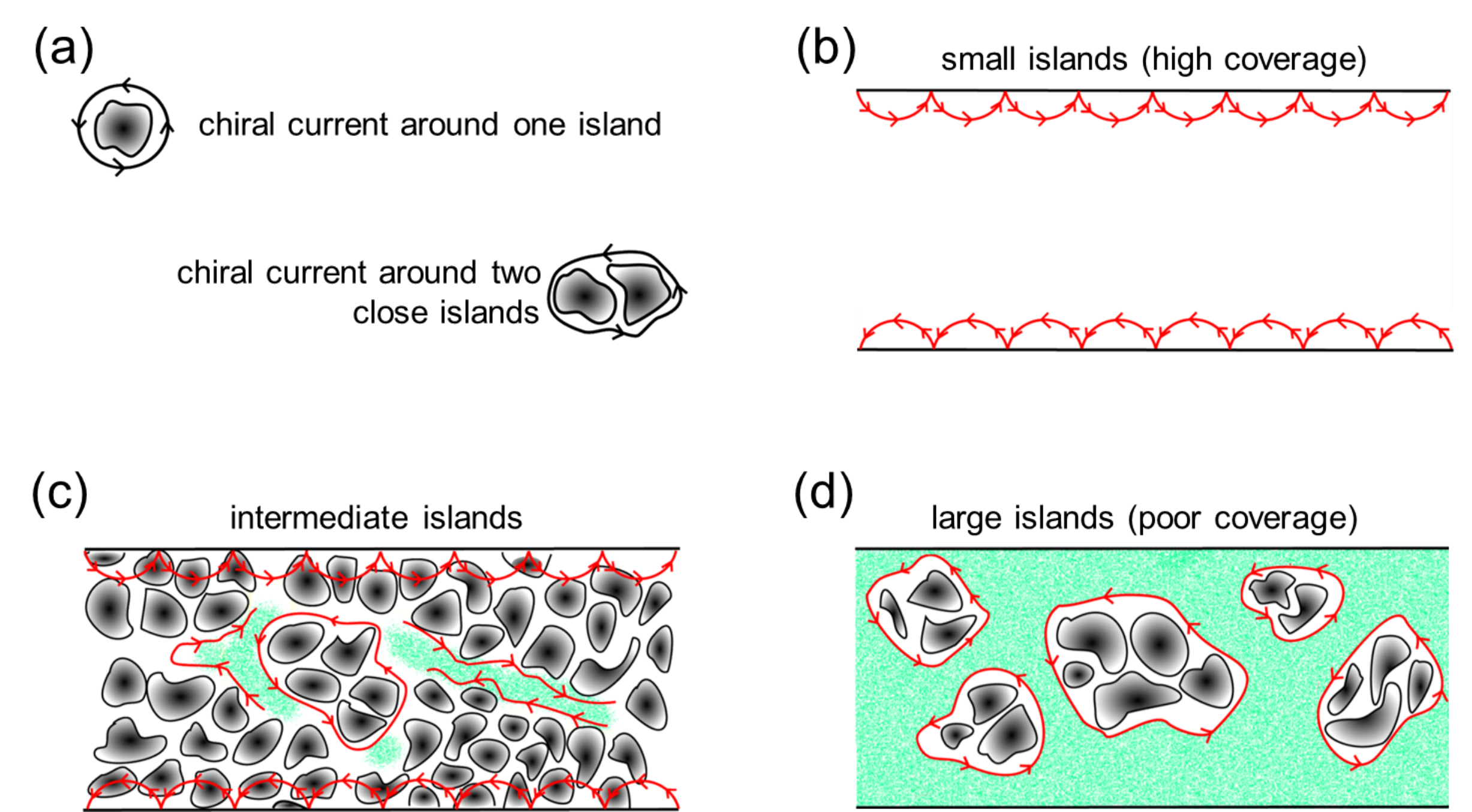}}
  \caption{(a) Sketch of the chiral currents (for given spin orientation) around an island of adatoms, and a group of two close islands. The current direction is indicated by the arrows. (b) Formation of edge channels for small island size. (c) For intermediate island size, edge currents, bulk chiral current around islands and weak nonchiral bulk currents (green regions) coexist. (d) For large island size, edge currents disappear and large non chiral bulk currents are present.}
  \label{fig:thallium_cartoon} 
  \end{center}
\end{figure}

We consider a 100 nm wide graphene ribbon with a concentration $n_\mathrm{ad}=15\%$ of thallium adatoms over a 100 nm long section. The adatoms are segregated into islands with given radius $r$. 
As shown in \ffref{fig:thallium_prl113}(a-e), for small islands with radius $r<1$ nm, we still observe a spin Hall phase with conductance plateau and edge polarization.
When increasing the island size to $r=1$ nm, see \fref{fig:thallium_prl113}(e), the plateau is reduced in width and an increase of the conductance above $2e^2/h$ at the edges of the topological gap is observed. To better understand this phenomenon, we look at the spin resolved spectral current density at $E=$100 meV, in correspondence of a conductance peak, see \ffref{fig:thallium_prl113}(f,g). We observe that the current starts flowing through the bulk and concentrates in two spatially separated regions. Here, the current forms vortices, whose propagation direction depends on the spin polarization of injected electrons \cite{CRE_PRL113}. From \fref{fig:thallium_prl113}(h) we can clearly see that edge polarization is preserved, but strongly polarized regions appear in correspondence of the current vortices.
For $r=1.5$ nm, see \ffref{fig:thallium_prl113}(i-l), the conductance plateaus is completely lost, and the conductance is above $2e^2/h$. This is consequence of the bulk current that develops around and between the islands. However, and very intriguingly, a residual spin accumulation at the sample edges is present, as in the SHE \cite{VIG_JSNM23,MAE_2011}. 
Indeed, giant SHE has been recently observed in graphene \cite{BAL_NC5} decorated with clustered transition metals, whose origin was attributed to resonant skew-scattering (SS) induced by adatoms \cite{FER_PRL112}.

The perturbation of the spin Hall phase resulting from segregation depends on the joint effect of the presence of large ``pristine'' regions between the clusters and the large potential shift under the islands. However, the signature of the SOC is still present, as evidenced by current vortices around the islands, whose chirality depends on the spin orientation.

We provide a heuristic explanation of these phenomena by means of \fref{fig:thallium_cartoon}.  Local chiral currents (i.e. flowing clockwise or counterclockwise depending on spin orientation) form around an isolated island of adatoms. When two islands are close, they form a single larger island and again the current turns around it, see \fref{fig:thallium_cartoon}(a).
For given adatom concentration, a small island size implies a large graphene coverage, which corresponds to an almost uniform distribution of the SOC. Therefore, as for the homogeneous distribution of adatoms, we observe the formation of chiral edge channels, as shown in \ffref{fig:thallium_prl113}(b,c,e) and illustrated in \fref{fig:thallium_cartoon}(b). 
When increasing the size of the islands, large regions of graphene are not covered by adatoms. As a consequence, we have the coexistence of chiral currents (at the edges and around the islands) and non-chiral bulk currents, indicated in green in \fref{fig:thallium_cartoon}(c). The conductance quantization is largely perturbed by the bulk currents. In particular, due to the opening of transport channels through the topological gap in the bulk, the conductance increases above $2e^2/h$. However, a strong spin accumulation at edges still holds, as visible in \ffref{fig:thallium_prl113}(i,l). 
Finally, for large island size, adatoms are clustered in isolated regions separated by large regions of pristine graphene, see \fref{fig:thallium_cartoon}(d). Transport is thus dominated by the bulk non-chiral currents, which completely destroy the conductance plateau. However, the local chiral currents around the island clusters still generate some spin accumulation, i.e. a non-quantum SHE, as illustrated in \sref{sec:she}.

%%%%% SPIN HALL EFFECT IN DISORDERED GRAPHENE %%%%%%%%%%%%%%%%%%%%%%%%%%%%%%%%%%%%%%%%%%%

\section{Spin Hall effect in disordered graphene} \label{sec:she}

\begin{figure}[b]
	\begin{center}
		\resizebox{9cm}{!}{\includegraphics{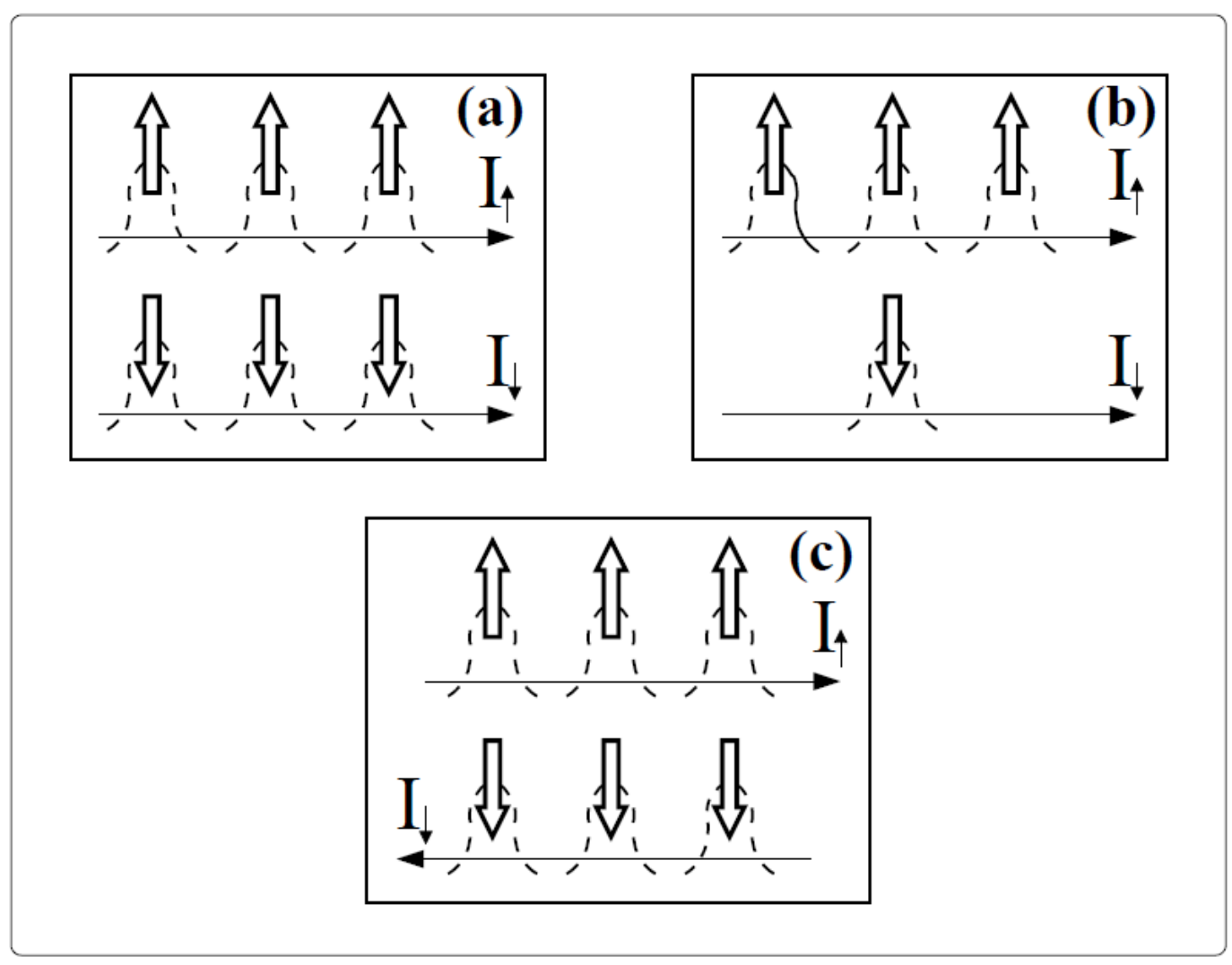}}
		\caption{Comparison of charge  $I$ and spin $I^{S_\alpha}$ currents associated with different combination of spin-resolved charge fluxes, $I^\uparrow$ and $I^\downarrow$, carrying  spin-$\uparrow$ and spin-$\downarrow$ electronic wave packets, respectively: (a) conventional unpolarized charge current is characterized by $I=I^\uparrow + I^\downarrow \neq 0$ and  $I^{S_\alpha}=I^\uparrow -I^\downarrow \equiv 0$; (b) spin-polarized  charge current $I \neq 0$ is accompanied also by non-zero 	spin current $I^{S_\alpha} \neq 0$; and (c) {\em pure} spin current $I^{S_\alpha}=I^\uparrow -I^\downarrow \neq 0$ arises when spin-$\uparrow$ electrons move in one direction, while an equal number of spin-$\downarrow$ electrons move in the opposite direction, so that net charge current is $I \equiv 0$.}
		\label{fig:spincurrents}
	\end{center}
\end{figure}

\begin{figure}[t]
	\begin{center}
		\resizebox{9cm}{!}{\includegraphics{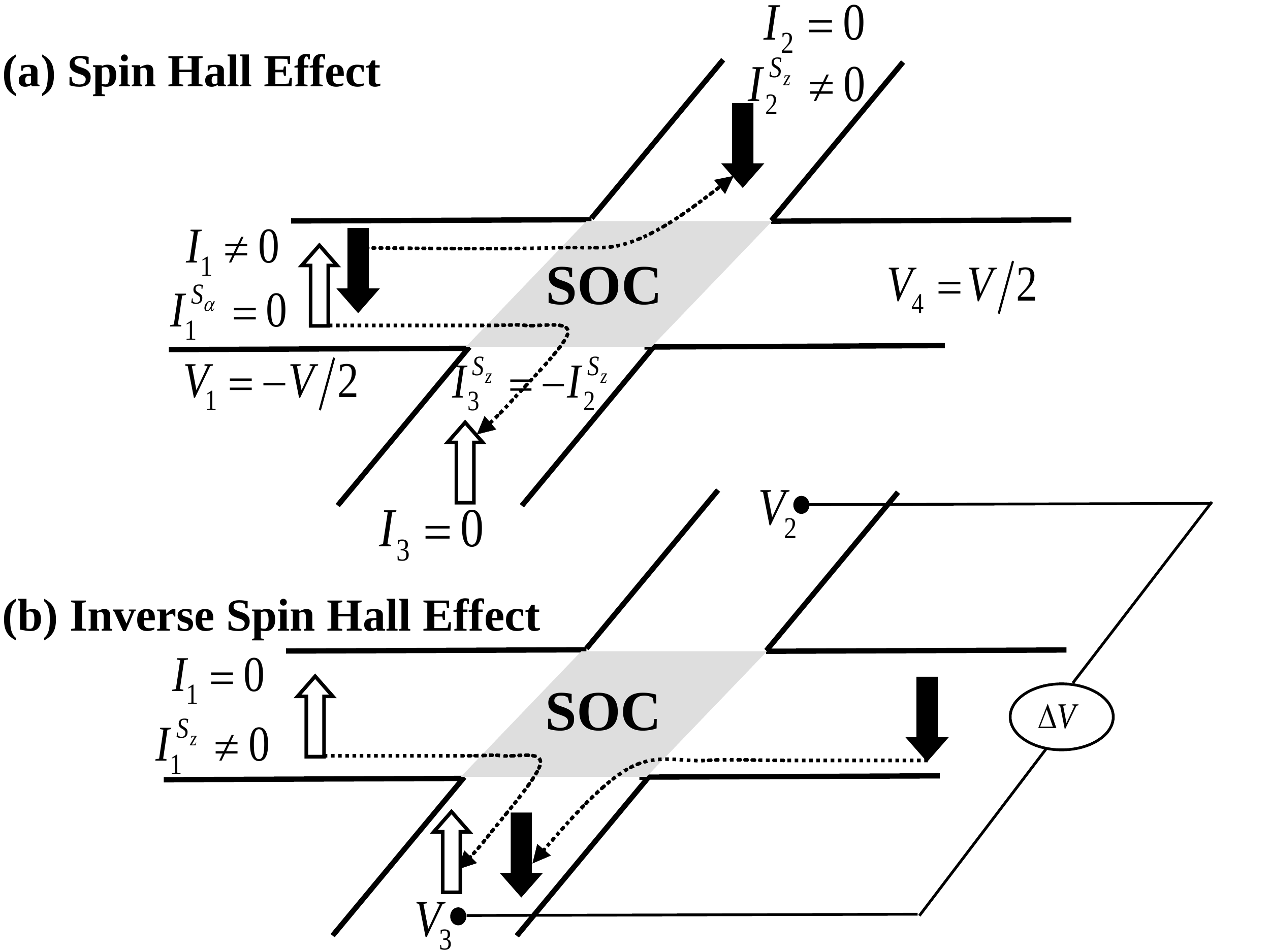}}
		\caption{Basic phenomenology of the direct and inverse SHE, assuming example of 2D system like graphene within the $xy$-plane: (a) in the direct SHE, conventional unpolarized charge current $I_1$ generates transverse pure spin current $I_2^{S_z}$  or spin accumulation (when transverse leads are removed) of opposite sign at opposite lateral edges; (b) in the inverse SHE, pure spin current $I_1^{S_z}$ generates transverse charge current $I_2$ or voltage $V_2 - V_3$ in an open circuit. Note that the usage of ideal transverse leads without SOC or other types of spin-dependent interactions, as assumed in (a), bypasses~\cite{Nikolic2006} the issue of how to unambiguously define the conserved spin current~\cite{SHI_PRL96,Sugimoto2006}.}
		\label{fig:she_ishe}
	\end{center}
\end{figure}

The SHE~\cite{SHE2,VIG_JSNM23} is a phenomenon where a conventional unpolarized charge current injected into a metal or a semiconductor generates a transverse pure spin current or a spin accumulation at the lateral sample boundaries. In the inverse SHE, an injected pure spin current generates a transverse charge current or a transverse voltage in an open circuit. The distinction between unpolarized charge, spin-polarized charge and pure spin current is explained in \fref{fig:spincurrents}. The two SHE effects, which are equivalent to each other due to Onsager reciprocity  relations~\cite{Hankiewicz2005}, are illustrated in \fref{fig:she_ishe} using four-terminal device geometry. While the SHE is analogous to the classical HE for charges, it occurs in the absence of externally applied magnetic field or magnetization that breaks the time-reversal symmetry. Instead, the spin separation requires SOC, which has emerged as one of the central resources for spintronics since, unlike cumbersome magnetic fields, SOC makes possible spin control on very short length and time scales {\it via} electric fields~\cite{Nitta1997}. We overview basics of SOC, as the manifestation of relativistic effects in solids and graphene in particular, in \aref{sec:soc}. 

The SHE was originally predicted in the early 1970s~\cite{dp1}, but remained largely unnoticed until its rediscovery in the 1990s~\cite{Hirsch1999,Zhang2000} and experimental confirmation in the early 2000s brought by advances in optical techniques for measuring spin accumulation in semiconductors~\cite{KAT_SCI306,WUN_PRL94}. In contrast to these early experiments, where usage of optical techniques has required semiconductor samples of $\sim 100$ $\mu$m size, later experiments have moved toward electrical detection at room temperature using much smaller (of size $\sim 1$ $\mu$m) metallic~\cite{Valenzuela2006,Hoffmann2013} and semiconductor~\cite{Ehlert2014} samples. Furthermore, the inverse SHE has become a ``standard detector'' of pure spin currents generated by variety of mechanisms other than direct SHE~\cite{Ando2011a}, and a number of SHE-based devices has been realized using different materials~\cite{Jungwirth2012}.

To compare efficiency of SHE-driven conversion of charge into spin in different materials, one uses the figure of merit known as the spin Hall (SH) angle $\theta_{\rm sH}$ defined by the ratio of driving charge current and resulting spin current. Using the example in \fref{fig:she_ishe}, for injected conventional charge current, $I_1 = I_1^\uparrow + I_1^\downarrow \neq 0$ and $I_1^{S_z} = I_1^\uparrow - I_1^\downarrow=0$, and generated pure spin current,  $I_2^{S_z} = I_2^\uparrow - I_2^\downarrow \neq 0$ and $I_2 = I_2^\uparrow + I_2^\downarrow=0$, the SH angle is defined by
\begin{equation}\label{eq:shadef}
    \theta_\mathrm{sH} = \frac{I_2^{S_z}}{I_1} \ .
\end{equation}
This quantity is dimensionless when using the same units for spin and charge currents, which are defined in terms of the spin-resolved charge currents $I^\uparrow$, $I^\downarrow$ carrying spins pointing along the $\alpha = \{ x,y,z\}$-axis, as illustrated in \fref{fig:spincurrents}. Note that separated spins are orthogonal to spin flux, as illustrated in \fref{fig:she_ishe} for 2D samples. In 3D samples~\cite{Wang2016a} charge current in the $x$-direction generates transverse spin currents in the $z$- and $y$-direction, which are polarized along the $y$- and $z$-axis, respectively, and have the same amplitudes. The definition in \eref{eq:shadef} is suitable for calculations based on the LB formula (see \aref{sec:qtalgorithms}) for current in the leads (as performed in \ssref{SHENLR} and \ref{sec:zshe}), while for calculations based on the Kubo formula for bulk conductivities (as performed in \sref{SHEKUBO}), we use   
\begin{equation} \label{eq:thsH2}
    \theta_{\rm sH}=\frac{\sigma_{xy}^{z}}{\sigma_{xx}} \ ,
\end{equation}
where $\sigma_{xy}^{z}$ is the SH conductivity and $\sigma_{xx}$ is the longitudinal charge conductivity.
Note that spin conductivity in two dimensions has the same unit as spin conductance and it is naturally given in the units of spin conductance quantum, $e/4\pi$. However, to make $\theta_\mathrm{sH}$ a dimensionless quantity, spin conductivity/conductance should be expressed in the same units as charge conductivity/conductance by replacing $e/4 \pi$ with charge conductance quantum $e^2/h=(2e/\hbar)(e/4\pi)$.

To date, measured values of $\theta_\mathrm{sH}$ range from $\sim 10^{-4}$ in semiconductors to $\sim 0.01$ for metals like Pt~\cite{Wang2016a} and $\sim 0.1$ for metals like \mbox{$\beta$-Ta} and \mbox{$\beta$-W}~\cite{Mellnik2014}. Thus, recent experiments~\cite{BAL_NC5,Balakrishnan2013} extracting surprisingly large $\theta_\mathrm{sH}\simeq 0.2$ from {\em all-electrical} measurements {\it via} combined direct and inverse SHE in multiterminal graphene devices (see \ffref{NewFig1} and ~\ref{Fig5S} for illustration) have attracted considerable attention. In order to enhance the minuscule SOC effects (see \aref{sec:soc}) in pristine graphene, these experiments have utilized heavy adatoms like Cu, Au, Ag~\cite{BAL_NC5} or even light adatoms like hydrogen~\cite{Balakrishnan2013} and fluorine~\cite{Avsar2015} in order to locally enhanced SOC in the graphene regions surrounding the adatoms.

Before delving into quantum transport modeling of $\theta_{\rm sH}$, as well as $R_\mathrm{NL}$ as the quantity actually measured in these experiments, we first overview in \sref{sec:extrintr} three distinct  microscopic mechanisms behind SHE stemming from different aspects of SOC in solids---SS and side jump (SJ) mechanisms associated with extrinsic  impurities, and intrinsic mechanism associated with uniform SOC affecting the band structure of the material. In addition, in \sref{sec:she_dis} we also overview recently provoked controversy due to the inability of some of the repeated measurements~\cite{Wang2015a,Kaverzin2015} on adatom-decorated graphene to unambiguously associate non-zero $R_\mathrm{NL}$ with spin-dependent transport.

\subsection{Physical mechanisms of the \SHE: Extrinsic versus intrinsic}\label{sec:extrintr}

The lab frame interpretation of SOC given in \aref{sec:soc} makes it easy to understand the Mott SS~\cite{Mott1929} off an impurity whose Coulomb field deflects a beam of spin-$\uparrow$ and spin-$\downarrow$ particles in opposite directions, thereby generating the SS contribution~\cite{SHE2,VIG_JSNM23} to extrinsic SHE. For example, if we look at spin-$\uparrow$ electron from behind moving along the $y$-axis, whose expectation value of the spin vector is oriented along the positive $z$-axis so that the corresponding magnetic dipole moment lies along the negative $z$-axis, then in the lab frame we also see its Lorentz transformed electric dipole moment ${\bf P}_{\rm lab}$ oriented along the negative $x$-axis. The electric dipole feels the force ${\bf F} = ({\bf P}_{\rm lab} \cdot \nabla) {\bf E}_{\rm lab}$, oriented in this case along the positive $x$-axis since gradient of the electric field ${\bf E}_{\rm lab}=-\nabla V_\mathrm{imp}(\mathbf{r})/e$ generated by the impurity is always negative outside of it. 

Note that this simple classical picture only explains one aspect of SOC-dependent interaction with impurity. The other one---the so-called SJ (i.e., sideways shift of the scattering wave packet)---requires fully quantum mechanical analysis. The elimination of the negative energy states in the Dirac equation, which leads to SOC Hamiltonian in \eref{eq:so}, can be viewed equivalently as a redefinition of the position operator~\cite{VIG_JSNM23,Sinitsyn} 
\begin{equation}\label{eq:rphys}
		\mathbf{r}_\mathrm{phys}= \mathbf{r} + \frac{\hbar}{4 m^2 c^2} \mathbf{p} \times \mathbf{s} \ ,
\end{equation}
in the positive energy subspace [i.e., by replacing canonical position operator by $\mathbf{r}_\mathrm{phys}$ in $V(\mathbf{r})$, and by expanding to first order in $\hbar/4 m^2 c^2$, one recovers \eref{eq:so}]. This leads to anomalous velocity operator in SO-coupled systems~\cite{VIG_JSNM23,Sinitsyn} 
\begin{equation}\label{eq:av}
		\mathbf{v} = \frac{d \mathbf{r}_\mathrm{phys}}{dt} = \frac{i}{\hbar} [H, \mathbf{r}_\mathrm{phys}] = \frac{\mathbf{p}}{m} - \frac{\hbar}{4m^2 c^2} \nabla V_\mathrm{imp}(\mathbf{r}) \times \mathbf{s} \ ,
\end{equation}  
for $H = \mathbf{p}^2/2m + V_\mathrm{imp}(\mathbf{r})$. During the collision of electronic wave packet with an impurity, the second term in \eref{eq:av} dominates because $\nabla V_\mathrm{imp}(\mathbf{r})/e= -d\mathbf{p}/dt$ is very large, so that the center of the wave packet will be displaced by $\Delta \mathbf{r}_\mathrm{phys} = \int dt \mathbf{v}(t) =  (\hbar/4m^2c^2) \Delta \mathbf{p} \times {\bf s}$. This includes the change of the internal structure of the wave packet in the course of scattering, and it is the origin of the SJ contribution~\cite{SHE2,VIG_JSNM23} to extrinsic SHE. 

The intrinsic SHE arises due to uniform SOC affecting electronic band structure. Such a SOC can also be viewed as the Zeeman interaction $-{\bf s} \cdot {\mathbf{B}}_n(\mathbf{p})$ with an internal effective magnetic field ${\mathbf{B}}_n(\mathbf{p})$ that depends on momentum (in order to preserve time-reversal invariance) and the band index $n$. For example, for the Rashba SOC in \eref{eq:rashba}, $\mathbf{B}_\mathrm{R} (\mathbf{p}) = \alpha (\hat{z} \times \mathbf{p})/\hbar$. The Rashba SOC Hamiltonian, $\hat{H}_\mathrm{R} = -{\bf s} \cdot \mathbf{B}_\mathrm{R} (\mathbf{p})$, then gives rise to the acceleration operator~\cite{Nikolic2005c,Adagideli2005}
\begin{eqnarray} \label{eq:soforce}
		\mathbf{a} = \frac{d^2 {\bf r}}{dt^2} = \frac{1}{\hbar^2} [\hat{H}_\mathrm{R},[\hat{\bf r},\hat{H}_\mathrm{R}]]  =  \frac{2 \alpha^2}{\hbar^3}  (\mathbf{p} \times \hat{z}) \otimes s_z \ .
\end{eqnarray}
Taking the expectation value of this operator in spin-polarized wave packet state shows that it deflects opposite spins in opposite direction, thus giving origin to the intrinsic SHE in finite-size samples~\cite{Nikolic2006,Adagideli2005,Nikolic2005b,Nikolic2009,Sheng2005a}. The spins along the $z$-axis will also precess in $\mathbf{B}_\mathrm{R}(\mathbf{p})$ (which leads to oscillations of the expectation value of ${\bf a}$ along the wire), and eventually dephase due to spin-orbit entanglement~\cite{Nikolic2005} (which leads to the decay of the expectation value of ${\bf a}$ along the wire). Thus, the strength of the intrinsic SHE is set by the competition between spin separation and spin dephasing induced concurrently by the same SOC mechanism~\cite{Nikolic2006,Nikolic2005b,Nikolic2009}. 

While impurities do not play an active role in the intrinsic SHE, they must be included in the calculation of intrinsic $\sigma_{xy}^z$ of macroscopic samples since normal (i.e., spin-independent) scattering off impurities is required to establish the steady-state transport regime in the presence of external electric field~\cite{VIG_JSNM23}. On the other hand, the SH conductance $G_\mathrm{sH}=I_2^{S_z}/(V_1-V_4)$ for the device in \fref{fig:she_ishe} or SH accumulation can be calculated~\cite{Nikolic2005b,Nikolic2009,Nikolic2005d} even if the SO-coupled central region is a perfectly clean sample in the ballistic transport regime. In such a situation, the electric field is zero and the charge current $I_1$ is driven by the electrochemical potential difference $eV_1-eV_4$ between the macroscopic Fermi liquid reservoirs responsible for dissipation~\cite{Schiro2009}.

Unlike metals with low mobility, where SJ contribution dominates over SS contribution and where it is difficult to differentiate SJ from the intrinsic mechanism~\cite{VIG_JSNM23}, due to the smallness of uniform SOC in graphene (see \aref{sec:soc}) the analysis of SHE in graphene with adatoms can be performed solely in terms of SJ versus SS contribution. For example, the recent SHE experiments on graphene decorated with Au or Cu adatoms were explained as a consequence of SS mechanism~\cite{BAL_NC5}, while experiments on graphene decorated with hydrogen adatoms have concluded that they are predominantly driven by SJ mechanism~\cite{Balakrishnan2013}. This is based on weak dependence of $\theta_\mathrm{sH}$ on the concentration of adatoms $n_\mathrm{ad}$ in the former case, and $\theta_\mathrm{sH} \propto n_\mathrm{ad}$  dependence in the latter case. Besides large SOC around the adatom position, strong perturbation of the potential around the adatom can also contribute to large magnitude of the extrinsic SHE~\cite{Gradhand2010}. 

\subsection{Controversies in interpretation of nonlocal transport measurements on adatom-decorated graphene} \label{sec:she_dis}

The failure of recent attempts~\cite{Wang2015a,Kaverzin2015} to confirm that the nonlocal resistance $R_\mathrm{NL}$ in multiterminal (see \fref{NewFig1}) adatom-decorated graphene is driven by spin transport has questioned the early interpretation in terms of SHE. For example, \cref{Wang2015a} found that the deposition of Au or Ir adatoms onto graphene did induce a large nonlocal signal, which, however, is insensitive to the applied in-plane magnetic field. Similarly, \cref{Kaverzin2015} found that nonlocal signal in hydrogenated graphene is also not sensitive to in-plane magnetic field, which should affect the nonlocal signal if spin current mediated it. The VHE and the contribution of long-range neutral valley currents (see \sref{sec:vhe}) as mediators of nonlocal signal can be excluded~\cite{Kaverzin2015}  due to absence of temperature dependence and broken inversion symmetry~\cite{GOR_SCI346,Song2015,JU_NAT520,KIR_PRB92,AND_JPSJ84}. Therefore, these experimental data call for in-depth scrutiny and clarification.

On the theoretical side, we note that semiclassical theories for $\theta_\mathrm{sH}$~\cite{FER_PRL112} and $R_\mathrm{NL}$~\cite{Abanin2009} suffer from many flaws since they utilize approximations known to be inaccurate~\cite{DIN_NP10,CHE_PRB85} at low energies close to CNP where the nonlocal signal is actually observed. While the Kubo formula~\cite{Streda1982} offers a fully quantum-mechanical treatment that can in principle capture all relevant effects, its standard analytical evaluations as perturbative expansion in disorder strength  neglect~\cite{Sinitsyn2006} interference terms generated by SS from pairs of closely spaced impurities~\cite{Ado2015,Milletari1,Milletari2}. Thus, the adatom clustering, which is usually observed experimentally~\cite{SUT_SS17}, should also impact $\theta_\mathrm{sH}$ as it introduces large variation of charge and spin transport characteristics~\cite{CRE_PRL113,PI_RPB80,KAT_PRB79}.

Section~\ref{SHEKUBO} presents the Kubo formula-based calculations of $\theta_{\rm sH}$ for square graphene sheet with periodic boundary conditions in the presence of randomly distributed adatoms as a function of the strength of SOC terms in \eref{eq:kanemele}, or adatom concentration covering both dilute and non-dilute regimes. This Section also presents calculations for realistic graphene samples with Au or Tl adatoms, which determine the strength of SOC terms (as extracted from DFT calculations, see \aref{sec:soc}), where we analyze the effect of randomly distributed versus clustered (as discussed also in \sref{sec:qshe}) adatoms. 

\Sref{SHENLR} presents the LB formula-based calculations of $\theta_{\rm sH}$ for Au-adatom-decorated multiterminal graphene. This approach makes it possible to obtain $\theta_\mathrm{sH}$ in both quasiballistic and diffusive transport regimes, as well as to compute the nonlocal resistance $R_\mathrm{NL}$ as the quantity directly measured in experiments, thereby revealing presence of non-SHE-related contributions to the nonlocal signal~\cite{TUA_arXiv:1603.03870}. This motivates us to propose a novel experimental setup (see \fref{Fig5S}), which can eliminate such background contributions and can make it possible to measure nonlocal resistance unambiguously connected to the SHE mechanism.

\subsection{Spin Hall angle in adatom-decorated graphene: Kubo formula approach} \label{SHEKUBO}

\begin{figure}[t!]
	\begin{center}
	  \resizebox{11cm}{!}{\includegraphics{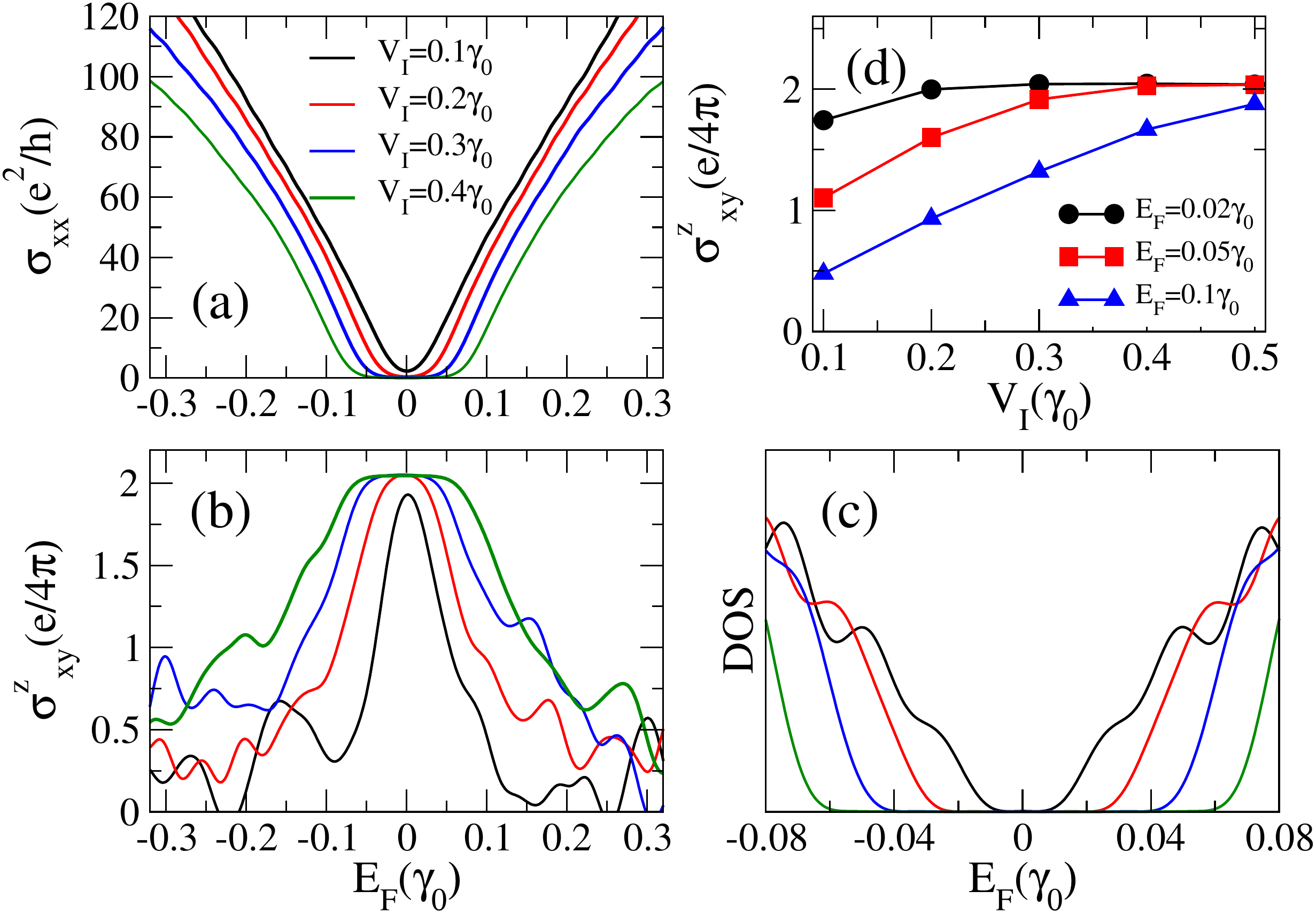}}	
		\caption{Spin and charge transport properties of graphene decorated with adatoms of concentration $n_\mathrm{ad}=20$\%. The adatoms are positioned at T-sites and they locally enhance only the intrinsic SOC---$V_\mathrm{I} \neq 0$; and $V_{\rm R}=V_{\rm PIA}=0$ in \eref{eq:kanemele}.  (a) Longitudinal conductivity and (b) SH conductivity as a function of $E_{\rm F}$ for increasing values of $V_\mathrm{I}$. (c) DOS and (d) SH conductivity as a function of $V_\mathrm{I}$ for different values of $E_{\rm F}$. We use $D= 2 \times 200 \times 200$ sites and $M=1600$ in the Kubo formula calculations (see \aref{sec:kubo-formula}). Adapted from \cref{Garcia2015}.}
		\label{sha-fig2}
	\end{center}
\end{figure}

\begin{figure}[t!]
	\begin{center}
	  \resizebox{11cm}{!}{\includegraphics{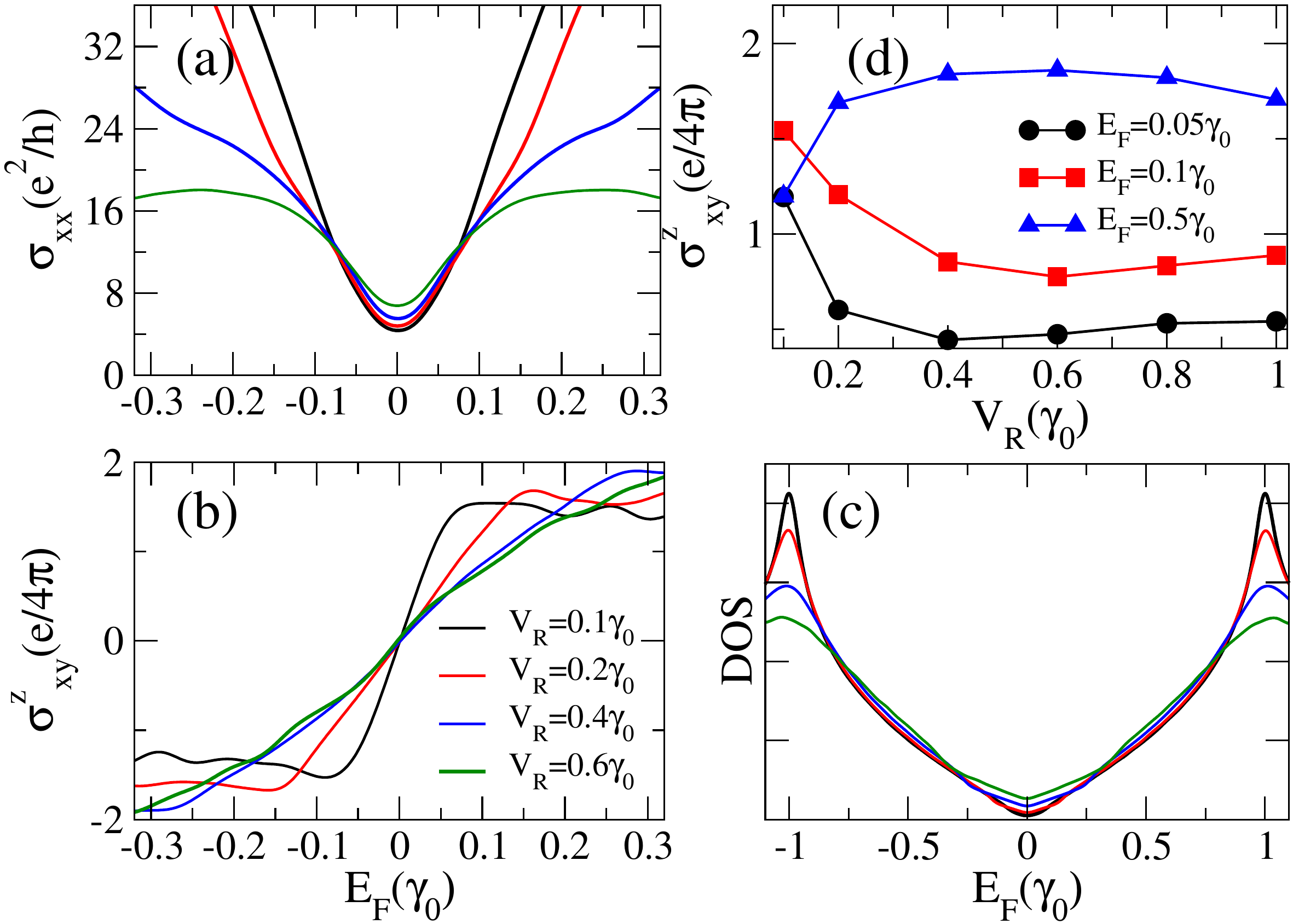}}
		\caption{Spin and charge transport properties of graphene decorated with a random distribution of adatoms of concentration $n_\mathrm{ad}=20$\%. The adatoms are positioned at T-sites and they locally generate only the Rashba SOC---$V_\mathrm{R} \neq 0$; and $V_{\rm I}=V_{\rm PIA}=0$ in \eref{eq:kanemele}.  (a) Longitudinal conductivity and (b) SH conductivity as a function of $E_{\rm F}$ for increasing values of $V_\mathrm{R}$. (c) DOS and (d) SH conductivity as a function of $V_\mathrm{R}$ for different values of $E_{\rm F}$. We use $D= 2 \times 200 \times 200$ sites and $M=800$ in the Kubo formula calculations (see \aref{sec:kubo-formula}).  Adapted from \cref{Garcia2015}.}
		\label{sha-fig4}
	\end{center}
\end{figure}

\begin{figure}[t!]
	\begin{center}
		\resizebox{11cm}{!}{\includegraphics{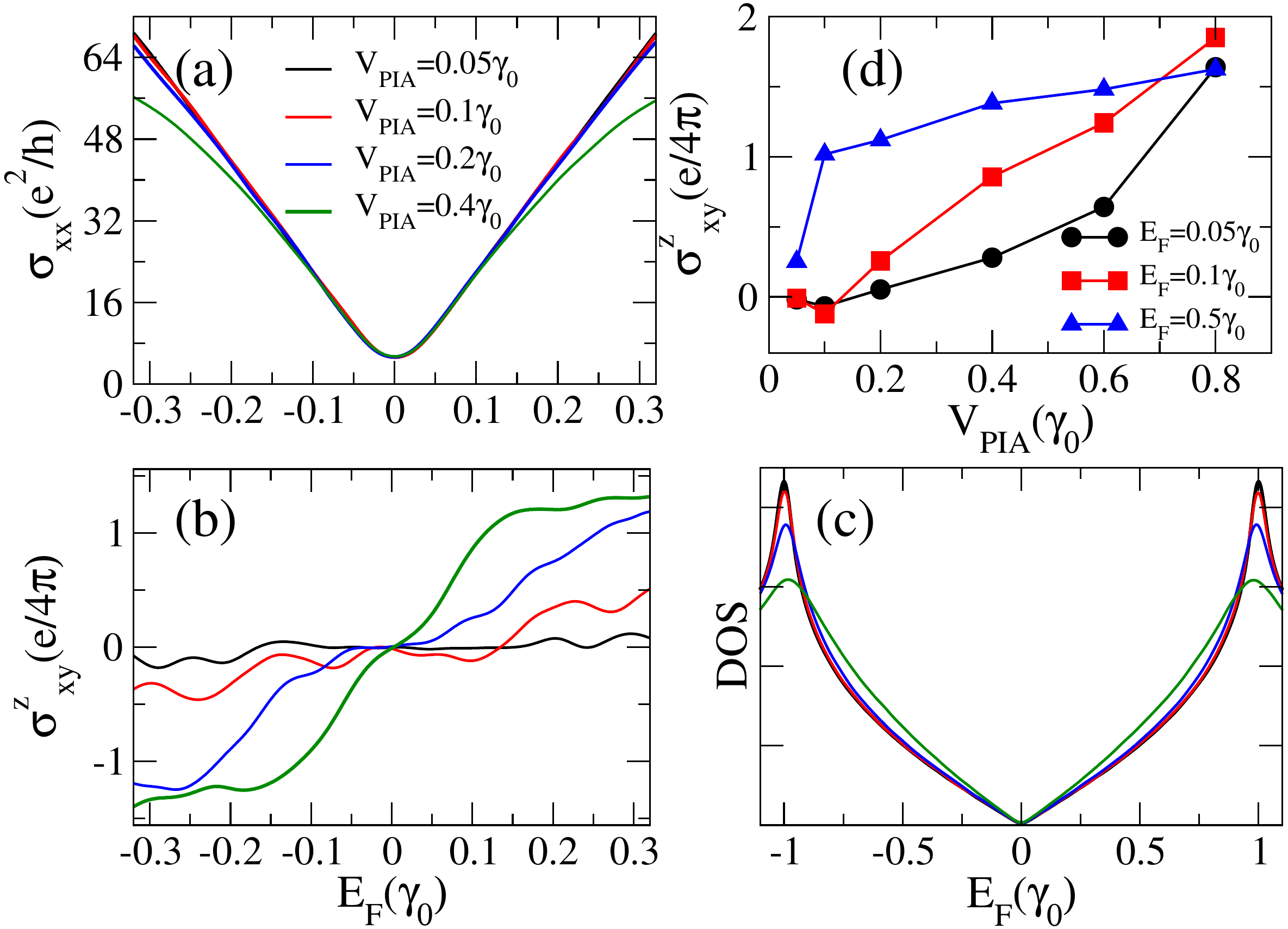}}
		\caption{Spin and charge transport properties of graphene decorated with a random distribution of adatoms of concentration $n_\mathrm{ad}=20$\%. The adatoms are positioned at T-sites and they locally generate only PIA  SOC---$V_\mathrm{PIA} \neq 0$; and $V_{\rm I}=V_{\rm R}=0$ in \eref{eq:kanemele}. (a) Longitudinal conductivity and (b) SH conductivity as a function of the $E_{\rm F}$ for increasing values $V_\mathrm{PIA}$. (c) DOS and (d) SH conductivity as a function of $V_\mathrm{PIA}$ for different values of the $E_{\rm F}$. We use $D= 2 \times 200 \times 200$ sites and $M=800$ in the Kubo formula calculations (see \aref{sec:kubo-formula}). Adapted from \cref{Garcia2015}.}
		\label{she-fig6}
	\end{center}
\end{figure}

\begin{figure}[t!]
	\begin{center}
		\resizebox{11cm}{!}{\includegraphics{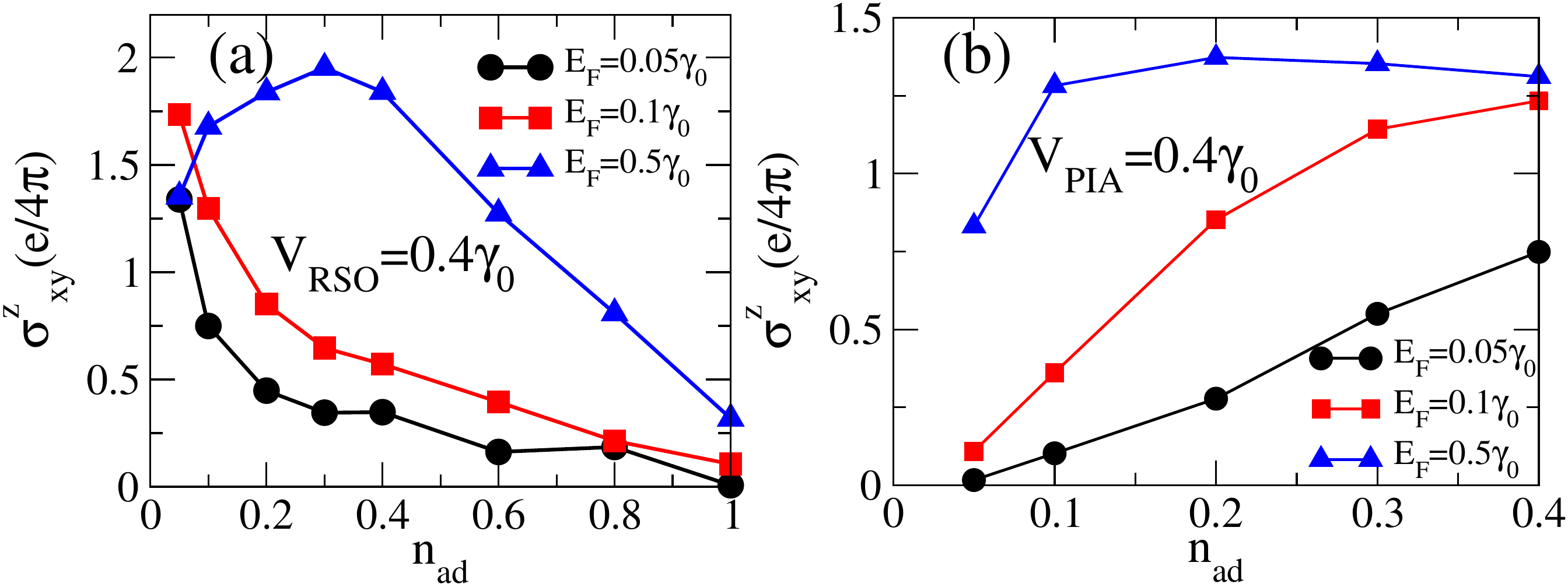}}
		\caption{Spin Hall conductivity of graphene decorated with a random distribution of adatoms as a function of their concentration $n_\mathrm{ad}$. The adatoms are positioned at T-sites and they locally generate only the Rashba SOC of strength $V_\mathrm{R}=0.4\gamma_0$ in (a) or PIA SOC of  $V_\mathrm{PIA}=0.4\gamma_0$ in (b).  We use $D= 2 \times 200 \times 200$ sites and $M=800$ in the Kubo formula calculations (see \aref{sec:kubo-formula}). Adapted from \cref{Garcia2015}.}
		\label{fig:rvspia}
	\end{center}
\end{figure}

Modeling SOC in graphene due to the adsorption of adatoms is a complex task, because it strongly depends on the interaction between the graphene and the impurity orbitals~\cite{WEE_PRX1,HU_PRL109,gmitrahydogenatedgraphene,Irmer2015,Zollner2016,Gmitra2016,KOC_arXiv2016}. Here, we employ minimal effective TB Hamiltonian in \eref{eq:kanemele} as an input for quantum transport calculations, whose construction and fitting of parameters ($V_\mathrm{I}$, $V_\mathrm{R}$, $V_\mathrm{PIA}$, and $\mu$) to DFT calculations is elaborated in \aref{sec:soc}. Below, we denote the position of adatoms as either H-site (see \fref{NewFig1} for illustration), which is at the center of a hexagonal ring of carbon atoms, or T-site, which is on the top of carbon atom. We neglect any on-site potential on the carbon atoms due to adatom, so that $\mu=0$ in \ffref{sha-fig2}, ~\ref{sha-fig4},~\ref{she-fig6}  and ~\ref{fig:rvspia}.

The SH conductivity is evaluated by computing the Kubo conductivity tensor using an efficient real-space method developed in \cref{Garcia2015}. Although this methodology was originally developed for charge conductivity, it has been recently extended to the calculation of spin conductivity \cite{Garcia2016,TUA_arXiv:1603.03870}. The methodology is delineated in \aref{sec:kubo-formula}.

To gain insight into the effect of three different SOC terms in \eref{eq:kanemele}, locally induced (such as $V_\mathrm{R}$ and $V_\mathrm{PIA}$) or enhanced (such as $V_\mathrm{I}$) by the presence of an adatom, we begin our analysis by scrutinizing the effect of purely intrinsic SOC---$V_\mathrm{I} \neq 0; V_{\rm R}=V_{\rm PIA}=0$. The presence of an energy gap $\Delta_\mathrm{I}$ is observed in the DOS in \fref{sha-fig2}(c), which results in an insulating behavior of the longitudinal charge conductivity shown in \fref{sha-fig2}(a). The zero charge conductivity  $\sigma_{xx}=0$ inside the gap of such a realization~\cite{WEE_PRX1,HU_PRL109,CRE_PRL113,CHA_NL14} of 2D TI phase (see also \sref{sec:qshe}) means that $\theta_\mathrm{sH}$ in \eref{eq:shadef} is ill-defined within the gap. So, our systematic analysis of the effect of different SOC terms on spin and charge transport is presented by plotting separately the SH conductivity $\sigma_{xy}^z$ and the longitudinal charge conductivity $\sigma_{xx}$.

By fitting the behavior for different concentrations and values of $V_\mathrm{I}$ we find that energy gap follows $\Delta_{\rm I} \propto V_{\rm I} n_\mathrm{ad}$, which is consistent with Kane-Mele~\cite{KAN_PRL95,KAN_PRL95b} model rescaled by the adatom concentration $n_\mathrm{ad}$ and it agrees with previous numerical calculations~\cite{Garcia2015,Liu2015,Shevtsov2012}. In \fref{sha-fig2}(b), quantized SH  conductivity is found for energies inside the gap $\Delta_\mathrm{I}$, as predicted by the Kane-Mele model~\cite{KAN_PRL95,KAN_PRL95b}, even in the presence of weak disorder~\cite{Sheng2005b}. Moreover, \fref{sha-fig2}(d) shows the robustness of the SH conductivity outside the gapped region, even at small strengths of the intrinsic SOC, which is important for experiments where $V_\mathrm{I}$ in \eref{eq:kanemele} is usually small and the gap $\Delta_{\rm I}$ is easily closed by disorder and/or temperature.

We proceed by analyzing the effect of adatoms that would generate purely Rashba SOC---$V_\mathrm{R} \neq 0$; and $V_{\rm I}=V_{\rm PIA}=0$ in \eref{eq:kanemele}. In the DOS shown in \fref{sha-fig4}(c), we notice the presence of new states at CNP. These states produce a slight increase in the minimum of $\sigma_{xx}$ for increasing values of SOC, as can be seen in \fref{sha-fig4}(a).  At the same time, the Rashba SOC strongly suppresses $\sigma_{xx}$ away from CNP, thus decreasing the mobility of graphene. The SH conductivity in \fref{sha-fig4}(b) changes sign as it crosses the CNP at which $\sigma_{xy}^z =0$ due to particle-hole symmetry~\cite{Nikolic2005b,Sheng2005a}. In the vicinity of CNP there is a rapid increase of the SH conductivity, saturating at $\approx\pm e/(2\pi)$, which is consistent with analytical calculations of $\sigma_{xy}^z$ in infinite homogeneous graphene with uniform Rashba SOC~\cite{Dyrdal2009}. The transition from negative to positive SH conductivity as a function of $E_{\rm F}$ gets more abrupt for small $V_{\rm R}$. Surprisingly, this translates into an increase of $\sigma_{xy}^z$ in the vicinity of CNP for decreasing $V_\mathrm{R}$, as shown in \fref{sha-fig4}(d). The Rashba SOC breaks the spin degeneracy and there are two non-degenerate bands for electrons at CNP. In the case of $n_\mathrm{ad}=100$\%, the sharp increase  of the SH conductivity occurs at $E_{\rm F}$ of the order of the Rashba splitting between the bands at energy scales where the DOS presents the contribution of a single band. A similar behavior is seen for $n_\mathrm{ad} = 20$\%  in \fref{sha-fig4} with the Rashba splitting scaled with the concentration.

Finally, we analyze the effect of adatoms that would generate purely PIA SOC---$V_\mathrm{PIA} \neq 0$; and $V_{\rm I}=V_{\rm R}=0$ in \eref{eq:kanemele}. The DOS plotted in \fref{she-fig6}(c) shows the emergence of new states in the vicinity of the CNP. These new states translate into a decrease of $\sigma_{xx}$ and mobility away from CNP, as shown in \fref{she-fig6}(a). \Fref{she-fig6}(b) shows that the SH conductivity reaches a maximum at high energies with a value that depends directly on $V_\mathrm{PIA}$. In contrast to the SH conductivity driven by the Rashba SOC in \fref{sha-fig4}(d), for PIA SOC $\sigma_{xy}^z$ at fixed $E_{\rm F}$ shown in \fref{she-fig6}(d) increases with the SOC strength, and it is very small in the limit $V_\mathrm{PIA} \rightarrow 0$. 

In \fref{fig:rvspia}, we compare the dependence on adatom concentration for the SH conductivity generated purely by the Rashba SOC versus PIA SOC. \FFref{fig:rvspia}(a) and ~\ref{sha-fig4}(d) suggests that Rashba SOC-generated contribution to the SH conductivity must be {\em extremely important} for experiments since it is larger for low adatom concentrations and weak $V_\mathrm{R}$. On the other hand, \ffref{fig:rvspia}(b) and ~\ref{she-fig6}(d) show that PIA SOC generated contribution to the SH conductivity will be negligible under realistic experimental conditions where $V_\mathrm{PIA}$ is weak~\cite{gmitrahydogenatedgraphene,Irmer2015,Zollner2016,Gmitra2016,KOC_arXiv2016} and concentration of adatoms is small~\cite{BAL_NC5,Balakrishnan2013}. 

\begin{figure}[t!]
	\begin{center}
		\resizebox{9cm}{!}{\includegraphics{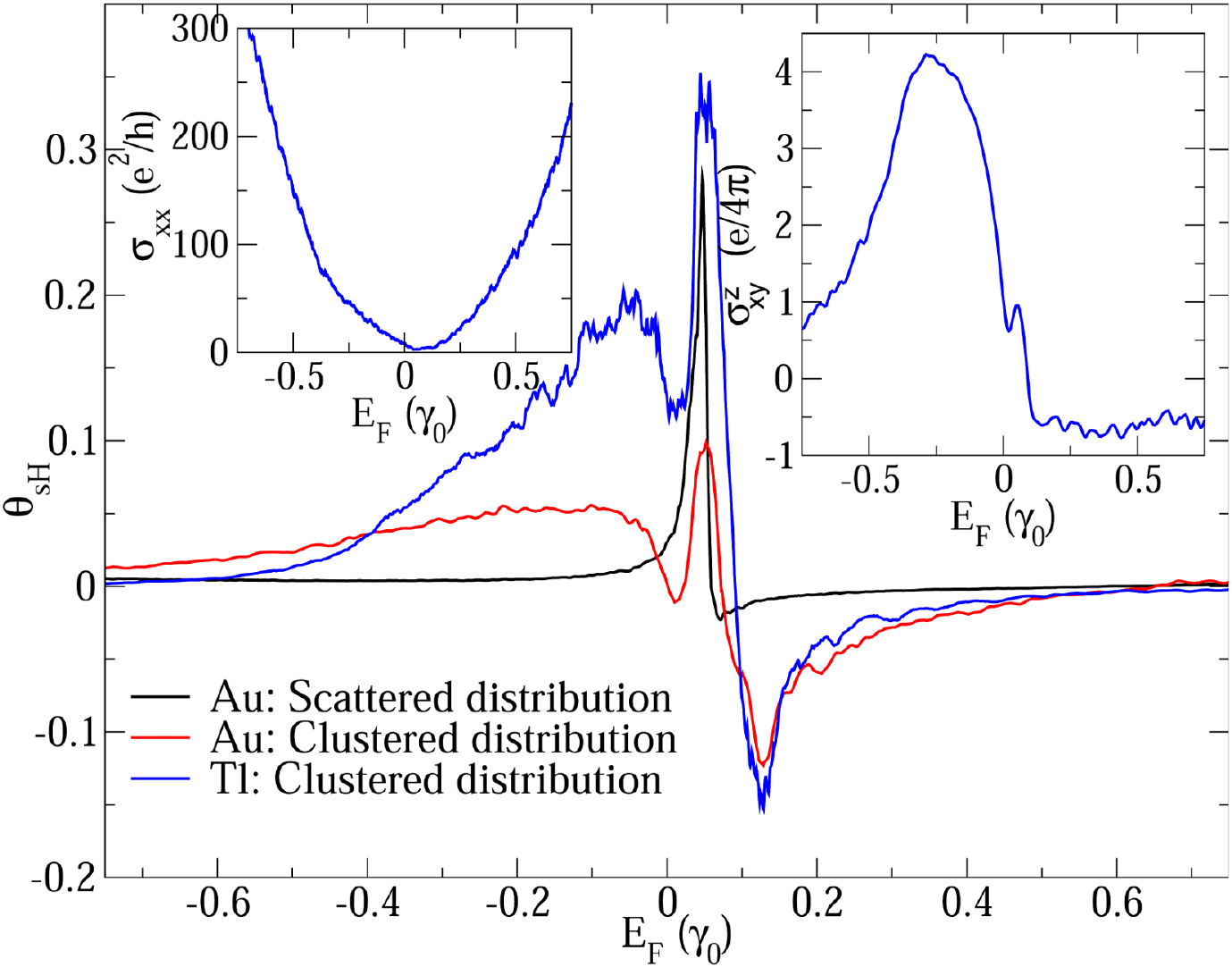}}
		\caption{The spin Hall angle $\theta_{\rm sH}$ for graphene with $n_\mathrm{ad} = 15$\% concentration of Au adatoms on H-sites, which are either scattered or clustered, as well as for graphene with $n_\mathrm{ad} = 15$\% concentration of clustered Tl adatoms. The two insets show the SH conductivity $\sigma_{xy}^z$ and the longitudinal charge conductivity $\sigma_{xx}$ corresponding to the same graphene decorated with Tl adatoms as in the main panel. Results are averaged over $400$ disorder realizations. Adapted from \cref{TUA_arXiv:1603.03870}.}
		\label{NewFigSAH}
	\end{center}
\end{figure}

We conclude insights from Kubo formula-based calculations by discussing spin and charge transport properties of realistic graphene samples encountered in experiments~\cite{BAL_NC5} where adatoms generate {\em both} the intrinsic and the Rashba SOC. Figure~\ref{NewFigSAH} plots the SH angle defined in \eref{eq:thsH2} as a function of $E_{\rm F}$ for $n_\mathrm{ad}=15$\% concentration of Au adatoms distributed in scattered fashion (i.e., isolated adatoms without any clustering, as illustrated in \fref{NewFig1}). This result is contrasted with samples where either Au or Tl adatoms are clustered into randomly distributed islands of radius \mbox{$\in [1,3]$ nm}. We use $V_\mathrm{I} = 0.007 \gamma_{0}$, $V_\mathrm{R}=0.0165 \gamma_{0}$, $V_\mathrm{PIA}=0$ and $\mu = 0.1 \gamma_{0}$ in \eref{eq:kanemele} for Au adatoms~\cite{DIN_NP10} residing on H-sites, and $V_\mathrm{I} = 0.02 \gamma_{0}$, $V_\mathrm{R}=V_\mathrm{PIA}=0$ and $\mu = 0$ for Tl adatoms residing on H-sites, where both sets of parameters are extracted (see also \sref{sec:soc}) from DFT calculations~\cite{WEE_PRX1,DIN_NP10}. 

Remarkably, $\theta_\mathrm{sH}$ shown in \fref{NewFigSAH} is very large close to CNP, reaching $0.1$--$0.3$ (depending on the type of adatom distribution), which is quite similar to the experimentally reported values~\cite{BAL_NC5} for Au adatoms. However, for Au adatoms a threefold decrease in $\theta_\mathrm{sH}$ is found when adatoms are segregated into islands with small radius, thus manifesting the detrimental effect of adatom clustering on SHE. This sharply contradicts the predictions of semiclassical transport theories, where $\theta_\mathrm{sH}$ increases with the radius of adatoms clusters~\cite{FER_PRL112}. Nevertheless, a rigorous comparison would require treating a system consisting of identical islands by both theories. Another discrepancy between our numerical exact results in \fref{NewFigSAH} and approximations made in semiclassical theories is that the latter predicts~\cite{FER_PRL112} how $\sigma_{xy}^z$ requires local enhancement of $V_\mathrm{I}$ while being little sensitive to $V_\mathrm{R}$. 

Our additional calculations in \fref{NewFigSAH} using heavier adatoms such as Tl point towards higher charge-to-spin conversion efficiency, even in the presence of adatom clustering. As discussed in \sref{sec:qshe} and ~\aref{sec:soc}, Tl adatoms locally enhance the intrinsic SOC while generating negligible Rashba SOC~\cite{WEE_PRX1}. For large $n_\mathrm{ad}$ and scattered distribution of Tl adatoms, 2D TI phase exhibiting QSHE is predicted~\cite{WEE_PRX1}. The impact of adatom clustering was studied in \cref{CRE_PRL113} where a crossover from QSHE to SHE was predicted upon Tl clustering, see \sref{sec:qshe_cluster}. \Fref{NewFigSAH} shows that clustered Tl adatoms lead to a $\theta_{\rm sH}$ that is similar in shape but larger than for Au adatoms, using the same $n_\mathrm{ad}$. Thus, our quantum transport calculations show that islands with nonzero $V_\mathrm{I}$ (as in the case of clustered Tl adatoms) are more efficient in generating extrinsic SHE than islands with nonzero $V_\mathrm{R}$ (as in the case of Au adatoms). 

In the experiments~\cite{BAL_NC5}, the density of gold clusters of diameter ranging from 20 to 40 nm is estimated to lie within $10^{10}{\rm cm}^{-2}$--$10^{11}{\rm cm}^{-2}$. This leads to \mbox{$n_\mathrm{ad} \simeq$ 2--3\%} assuming that clusters are two-dimensional. The values of $\theta_\mathrm{sH}$ obtained from the Kubo formula calculations assume larger adatom concentration $n_\mathrm{ad}=15\%$. Because of too large mean free paths (above the micrometer) for few percent adatom densities, we cannot (within our present computational capability) reach the diffusive regime in which the Kubo conductivities could be safely estimated. An estimate of $\theta_\mathrm{sH}$ for much lower density is actually not straightforward because the scaling of $\sigma_\mathrm{sH}$ with $n_\mathrm{ad}$ is predicted to ultimately depend on the mechanism dominating the SHE~\cite{Milletari1,Milletari2}. 

\begin{figure}[t!]
	\begin{center}
		\leavevmode
		\resizebox{9cm}{!}{\includegraphics{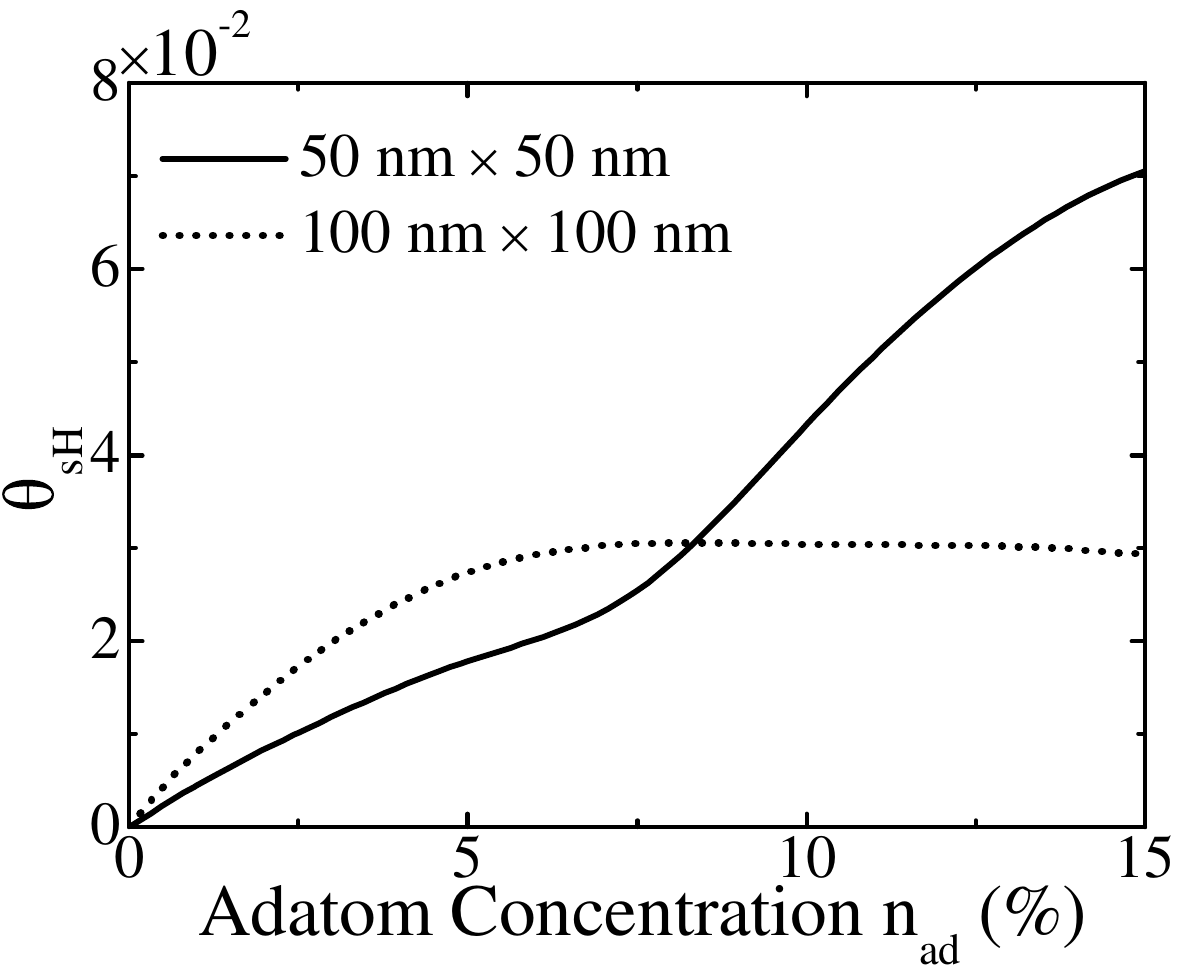}}
		\caption{Spin Hall angle as a function of the concentration of randomly scattered Au adatoms. These results are obtained from the LB formula applied to four-terminal graphene devices whose central square-shaped region of the size \mbox{$50$ nm $\times$ $50$ nm} or \mbox{$100$ nm $\times$ $100$ nm} is attached to four semi-infinite leads of the same respective width. The values of $\theta_\mathrm{sH}$ are averaged over the $E_{\rm F}$ interval $[-0.01 \gamma_0,0.01\gamma_0]$.}
		\label{conc}
	\end{center}
\end{figure}

Neglecting localization effects, the scaling of longitudinal conductivity should follow the Fermi golden rule, $\sigma_{xx} \sim 1/n_\mathrm{ad}$. Similarly, the spin Hall conductivity follows $\sigma_{xy}^z\sim  1/n_\mathrm{ad}$, but only when the SS mechanism predominates the extrinsic SHE~\cite{FER_PRL112,Milletari1,Milletari2}, as confirmed numerically in \fref{fig:rvspia}(a). As discussed in \ccref{Milletari1,Milletari2}, $\sigma_\mathrm{sH}$ should be dominated by a $n_\mathrm{ad}$-independent value for the quantum SJ mechanism, whereas higher order quantum interference terms between scattering paths could lead to $n_\mathrm{ad}^{\alpha}$ dependence (where $\alpha=1,2,...$)~\cite{FER_PRL112,Milletari1,Milletari2,Aires3}. 

Therefore in the limit of small $n_\mathrm{ad}$, $\theta_\mathrm{sH}$ is expected to be either constant or $\propto n_\mathrm{ad}$ ($\propto n_\mathrm{ad}^{\alpha+1}$). The value $n_\mathrm{ad}=15\%$ used in our Kubo formula calculations lies outside the dilute adatom regime where such theories have been developed~\cite{Garcia2016,Milletari1,Milletari2}, but based on the arguments above we extrapolate that for few percent of Au adatom concentration, the maximum value should range within \mbox{$\theta_\mathrm{sH} \simeq$ 0.01--0.1}, where the lower limit is for adatom clusters. Thus, our estimate is about one order of magnitude lower than the value reported in \cref{BAL_NC5}. Finite temperatures and larger clusters will lead to even lower spin Hall angles.

A brute-force calculation of $\theta_\mathrm{sH}$ for {\em arbitrary} adatom concentration in finite-size samples is possible using the multiterminal LB formula approach discussed in \sref{SHENLR}. \Fref{conc} shows that $\theta_\mathrm{sH}$ does increase with the adatom concentration in the limit of low $n_\mathrm{ad}$, with values agreeing with estimates made above. Comparing our results in \fref{conc} with those in \ccref{Garcia2016,Milletari1,Milletari2} suggests that SJ and anomalous quantum processes could {\it dominate} the physics of SHE in graphene decorated with {\it low} concentration of adatoms. 

\subsection{Spin Hall angle and nonlocal resistance in multiterminal adatom-decorated graphene: Landauer-B\"{u}ttiker approach} \label{SHENLR}

\begin{figure}[t!]
	\begin{center}
		\leavevmode
		\resizebox{14cm}{!}{\includegraphics{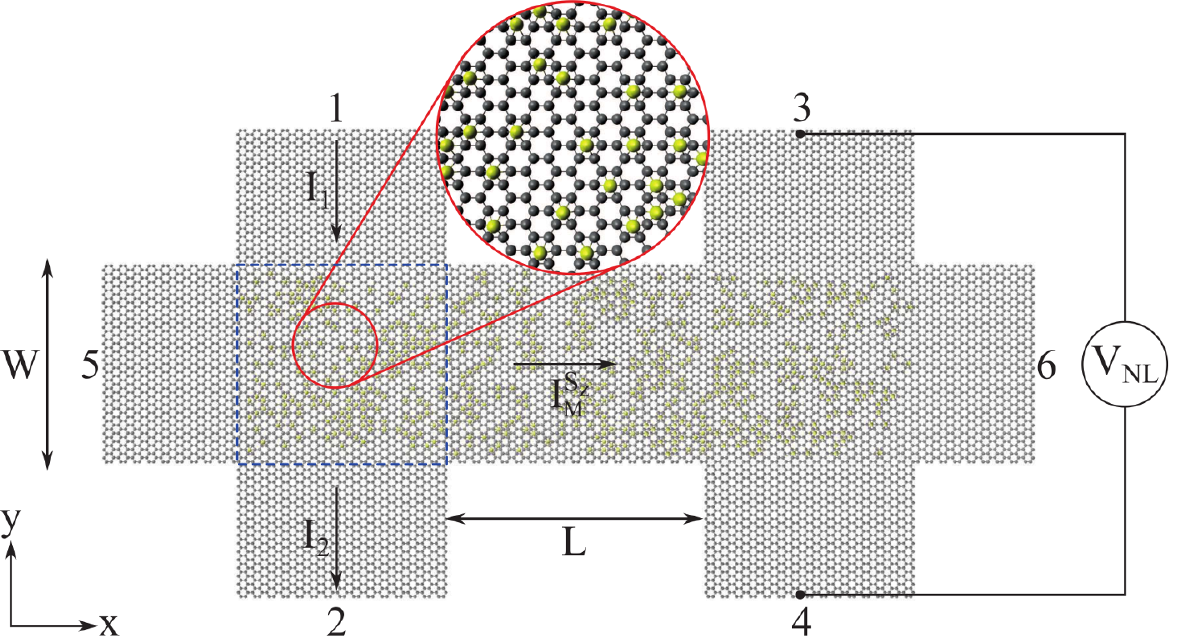}}
		\caption{Schematic view of a six-terminal graphene device used for computing SH angle $\theta_\mathrm{sH}$  and nonlocal resistance $R_{\rm NL}$ {\it via} the LB formula. The central channel is AGNR of width $W=50$ nm (composed of $3m+2$ dimer lines, so that its electronic structure resembles that of large-area graphene \cite{CRE_NR1}) and variable length $L=10$--$300$ nm, while the attached ideal leads are modeled as pristine ZGNRs. In the zoom, black circles represent carbon atoms and yellow circles label positions of scattered or clustered Au adatoms. The dashed square denotes the sample (of the size \mbox{$400$ nm $\times$ $400$ nm} and with periodic boundary conditions) used in the calculations of spin and charge conductivities {\it via} the Kubo formula in \sref{SHEKUBO}. }
		\label{NewFig1}
	\end{center}
\end{figure}

To describe nonlocal transport in SHE experiments~\cite{BAL_NC5,Balakrishnan2013}, we simulate six-terminal graphene devices, as depicted in \fref{NewFig1}, by using the LB formula (see \aref{sec:qtalgorithms}) as efficiently implemented in the KWANT package~\cite{Groth2014}. In the SHE-based explanation for the origin of nonlocal signal, the injected transverse charge current between leads 1 and 2 generates the longitudinal spin current $I_5^{S_z}$ in lead 5, as well as the putative mediative spin current $I_\mathrm{M}^{S_z}$. The conversion of $I_\mathrm{M}^{S_z}$ {\it via} the inverse SHE into the voltage $V_{\rm NL}=V_3-V_4$ between the leads 3 and 4 then gives nonzero $R_{\rm NL}$. Similarly to \eref{eq:shadef}, the SH angle for the device geometry in \fref{NewFig1} is obtained from $\theta_{\rm sH}=I_{5}^{S_{z}}/I_{1}$, and the nonlocal resistance is given by $R_{\rm NL}=V_{\rm NL}/I_1=(V_3-V_4)/I_1$.

To understand the different mechanisms, including those not related to spin transport, that can contribute to $R_{\rm NL}$, or the importance of resonant impurity scattering, we analyze three different situations. First, we consider the case for which no adatom is present in the central region of the device in \fref{NewFig1}. Second, we consider the device in \fref{NewFig1} where a homogeneous Rashba SOC is present within the entire central region. Finally, we consider central region with random distribution of Au adatoms and with concentration $n_\mathrm{ad}=15\%$, which can be either scattered or clustered into islands of radius \mbox{$\in [1,3]$ nm}, in complete analogy with the Kubo formula calculations presented in \sref{SHEKUBO} and using the same parameters $V_\mathrm{I} = 0.007 \gamma_{0}$, $V_\mathrm{R}=0.0165 \gamma_{0}$, $V_\mathrm{PIA}=0$ and $\mu = 0.1 \gamma_{0}$ in \eref{eq:kanemele} for scattered Au adatoms~\cite{DIN_NP10} residing on H-sites or $\mu = 0.3 \gamma_{0}$ for clustered Au adatoms.

\begin{figure}[t!]
	\begin{center}
		\leavevmode
		\resizebox{9cm}{!}{\includegraphics{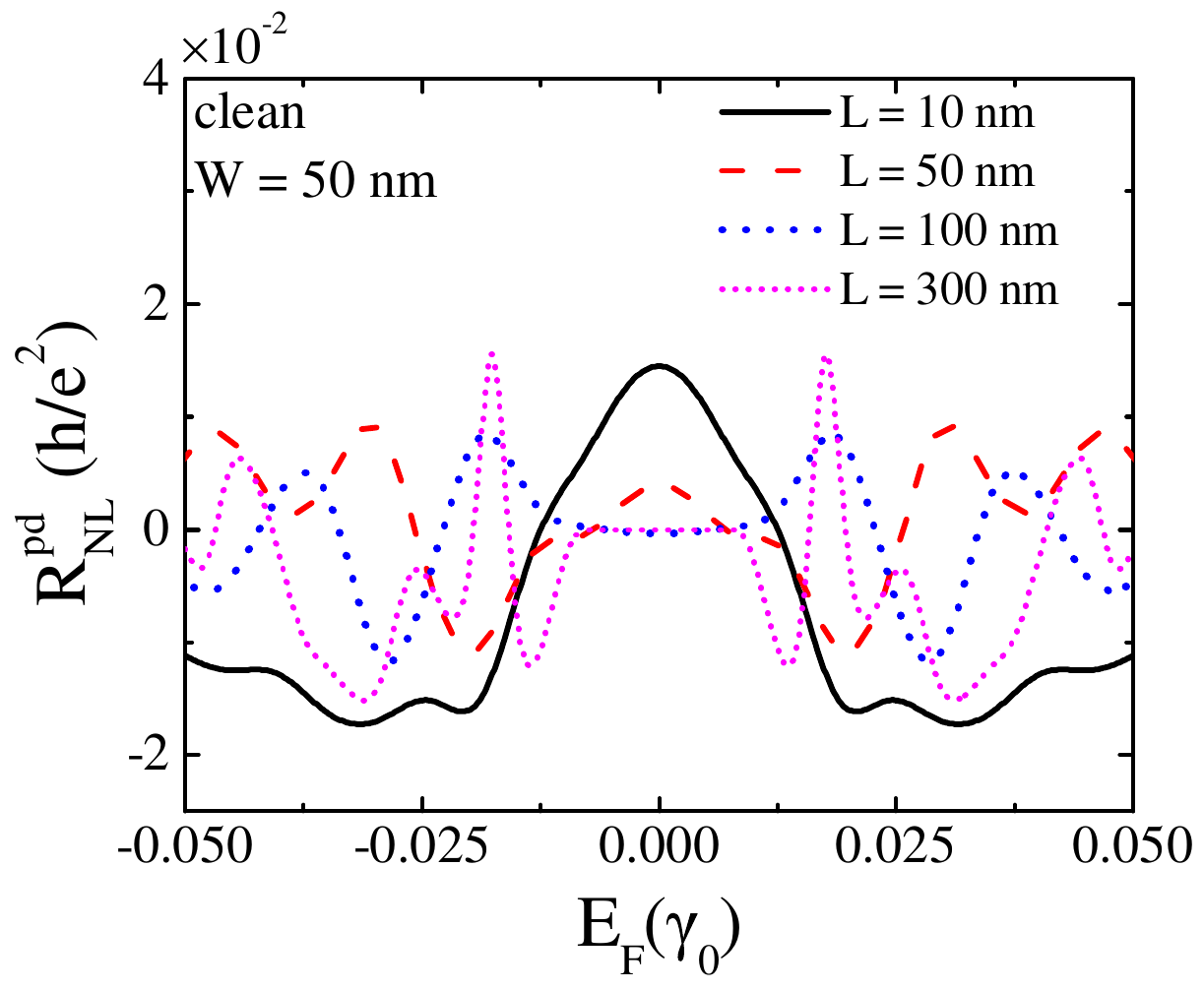}}
		\caption{Nonlocal resistance $R_\mathrm{NL}$ computed {\it via} the LB formula for six-terminal graphene device in \fref{NewFig1} which is perfectly clean (i.e., no adatoms, defects or impurities in either its central region or the attached six leads). Adapted from \cref{TUA_arXiv:1603.03870}.}
		\label{NewFig4S}
	\end{center}
\end{figure}

\Fref{NewFig4S} shows the scaling of $R_\mathrm{NL}$ with the length $L$ (at fixed width $W$) for pristine graphene device in \fref{NewFig1}. Such a positive nonlocal signal is specific to Dirac electron systems, like graphene~\cite{Tworzydlo2006} or 3D TI metallic surfaces~\cite{CHA_NL14}, where evanescent wave functions penetrate through zero gap of the Dirac cone to generate the  ``pseudodiffusive'' transport close to CNP. The pseudodiffusive transport regime is characterized by Ohmic-like two-terminal conductance $G \propto 1/L$~\cite{CHA_NL14,Tworzydlo2006,CRE_PRB76}, even though the device is perfectly clean. This mechanism is expected to provide background contribution $R_\mathrm{NL}^\mathrm{pd}$ to total $R_\mathrm{NL}$, as long as $W > L$,  as confirmed in \fref{NewFig4} for Au-adatom-decorated graphene.

\begin{figure}[t!]
	\begin{center}
		\leavevmode
		\resizebox{13cm}{!}{\includegraphics{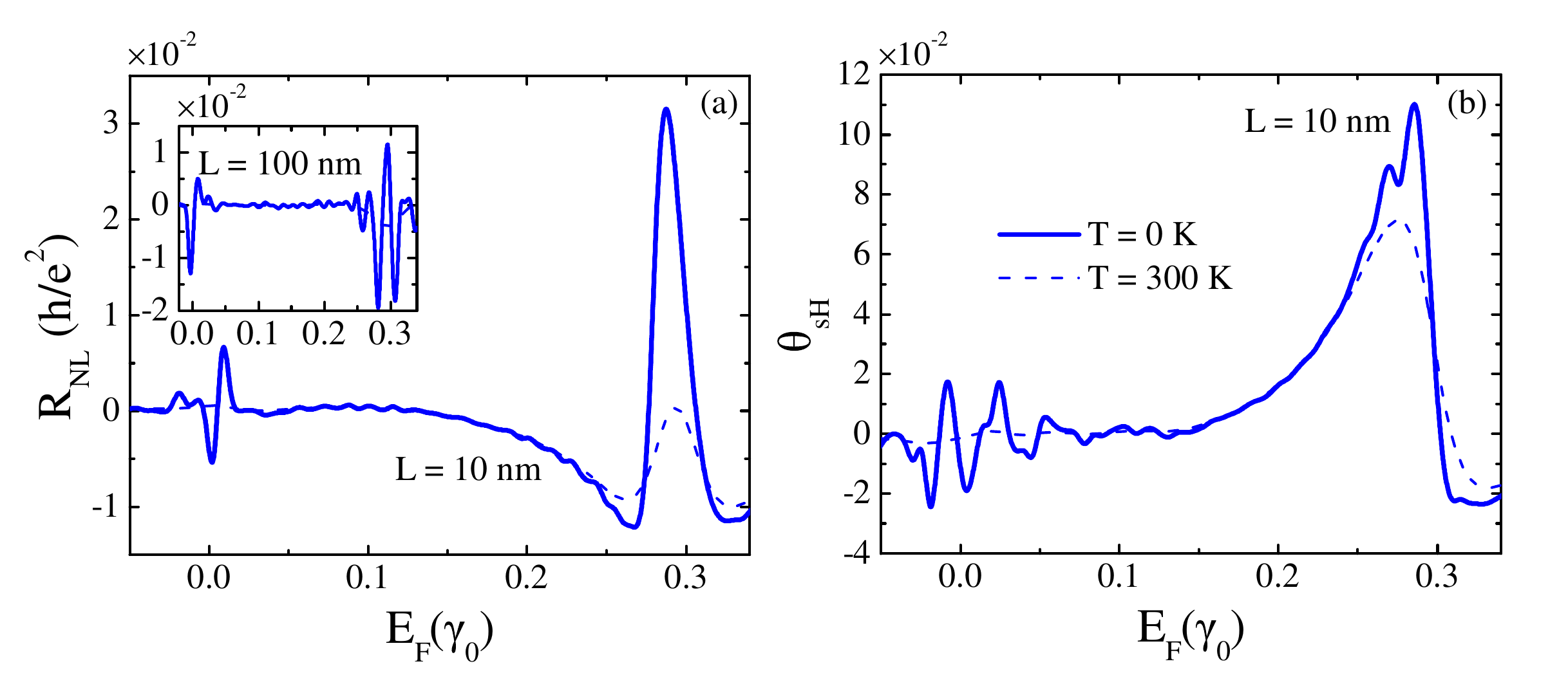}}
		\caption{(a) Nonlocal resistance $R_{\rm NL}$ and (b) SH angle as a function of the $E_{\rm F}$ in six-terminal graphene device with uniform distribution of Au adatoms. Since every hexagon within the central region in \fref{NewFig1} is covered by one Au adatom, this setup is described by homogeneous Rashba SOC term in \eref{eq:kanemele}. Adapted from \cref{TUA_arXiv:1603.03870}.}
		\label{NewFig3S}
	\end{center}
\end{figure}

\begin{figure}[t!]
	\begin{center}
		\leavevmode
		\resizebox{13cm}{!}{\includegraphics{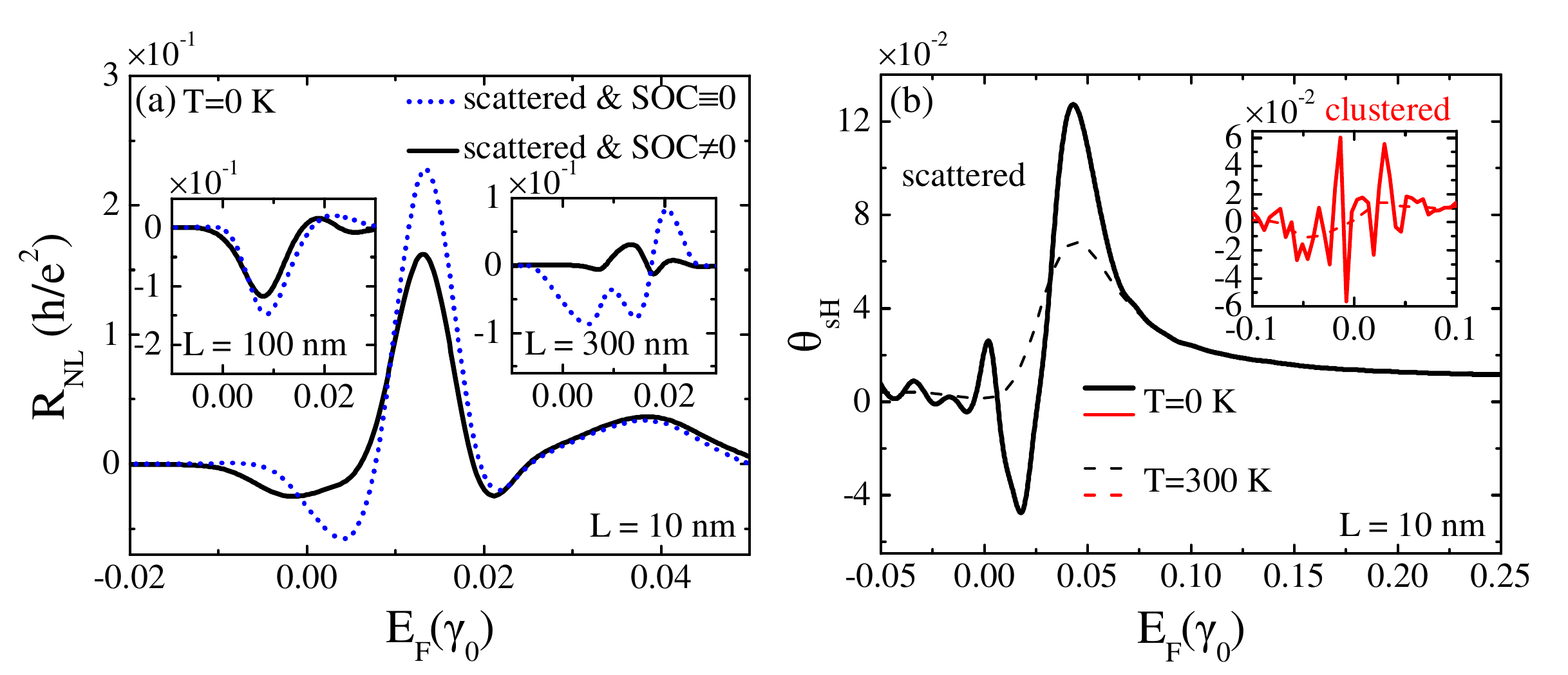}}
		\caption{(a) Nonlocal resistances $R_\mathrm{NL}$ as a function of $E_{\rm F}$ for various channel lengths---$L = 10$ nm (main frame); $L = 100$ nm (left inset); and $L = 300$ nm (right inset)---at fixed channel width $W = 50$ nm of six-terminal graphene device in \fref{NewFig1} with $n_\mathrm{ad}=15$\%  concentration of scattered Au adatoms. 
			$R_\mathrm{NL}$ of non-SHE (SOC $\equiv 0 \Leftrightarrow V_\mathrm{I} = V_\mathrm{R} = 0$) origin is plotted as dotted line. (b) The SH angle $\theta_\mathrm{sH}$ for the same $n_\mathrm{ad}$ as in panel (a), where adatoms are scattered (main frame) or clustered (inset).  All curves are averaged over 10 adatom configurations. Adapted from \cref{TUA_arXiv:1603.03870}.}
		\label{NewFig4}
	\end{center}
\end{figure}

\Fref{NewFig3S} shows $R_{\rm NL}$ and $\theta_{\rm sH}$ for the case of a  uniform distribution of Au adatoms, where each hexagon within the central region hosts one Au adatom. Both quantities are calculated at temperatures $T=0$ K and $T=300$ K, where the latter includes thermal broadening effects in the LB calculations. The uniform Rashba SOC generates the intrinsic SHE in multiterminal devices, akin to the one found in multiterminal 2DEGs~\cite{Nikolic2006,Nikolic2005b,Nikolic2009,Sheng2005a}. The large value of the nonlocal signal and $\theta_{\rm sH}$ is observed away from CNP due to doping of graphene by $\mu = 0.3 \gamma_{0}$ (chosen by viewing central region as a single large cluster) in \eref{eq:kanemele}. The SH angle and $R_{\rm NL}$ due to such an intrinsic SHE are actually smaller than the same quantities observed for scattered Au adatoms in \fref{NewFig4} (especially for the ``pure SHE'' device setup in \fref{Fig5S} and the corresponding results in \fref{NewFig6S}). This confirms the importance of resonant scattering off adatoms for enhancing the extrinsic SHE, a conclusion reached also in semiclassical transport theories~\cite{FER_PRL112}.

The SH angle for graphene with scattered or clustered Au adatoms is presented in \fref{NewFig4}(b), and can be compared with the corresponding Kubo formula results in \fref{NewFigSAH}. The value of $\theta_{\rm sH}\sim 0.1$ in the scattered case, as well as decrease of $\theta_{\rm sH}$ from scattered to clustered Au adatom distribution, are in full accord with the conclusions obtained from the Kubo formula calculations. We also find that thermal broadening reduces $\theta_{\rm sH}$. 

\begin{figure}[t!]
	\begin{center}
		\leavevmode
		\resizebox{15cm}{!}{\includegraphics{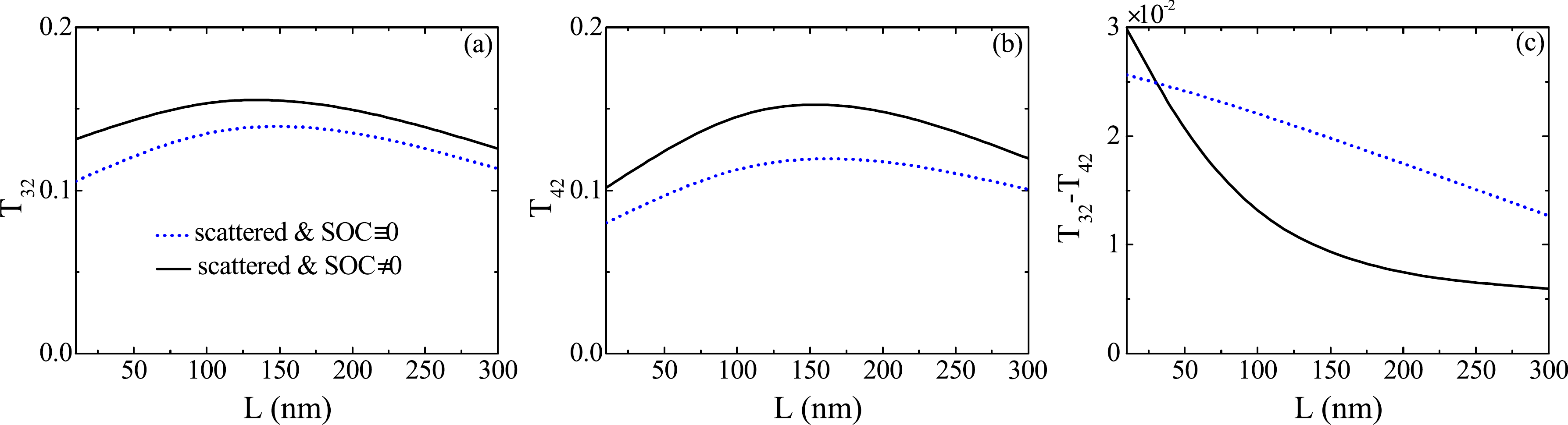}}
		\caption{Scaling with the length $L$ of the transmission functions $T_{pq}$ from lead $q$ to lead $p$ in six-terminal graphene device of width $W=50$ nm in \fref{NewFig1}: (a) $T_{32}$; (b) $T_{42}$; and (c) their difference, $T_{32}-T_{42}$. The central region is covered by scattered distribution of Au adatoms of concentration $n_\mathrm{ad}=15\%$, while their SOC is switched on (solid lines) or off (dotted lines). All curves are obtained by averaging $T_{pq}(E)$ over the energy interval $[-0.01 \gamma_0,0.01\gamma_0]$.}
		\label{FigS32}
	\end{center}
\end{figure}

Surprisingly, we observe large $R_{\rm NL}$ in \fref{NewFig4}(a) even when all SOC terms are artificially switched off ($V_\mathrm{R}=V_\mathrm{I}=0$), while keeping random on-site potential $\mu \neq 0$ due to Au adatoms. In addition, we find a complex sign change of $R_{\rm NL}$ with varying the channel length $L$. The change of sign of $R_{\rm NL}$ with increasing channel length from \mbox{$L=10$ nm} to \mbox{$L=300$ nm} suggests the following interpretation. The total $R_{\rm NL}$ has in general {\em four} contributions 
\begin{equation}\label{eq:totalrnl}
R_\mathrm{NL}=R_\mathrm{NL}^\mathrm{SHE}+R_\mathrm{NL}^\mathrm{\mathrm{Ohm}}+R_\mathrm{NL}^\mathrm{qb}+R_\mathrm{NL}^\mathrm{pd},
\end{equation}
assuming they are additive after disorder averaging. For unpolarized charge current injected from lead 1 (i.e., electrons injected from lead 2): 
\begin{itemize}
\item $R_\mathrm{NL}^\mathrm{SHE}$ due to the combined direct and inverse SHE has a positive sign; 
\item $R_\mathrm{NL}^\mathrm{Ohm}$ is trivial Ohmic contribution due to classical diffusive charge transport~\cite{Abanin2009} and has a positive sign; 
\item $R_\mathrm{NL}^\mathrm{qb}$ is the negative quasiballistic contribution arising due to direct transmission $T_{32} \neq 0$ from lead 2 to lead 3, as observed previously in SHE experiments on multiterminal gold devices~\cite{Mihajlovic}; 
\item $R_\mathrm{NL}^\mathrm{pd}$ is a positive contribution due to pseudodiffusive transport specific to graphene, as explained in \fref{NewFig4S}. 
\end{itemize}

\begin{figure}[t!]
	\begin{center}
		\leavevmode
		\resizebox{14cm}{!}{\includegraphics{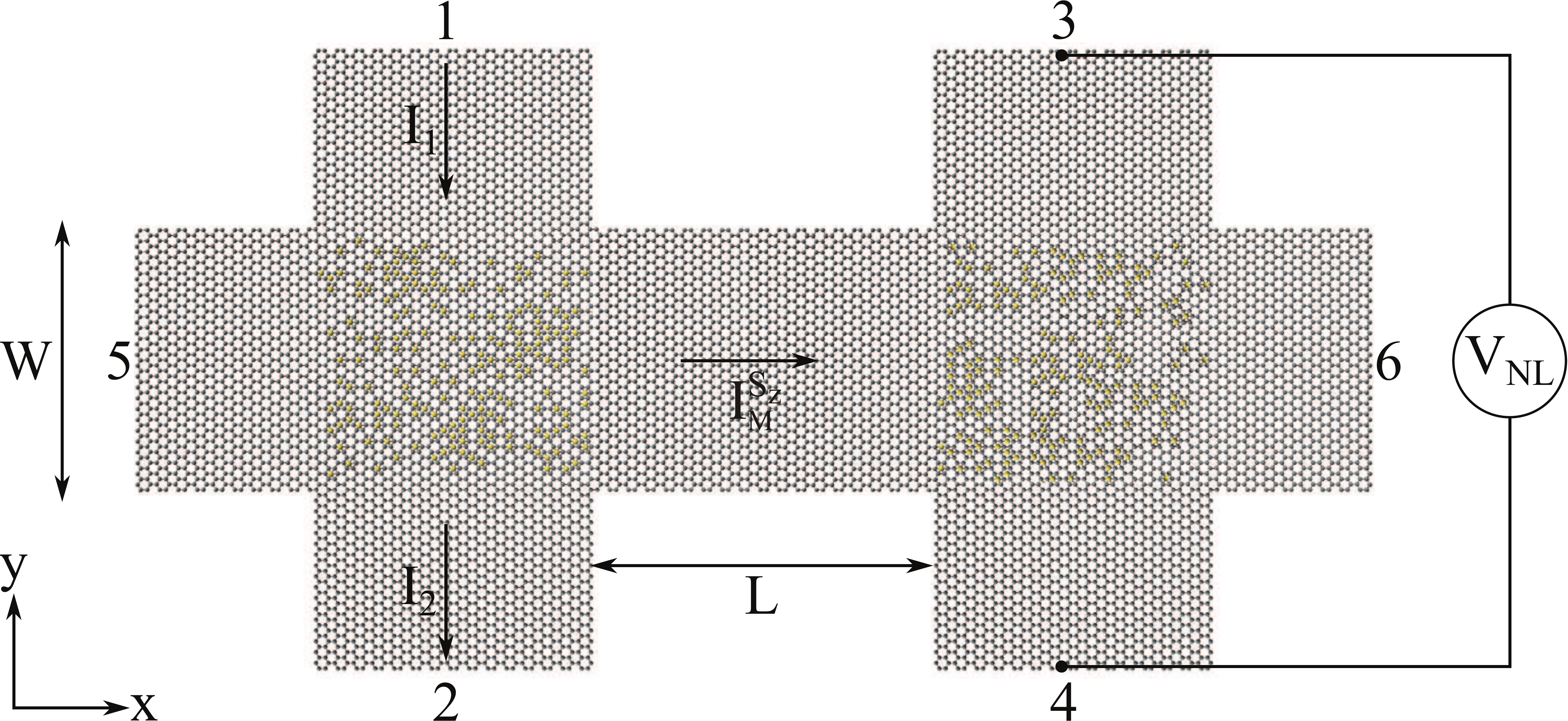}}
		\caption{Schematic view of the proposed six-terminal graphene device where adatoms in the channel connecting two crossbars are removed in order to isolate $R_{\rm NL}^{\rm SHE}$ by bringing other three contributions to total $R_\mathrm{NL}$ in \eref{eq:totalrnl} to zero. The concentration of Au adatoms is $n_\mathrm{ad}=15$\% within the left and the right crossbar area.}
		\label{Fig5S}
	\end{center}
\end{figure}

\begin{figure}[h!]
	\begin{center}
		\leavevmode
		\resizebox{12cm}{!}{\includegraphics{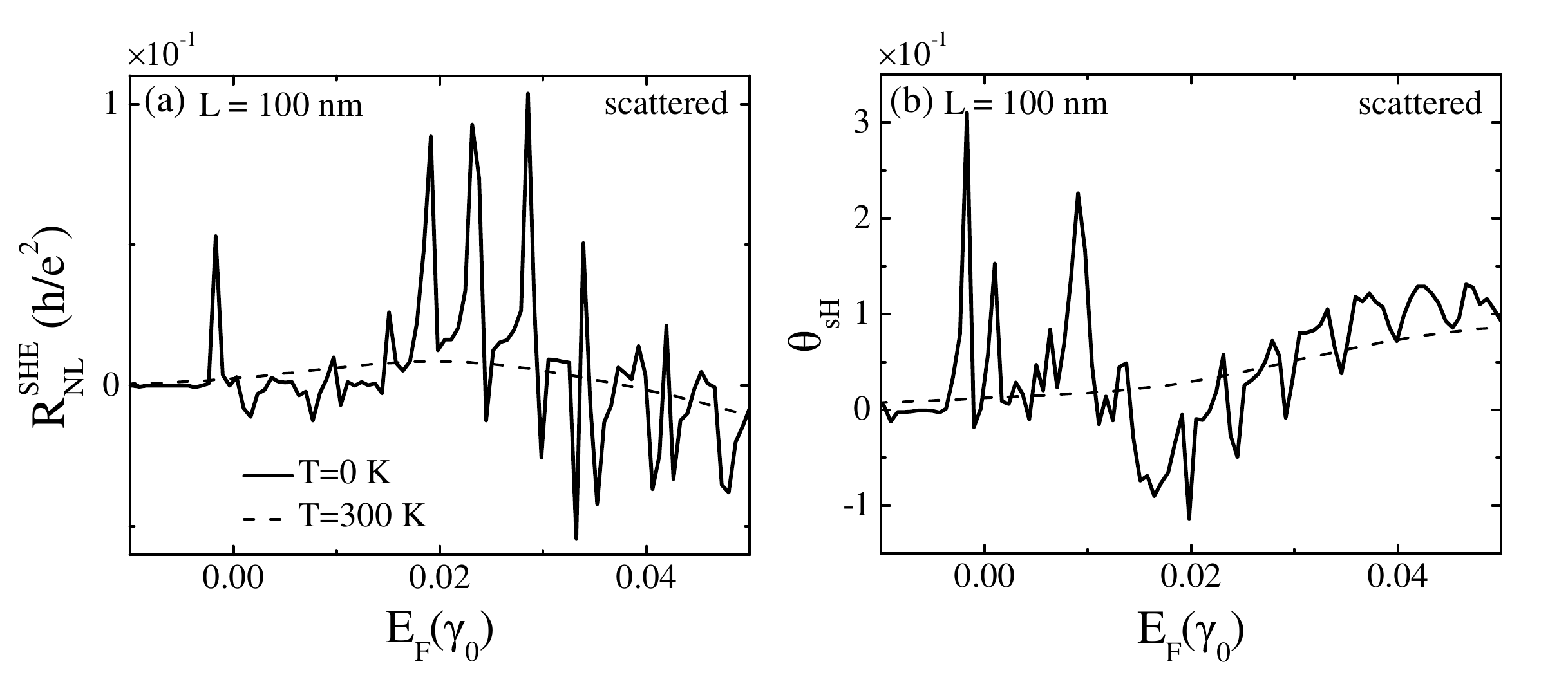}}
		\caption{ (a) Nonlocal resistance $R_{\rm NL}^{\rm SHE}$ and (b) SH angle $\theta_{\rm sH}$ for six-terminal device shown in \fref{Fig5S}. Adapted from \cref{TUA_arXiv:1603.03870}.}
		\label{NewFig6S}
	\end{center}
\end{figure}

In device geometries with $W>L$, such as for $W=50$ nm and $L=10$ nm case in the main frame of \fref{NewFig4}(a), positive sign $R_\mathrm{NL}$ is dominated by $R_\mathrm{NL}^\mathrm{pd}$ akin to \fref{NewFig4S}. 
However, due to scattering of impurities of uniform strength at CNP \cite{Titov2007}, $R_\mathrm{NL}^\mathrm{pd}$ in the main frame of \fref{NewFig4}(a) can be larger than in the case of perfectly clean multiterminal graphene in \fref{NewFig4S}.
The negative sign of $R_\mathrm{NL}$ in the two insets in \fref{NewFig4} in the absence of SOC, $V_\mathrm{R}=V_\mathrm{I}=0$, and for $L>W$ suggests that $R_\mathrm{NL}^\mathrm{Ohm}$ can be safely neglected in our samples due to small concentration of adatoms. That is, we can estimate the mean free path $\ell$ for $n_\mathrm{ad}=15\%$ concentration of Au adatoms to be between 300 nm and 400 nm, so that when diffusive transport regime sets in for $\ell < L$  the Ohmic contribution scaling as $R_\mathrm{NL}^\mathrm{Ohm} \propto \exp(-\pi L/W)$~\cite{Kaverzin2015,Abanin2009} is already negligible due to $L/W \gg 1$. 

Therefore, for $L>W$ the main competition is between $R_\mathrm{NL}^\mathrm{qb}$ with negative sign and $R_\mathrm{NL}^\mathrm{SHE}$ with positive sign, as found in the two insets of \fref{NewFig4}(a). 
\Fref{FigS32} shows the scaling of the transmission function $T_{pq}=\mathrm{Tr}[\mathbf{t}_{pq} \mathbf{t}_{pq}^\dagger]$ in \eref{eq:gpq} with the length $L$ (at fixed width $W$) for electron paths from lead $2 \rightarrow 3$ and lead $2 \rightarrow 4$, as well as their difference, in six-terminal graphene device in \fref{NewFig1}. The difference $T_{32} - T_{42}$ being positive means that more electrons arriving into lead 3 than in lead 4 will cause negative $R_\mathrm{NL}^\mathrm{qb}$ at some intermediate length scales. The slow decay of quantities in \fref{FigS32} characterizing the quasiballistic transport regime can manifest as long as the channel length $L$ is smaller than the mean free path.
Thus, the existence of contributions to $R_\mathrm{NL}$ that do not originate from SHE and can be much larger than $R_\mathrm{NL}^\mathrm{SHE}$ could account for the insensitivity of the total $R_\mathrm{NL}$ in \eref{eq:totalrnl} to the applied external in-plane magnetic field observed in some experiments \cite{Wang2015a,Kaverzin2015}.

The difficulty in clarifying the dominant contribution to $R_\mathrm{NL}$ could be resolved by detecting its sign change as a function of the channel length $L$ in \fref{NewFig1}. An alternative is to design a setup where $R_\mathrm{NL}^\mathrm{\mathrm{Ohm}}$, $R_\mathrm{NL}^\mathrm{qb}$, and $R_\mathrm{NL}^\mathrm{pd}$ are negligible so that $R_\mathrm{NL}^\mathrm{SHE}$ can be isolated. We propose such setup in \fref{Fig5S} where adatoms are completely removed from the channel. When such perfectly clean channel is sufficiently long, $R_\mathrm{NL}^\mathrm{pd}=0$ due to $L>W$ and $R_\mathrm{NL}^\mathrm{Ohm}, \, R_\mathrm{NL}^\mathrm{qb} \rightarrow 0$ due to the absence of impurity scattering in the channel, so that mediative spin current $I^{S_z}_\mathrm{M}$ generated by direct SHE in the first crossbar arrives conserved~\cite{Nikolic2006} at the second crossbar where it is converted into $V_\mathrm{NL}$ by the inverse SHE.  Indeed, \fref{NewFig6S} demonstrates that $R_\mathrm{NL}$ and $\theta_\mathrm{sH}$ in this setup are {\it unambiguously} related since they both display sharp peak at virtually the same $E_F$ very close to CNP.  

%%%%%% ZEEMAN SPIN HALL EFFECT IN MULTITERMINAL GRAPHENE %%%%%%%%%%%%%%%%%%%%%%%%%%%%%%%%

\section{Zeeman spin Hall effect in multiterminal graphene}\label{sec:zshe}

\begin{figure}
	\center{\includegraphics[scale=0.35,angle=0]{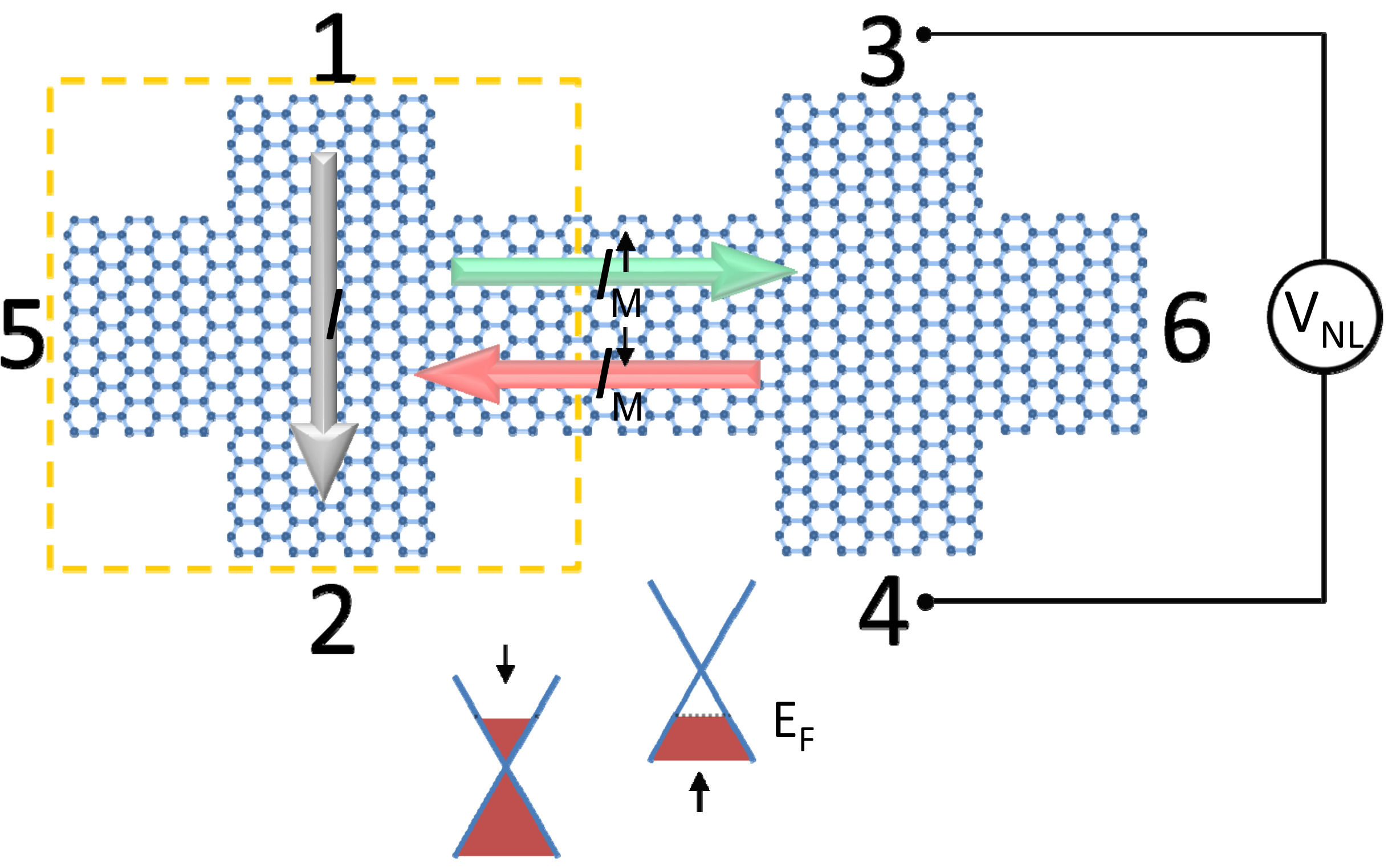}}
	\caption{Schematic view a six-terminal graphene geometry used for computing the nonlocal voltage $V_\mathrm{NL}=V_3-V_4$ and the corresponding nonlocal resistance $R_\mathrm{NL}=(V_3-V_4)/I_1$ due to ZSHE. The central region of the device consists of an AGNR (of width $W$) in longitudinal direction and a portion of transverse ZGNR leads attached to it. Dashed square encloses the device used in four-terminal quantum transport simulations. For simplicity, out-of-plane external magnetic field and many-body interactions causing dephasing are assumed to be present only within the central region of the four-terminal or six-terminal devices, while the attached ZGNR leads are assumed to be ideals ones (i.e.,  free of spin and charge interactions).}
	\label{fig:fig1_zshe}
\end{figure}

The Zeeman SHE (ZSHE) is a phenomenon where an injected unpolarized longitudinal charge current generates a transverse spin current in graphene under an out-of-plane magnetic field~\cite{CHE_PRB85,Abanin2011,Renard2014,Wei2016,Abanin2011a}. Unlike the conventional SHE discussed in \sref{sec:she}, ZSHE does not require SOC. Instead, out-of-plane magnetic field splits the Dirac cone of graphene by the Zeeman interaction, where electron- and hole-like carriers acquire opposite spins near CNP, as illustrated by the inset in \fref{fig:fig1_zshe}. 

The ZSHE was originally discovered~\cite{Abanin2011} by detecting a nonlocal resistance in multiterminal graphene devices placed in an external out-of-plane magnetic field.  The nonzero $R_\mathrm{NL}$ can be explained using \fref{fig:fig1_zshe}, where an injected unpolarized charge current flowing between leads 1 and 2 generates {\it via} ZSHE the longitudinal spin current $I_5^{S_z}$ in lead 5 and the mediative pure spin current $I^{S_z}_\mathrm{M} =I^\uparrow_\mathrm{M} - I^\downarrow_\mathrm{M}$ flowing toward lead 6. The mediative spin current is then converted {\it via} the inverse ZSHE into the voltage drop $V_\mathrm{NL}$ between the leads 3 and 4, and the corresponding nonlocal resistance is defined by $R_{\rm NL} = (V_3-V_4)/I_{1}$, in complete analogy with nonlocal signal discussed in \sref{SHENLR}. The distance between the pairs of contacts 1 and 2 and the pairs of contacts 3 and 4 in experiments is few microns~\cite{Abanin2011,Renard2014,Wei2016}. 

The nonlocality in electronic transport, signified by the voltage $V_\mathrm{NL}$ in \ffref{fig:fig1_zshe} and \ref{NewFig1} between contacts that are far from the classical path of injected charge current, is a rare and highly non-trivial effect.  It has been previously associated with phase coherence of single electrons (such as in systems exhibiting QHE~\cite{McEuen1990,Haug1993} and QSHE~\cite{Roth2009}), or long-range order in interacting many-electron systems (such as charge density waves and superconductors). For example, in sufficiently high magnetic field and at sufficiently low temperatures, phase-coherent transport of independent electrons through edge states~\cite{Li2013} of QHE will generate peaks in $R_\mathrm{NL}$ at specific charge densities~\cite{Abanin2011} (e.g., in 2DEGs in QHE regime such peaks have been observed even for distances of $\sim 1$ mm between the pairs of contacts 1 and 2  and the pairs of contacts 3 and 4~\cite{McEuen1990,Haug1993}). However, the peak of $R_\mathrm{NL}$ at CNP is observed~\cite{Abanin2011} even in weak magnetic fields $B \simeq 1$ T and at room temperature $T=300$ K, which is clearly outside of the QHE regime.

Recently, the ZSHE induced $R_\mathrm{NL}$ was enhanced by an order of magnitude by replacing external out-of-plane magnetic field with magnetic exchange field ($>14$ T, with the potential to reach hundreds of Tesla) from a ferromagnetic insulator overlayer covering graphene, which points to possible spintronic  applications~\cite{Wei2016}. For example, in contrast to conventional spin injection of spin-polarized electrons into semiconductors and 2D materials, where tunneling through a barrier is typically employed, the ZSHE directly spin-polarizes electrons.

The same multiterminal geometry for measuring $R_\mathrm{NL}$ associated with ZSHE has been later employed to measure $R_\mathrm{NL}$ associated with SHE in graphene with adatom-induced SOC (\sref{sec:she}) or VHE (\sref{sec:vhe}). The seminal experiments~\cite{Abanin2011} on ZSHE have also provided a blueprint on how to make graphene devices with high mobility (between $5\times 10^4$ and $1.5\times 10^5$ cm$^2$/Vs for carrier concentrations $n \sim 10^{11}$ cm$^{-2}$) by using thin crystals of h-BN as a substrate. Such devices exhibit $R_\mathrm{NL}$ that is 10 to 100 times larger than in conventional devices, where graphene is placed on top of an oxidized Si wafer~\cite{Abanin2011}. The usage of atomically flat h-BN substrate rules out Rashba SOC~\cite{Winkler2003} (due to charge impurities from the substrate~\cite{ERT_PRB80} or lattice distortion by adatoms~\cite{CastroNeto2009}) that could otherwise generate $R_\mathrm{NL}$ {\it via} combined direct and inverse SHE discussed in \sref{SHENLR}. 

Note that some ZSHE experiments~\cite{Renard2014} have also revealed non-spin-related effects (such as thermomagnetic ones) contributing to $R_\mathrm{NL}$, whose signal can be even larger than due to ZSHE. This highlights the same issue discussed in \sref{sec:she}---{\em that care must be taken when associating measured $R_\mathrm{NL}$ to putatively dominant microscopic mechanism}.

An intuitive picture of ZSHE can be constructed simply by using classical Newtonian dynamics of massless Dirac fermions where charge of the electron behaves incoherently but its spin behaves coherently and is, therefore, described by quantum mechanics. The classical Hamiltonian of low-energy quasiparticles close to CNP is given by $H^{\pm }({\bf p})=\pm v_{F}\sqrt{p_{x}^2+p_{y}^2}$, which in the weak  external magnetic field ${\bf B}=\nabla \times {\bf A}$ becomes $H^{\pm}({\bf p})=\pm v_{F}\sqrt{(p_{x}-eA_{x})^2+(p_{y}-eA_{y})^2}$. The classical velocity is then given by $v_{x,y}^\pm = \partial H^{\pm}/\partial p_{x,y}=\pm v_{F} {\Pi}_{x,y}/\sqrt{{\bm \Pi}^2}$, where ${\bm \Pi}={\bf p} - e{\bf A}$, and the corresponding acceleration is
\begin{equation}
{\bf a}^\pm = \frac{d {\bf v}}{d t} = \pm \frac{e v_{F} {\bf v}^\pm \times {\bf B}}{\sqrt{{\bm \Pi}^2}} =  \frac{e v^2_{F} {\bf v}^\pm \times {\bf B}}{E^\pm} \ .
\end{equation}
Thus, the quasiparticles with energy $E^+$ above CNP (or below with energy $E^-$) moving in a weak (i.e., non-quantizing) perpendicular magnetic field will experience a transverse force, which deflects them to the left (right). Furthermore, when $E^\pm$ is very close to CNP such a deflecting force will be very large.

Although the Zeeman splitting $\Delta_{\rm Z}$ in 2DEG is typically small in a weak external magnetic field~\cite{Winkler2003}, it can play an essential role in graphene for temperatures $k_BT < \Delta_{\rm Z}$ by shifting the Dirac cones for opposite spins to induce two types of carriers illustrated in the inset in \fref{fig:fig1_zshe}. The quasiparticles with energy $E^-$ are \mbox{spin-$\uparrow$} polarized, while those with energy $E^+$ are \mbox{spin-$\downarrow$} polarized. These two effects, classical for charge and quantum for spin, conspire to generate transverse spin current in response to longitudinal charge current, as illustrated in \fref{fig:fig1_zshe}. Such a phenomenology is similar to SHE in multiterminal graphene (see \sref{SHENLR}) and 2DEG devices~\cite{Nikolic2005b,Brune2010}, even though no SOC is involved to provide the deflecting force~\cite{Nikolic2005c} of opposite direction for opposite spins, as exemplified by \eref{eq:soforce}.

These simple arguments for the existence of ZSHE in graphene can be converted into a semiclassical transport theory based on the Boltzmann transport equation (BTE)~\cite{Abanin2011a}. However, the BTE approach is known to give incorrect results for transport properties close to 
CNP~\cite{Adam2009,Klos2010}. When applied to ZSHE, it is valid in high-$T$ and weak-$B$ regime, while experiments~\cite{Abanin2011} have observed increasingly more profound nonlocality in the low-$T$ and/or strong-$B$ regime, so that a unified theory is called for that can cover such a wide range of parameters. For example, such a theory should explain the nonlocal voltage in strong (i.e., quantizing) external magnetic field, as well as at intermediate temperatures where edge-state transport mechanism is removed.

The {\em fully} quantum transport theory of ZSHE in multiterminal graphene devices was formulated in \cref{CHE_PRB85}. It is based on NEGF formalism~\cite{Stefanucci2013}, where its combination with phenomenological~\cite{Golizadeh-Mojarad2007a} many-body self-energies that take into account dephasing processes involving simultaneous phase and momentum relaxation is delineated in \aref{sec:qtalgorithms}. This approach intrinsically accounts for the contributions of both electrons and holes, which is crucial to describe transport near CNP~\cite{Abanin2011a}. It can also handle arbitrary scattering processes, in contrast to semiclassical theories of charge~\cite{Adam2009,Klos2010} and spin~\cite{DIN_NP10} transport, which are known to break down close to CNP. Finally, it yields the spin Hall angle $\theta_\mathrm{sH}$, as the ratio of transverse spin Hall and longitudinal charge currents characterizing the strength of any SHE~\cite{SHE2,VIG_JSNM23,Hoffmann2013}, as well as experimentally measured nonlocal resistance $R_\mathrm{NL}$. In contrast, semiclassical theories~\cite{FER_PRL112,Abanin2011a} of SHE in graphene are typically focused on computing only $\theta_\mathrm{sH}$, which, however, is not directly measurable quantity due lack of ``spin current ammeter''~\cite{Adagideli2006}. 

The results of quantum transport simulations of ZSHE in multiterminal graphene devices presented in \ffref{fig:fig2_zshe}, \ref{fig:fig3_zshe} and \ref{fig:fig4_zshe} demonstrate how this approach interpolates smoothly between the phase-coherent transport regime at low temperatures and in the quantizing external magnetic field and the semiclassical transport regime at higher temperatures. In particular, dephasing by many-body interactions destroys features found at low-$T$ while leaving peaks (of reduced magnitude) in the spin Hall conductance and nonlocal voltage around CNP, in full accord with experiments~\cite{Abanin2011,Renard2014,Wei2016}.

Close to the CNP, graphene in an external magnetic field can be described by the TB Hamiltonian with a single $2p_z$ orbital {\it per} site
\begin{equation}\label{eq:hamilton_zshe}
H = \sum_{i} (\varepsilon_i + g \mu_B \sigma B )c_{i \sigma}^\dagger c_{i \sigma} - \gamma_0 \sum_{\langle ij \rangle, \sigma} e^{i \phi_{ij}} c_{i \sigma}^\dagger c_{j \sigma} \ .
\end{equation}
Here $c_{i \sigma}^\dag$ ($c_{i \sigma}$) creates (annihilates) electron with spin $\sigma$ in the $2p_z$ orbital located on site $i$; $\varepsilon_i$ is the on-site energy;  $\sigma = +1$ for spin-$\uparrow$ electron and $\sigma = -1$ for spin-$\downarrow$ electron so that Zeeman splitting is given by $\Delta_{\rm Z}=2 g \mu_B B$ with $g=2.0$; and $\gamma_0$ is the nearest-neighbor hopping parameter. 
The external magnetic field enters through the Peierls phase factor $\phi_{ij}$, see \eref{eq:TBH} in \sref{sec:qhe}.

The active region of the device in \fref{fig:fig1_zshe} consists of an AGNR and a portion of semi-infinite ideal leads modeled as ZGNRs. The electronic structure and the DOS of an AGNR composed of $3m+2$ dimer lines resemble~\cite{CRE_NR1} (if we assume that only the nearest-neighbor hopping $\gamma_0$ is non-zero) those of large-area graphene employed experimentally. Although ZGNRs are insulating at very low temperatures due to one-dimensional spin-polarized edge states coupled across the width $W$ of the GNR, such an unusual magnetic ordering and the corresponding band gap is easily destroyed above \mbox{$T \gtrsim 10$ K} so that we employ them as a model for ideal metallic leads~\cite{Yazyev2008,Kunstmann2011}. The weak vs. strong magnetic field is tuned using the ratio $W/\ell_B$, where $W$ is the width of the AGNR channel in \fref{fig:fig1_zshe}. All graphene devices simulated in \ffref{fig:fig2_zshe}, ~\ref{fig:fig3_zshe} and ~\ref{fig:fig4_zshe} are placed in quantizing external magnetic field, $W/\ell_B>1$.

We employ the momentum-relaxing model within NEGF formalism, discussed in \aref{sec:qtalgorithms}, to account for the local and simultaneous phase and momentum relaxation. This model can be physically interpreted as a highly simplified version (valid in the high-temperature limit) of the self-consistent Born approximation for electron-phonon interaction~\cite{Frederiksen2007,Mahfouzi2014}. We note that the momentum-relaxing model has been previously used to study dephasing effects in the integer QHE~\cite{Cresti2008a} where phenomenological dephasing length is often invoked~\cite{Pryadko1999} to account for electron-electron and electron-phonon scattering without delving into the microscopic details of such interactions.

For phase-coherent multiterminal transport of charge and spin described by \eeref{eq:mlbcharge} and ~\eqref{eq:mlbspin}, respectively, the recently developed algorithms~\cite{Groth2014} make it possible to simulate devices containing $\sim 10^6$ atomic orbitals (as demonstrated in \sref{sec:she}) with modest computational resources by exploiting sparse nature of the Hamiltonian matrix.  However, in the presence of dephasing, one needs to manipulate dense matrices in the formulas of \aref{sec:qtalgorithms}, which becomes prohibitively expensive for multiterminal graphene devices of the size employed in ZSHE experiments~\cite{Abanin2011}. Therefore, since we have to perform such a computation on a grid of energy points, we select much smaller size for the active region of the device in \fref{fig:fig1_zshe}---\mbox{$W \simeq 2.7$ nm} for 4-terminal devices and \mbox{$W \simeq 2.0$ nm} for 6-terminal devices. Given the very small device size in our simulation, we have to apply unrealistically large external magnetic fields in order to bring the device into the quantizing regime $W/\ell_B>1$. Nevertheless, the important parameter for comparing our results with experiments is not the absolute value of $W$ or $B$ but the ratio $W/\ell_B$.

\subsection{\ZSHE\ in four-terminal graphene}\label{sec:zshe_4terminal}

\begin{figure}[b!]
	\center{\includegraphics[scale=0.32,angle=0]{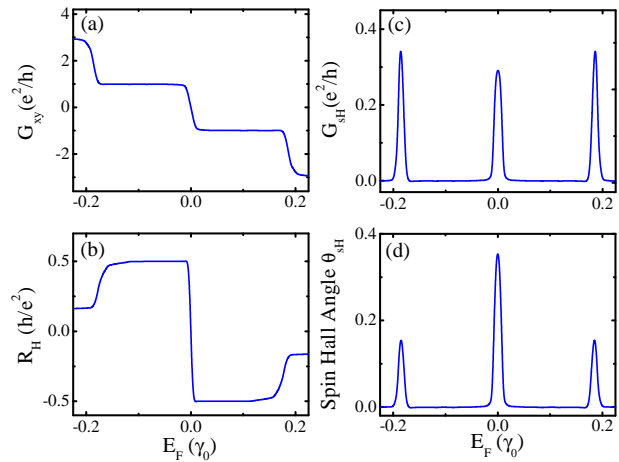}}
	\caption{The charge and spin transport quantities in four-terminal graphene devices illustrated in \fref{fig:fig1_zshe}: (a) charge Hall conductance $G_{xy}=I_5/(V_1-V_2)$; (b) charge Hall resistance $R_{\rm H}=(V_5 - V_6)/I_{1}$; (c) spin Hall conductance $G_{\rm sH}=I_5^{S_z}/(V_1-V_2)$; and (d) SH angle $\theta_{\rm sH}=I_5^{S_z}/I_1$. The width of AGNR channel is $W/\ell_B=3.42$ in the units of the magnetic length $\ell_B$ and a small momentum-relaxing dephasing $d_m=0.04 \gamma_0$ (see \aref{sec:qtalgorithms}) is introduced into the central region shown in \fref{fig:fig1_zshe}. Adapted from \cref{CHE_PRB85}.}
	\label{fig:fig2_zshe}
\end{figure}

In the analysis of the four-terminal graphene device in \fref{fig:fig1_zshe}, voltage $V/2$ is applied to lead 1 and $-V/2$ to lead 2, while voltages on leads 5 and 6 are set to zero. \Fref{fig:fig2_zshe} shows that in the quantizing external magnetic field $W/\ell_B>1$ the four-terminal device generates large spin Hall conductance 
\begin{equation}\label{eq:shconductance}
G_{\rm sH} = \frac{I_5^\uparrow - I_5^\downarrow}{V_1-V_2} \ ,
\end{equation}
and the corresponding SH angle 
\begin{equation}\label{eq:shangle}
\theta_{\rm sH} = \frac{I_5^{S_z}}{I_1} = \frac{G_{\rm sH}}{G_L} \ ,
\end{equation}
where $G_L=I_1/(V_1-V_2)$ is the longitudinal charge conductance. The spin current $I_5^{S_z} = I_5^\uparrow - I_5^\downarrow$ is the difference of spin-resolved charge currents composed of spin-$\uparrow$ or spin-$\downarrow$ electrons polarized along the \mbox{$z$-axis}, which is orthogonal to the plane of graphene.

\begin{figure}
	\center{\includegraphics[scale=0.32,angle=0]{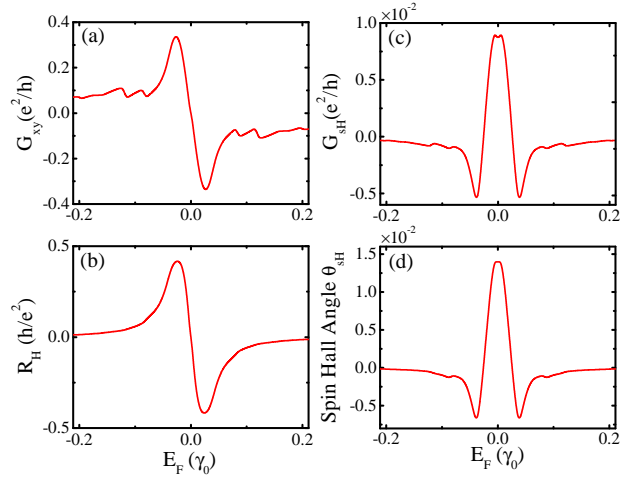}}
	\caption{The charge and spin transport quantities in  four-terminal graphene devices illustrated in \fref{fig:fig1_zshe}: (a) charge Hall conductance $G_{xy}$; (b) charge Hall resistance $R_{\rm H}$; (c) spin Hall conductance $G_{\rm sH}$; and (d) spin Hall angle $\theta_{\rm sH}$. The width of the AGNR channel is $W/\ell_B=1.53$ in the units of the magnetic length $\ell_B$ and large momentum-relaxing dephasing $d_m=0.4 \gamma_0$ (see \aref{sec:qtalgorithms}) is introduced into the central region shown in \fref{fig:fig1_zshe}. Adapted from \cref{CHE_PRB85}.}
	\label{fig:fig3_zshe}
\end{figure}

The value of $G_{\rm sH}$ shown in \fref{fig:fig2_zshe}(c) is comparable to the one found in \sref{sec:she} for graphene with SOC due to adatoms. Unlike intrinsic SHE~\cite{Nikolic2005b,Nikolic2009} in finite-size 2DEGs, where Rashba SOC induces both the transverse spin deflection~\cite{Nikolic2005c} and spin dephasing that compete against each other in the processes of generating pure spin current, in the ZSHE spin precession is absent and transverse spin current is pure only at CNP [i.e., total charge current $I_{5,6}=I_{5,6}^\uparrow + I_{5,6}^\downarrow$ becomes zero at CNP in \fref{fig:fig2_zshe}(a)]. This could be advantageous for spintronic applications since spin dephasing in the course of spin precession is evaded, as demonstrated by the experimental detection of $R_\mathrm{NL}$ even at distances $\sim 10$ $\mu$m away from the region where SH current was generated~\cite{Abanin2011}. We note that for very strong magnetic field, as could be achieved in graphene covered by a ferromagnetic insulator overlayer~\cite{Wei2016}, the peaks of $G_{\rm sH}$ in \fref{fig:fig2_zshe}(c) would become quantized~\cite{Sun2010} as a realization of the QSHE~\cite{QI_RMP83} in the absence of SOC.

The introduction of dephasing processes into four-terminal devices, which relax both \cite{Golizadeh-Mojarad2007a} the phase and the momentum of quasiparticles propagating through the active region, destroys the quantization of the charge Hall conductance $G_{xy}=I_5/(V_1-V_2)$ or charge Hall resistance  $R_{\rm H}$ and underlying chiral edge states, as demonstrated by the transition from \fref{fig:fig2_zshe}(a) to \fref{fig:fig3_zshe}(a) for $G_{xy}$ and from \fref{fig:fig2_zshe}(b) to \fref{fig:fig3_zshe}(b) for $R_{\rm H}$. The charge Hall resistance in four-terminal devices is defined as $R_{\rm H} = (V_5-V_6)/I_1$ for the measuring geometry, where current $I_1$ is injected into lead 1 and voltages $V_5$ and $V_6$ develop as the response to it. The SH conductance and SH angle are concurrently reduced by two orders of magnitude, which are values similar to those found in semiclassical approaches~\cite{Abanin2011a} in the temperature range \mbox{$T= \text{200--300 K}$}. Thus, \fref{fig:fig3_zshe} can be used to tune phenomenological parameters (see \aref{sec:qtalgorithms}) controlling the strength of momentum-relaxing dephasing.

\subsection{\ZSHE\ induced nonlocal resistance in six-terminal graphene}\label{sec:zshe_6terminal}

\begin{figure}[t!]
	\center{\includegraphics[scale=0.32,angle=0]{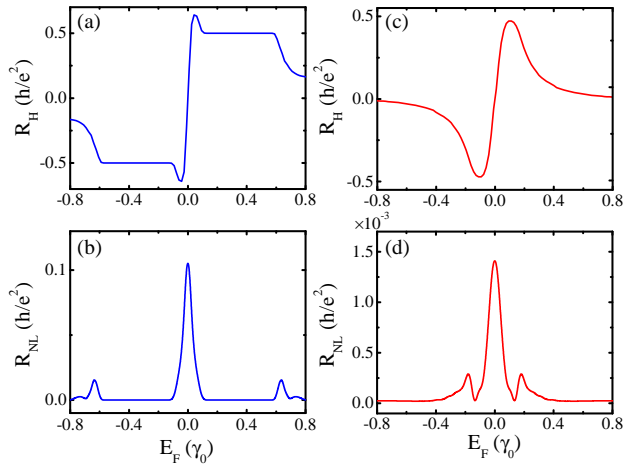}}
	\caption{Panels (a) and (c) plot charge Hall resistance $R_{\rm H} = (V_1 - V_2)/I_{5}$, while panels (b) and (d) plot nonlocal resistance $R_{\rm NL} = (V_3 - V_4)/I_{1}$ as the central quantity measured in the ZSHE experiments~\cite{Abanin2011,Renard2014,Wei2016} on six-terminal graphene devices. The quantum coherence is retained in panels (a) and (b) where only a small momentum-relaxing dephasing $d_m=0.02\gamma_0$ (see \aref{sec:qtalgorithms}) is present in the central region of the device, while much larger dephasing $d_m=0.5\gamma_0$ is used for panels (c) and (d). The width of the AGNR channel in \fref{fig:fig1_zshe} is $W/\ell_B=3.42$ in panels (a) and (b) and $W/\ell_B=1.53$ in panels (c) and (d) in the units of the magnetic length $\ell_B$. Adapted from \cref{CHE_PRB85}.}
	\label{fig:fig4_zshe}
\end{figure}

In the analysis of six-terminal graphene devices in \fref{fig:fig1_zshe}, a charge current $I_1$ is injected through lead 1 and a current $-I_1$ flows through lead $2$, while $I_\alpha \equiv 0$ in all other leads. We then compute voltages that develop in the leads $\alpha=3,4,5,6$ labeled in \fref{fig:fig1_zshe} in response to injected current $I_1$. \Fref{fig:fig4_zshe}(b) shows peaks in the nonlocal resistance within the phase-coherent transport regime, which closely resemble the CNP and side peaks observed experimentally in strong (quantizing) external magnetic field~\cite{Abanin2011}. We note that peaks of both $R_\mathrm{NL}$ in \fref{fig:fig4_zshe}(b) and of $\theta_\mathrm{sH}$ in \fref{fig:fig2_zshe}(d) reside within the transition regime between two QH plateaus where edge states actually delocalize and ``percolate'' through the bulk~\cite{Richter1994}.

The transition of $R_{\rm NL}$ from \fref{fig:fig4_zshe}(b) to \fref{fig:fig4_zshe}(d) shows how dephasing removes both side peaks while leaving the nonlocal voltage around CNP, which becomes two orders of magnitude smaller than in the phase-coherent regime. The charge Hall resistance in six-terminal devices, $R_{\rm H} = (V_1-V_2)/I_5$ defined for injected current $I_5$ and voltages measured between leads 1 and 2 (for $I_1=I_2=0$), changes smoothly from \fref{fig:fig4_zshe}(a) to \fref{fig:fig4_zshe}(c) as dephasing in increased, where the curve in \fref{fig:fig4_zshe}(c) looks remarkably similar to those observed experimentally~\cite{Abanin2011a} for \mbox{$T=250$ K} and $B= \text{1--12 T}$ or in semiclassical transport theories~\cite{Abanin2011a} (where steep change of $R_\mathrm{H}$ amplifies $R_\mathrm{NL}$~\cite{Wei2016}).

%%%%% VALLEY HALL EFFECT IN GRAPHENE %%%%%%%%%%%%%%%%%%%%%%%%%%%%%%%%%%%%%%%%%%%%%%%%%%%%

\section{Valley Hall effect in graphene} \label{sec:vhe}

Degenerate valleys of energy bands well separated in momentum space constitute a discrete degree of freedoms for low-energy carriers with long relaxation time. The valley index may be used as a non-volatile information carrier provided that it can be coupled to external probes. In graphene with broken inversion symmetry, the valley index can play a similar role as the spin degree of freedom in phenomena such as Hall transport, magnetization, optical transition selection rules, and chiral edge modes \cite{LU_PRB81,YAO_PRL102,CHE_PRB77,LIU_AP59}. As a result, proposals are made to control of valley dynamics by magnetic, electric, and optical means, in the quest for valley based information processing (``valleytronics'').

Recently, generalizations to monolayers of MoS$_2$ and other group VI transition metal dichalcogenides have also been experimentally achieved \cite{Mak1489,Mak222}. Those materials are direct bandgap semiconductors with band edges located at $K$ points. The low energy electrons and holes are well described by massive Dirac fermions with strong spin-valley coupling. Valley and spin dependent optical transition selection rules were reported as well as coexistence of VHE and SHE. This suggests possible control of both valley and spin degrees of freedom for potential integrated spintronics and valleytronics applications based on hybrid two-dimensional materials~\cite{Xu_2014}.

\subsection{Concept and new device principles} \label{sec:vhe_concept}

The control of the valley degree of freedom in graphene has a decade long history~\cite{RYC_NP3,XIA_PRL99}. 
Several proposals have been made to obtain a valley valve in order to generate a valley polarized current or to filter electrons with given valley polarization. 

\begin{figure} [t!]
	\begin{center}
	\resizebox{12cm}{!}{\includegraphics{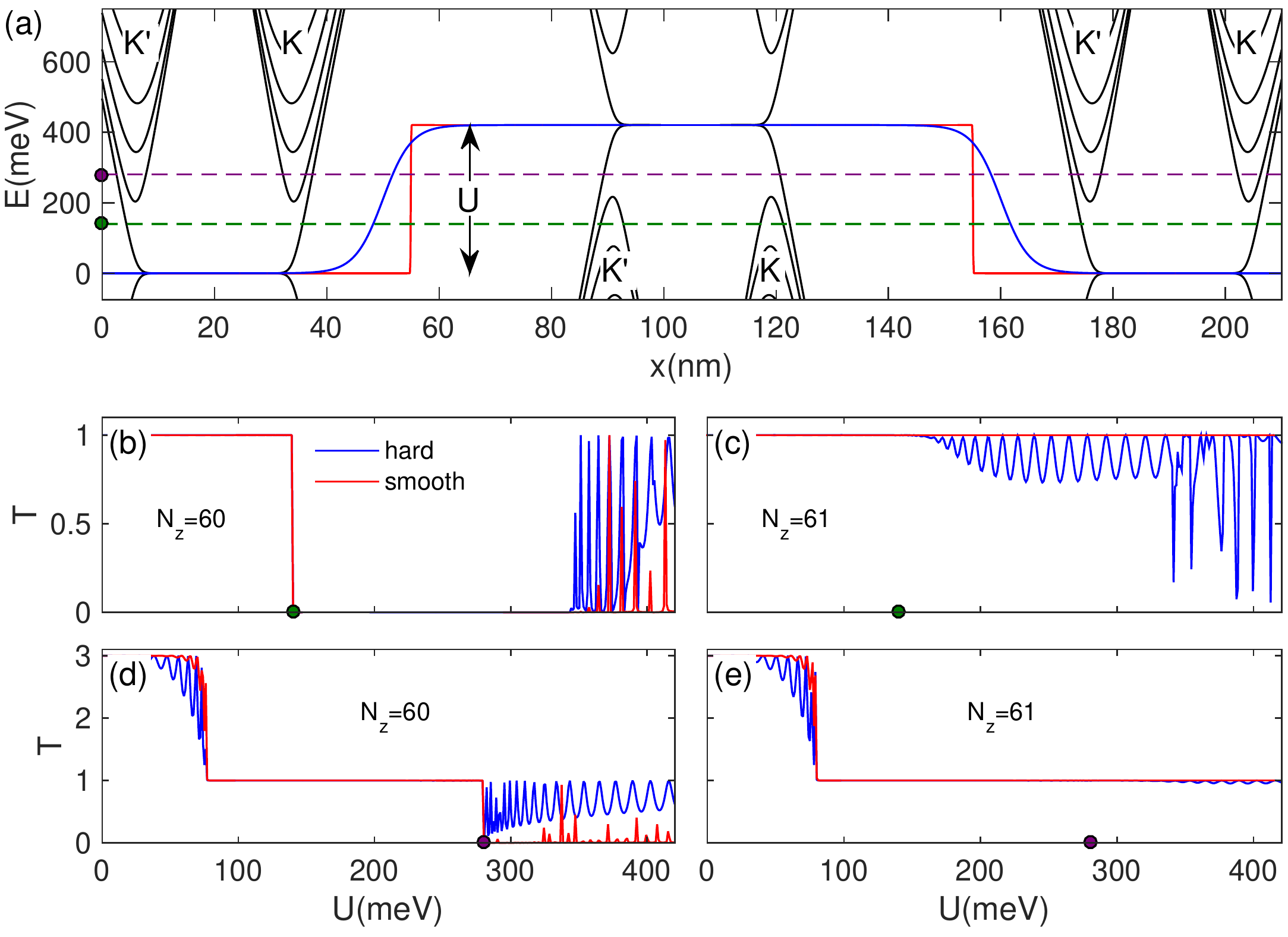}}
	\caption{(a) Potential profile of the smooth and sharp barriers (with height $U$) considered in the simulations as a function of the position along the ribbon axis. The band structure of a ZGNR is reported in the different regions. The green and (purple) dashed lines correspond to $E_{\rm F}$ between the bottom of the first and second (second and third) conduction bands considered in the simulations. (b) Transmission coefficient for a ZGNR composed of $N_z=60$ carbon chains as a function of the sharp/smooth potential barrier height $U$ when $E_{\rm F}$ is between the bottoms of the first and the second conduction bands (indicated by a green circle). (c) Same as (b) for a ribbon composed of $N_z=61$ chains. (d) Same as (b) for $E_{\rm F}$ between the bottoms of the second and the third conduction bands (indicated by a purple circle). (e) Same as (d) for a ribbon composed of $N_z=61$ chains.}
	\label{fig:valley_valve}
	\end{center}
\end{figure}

The first idea \cite{RYC_NP3,ZHA_APL93} was developed on the basis of a result obtained by Wakabayashi and Aoki \cite{WAK_IJMPB16}, where a potential barrier creating a $pn$ junction in a ZGNR is able to block the current at energies close to the Fermi level. In the band structure of a ZGNR, the $K$ and $K'$ valley are separated and an almost flat band (corresponding to states localized at the edges) extends between the two valleys.
\Fref{fig:valley_valve}(a) shows such a band structure in the three regions of a ribbon composed of $N_z=60$ zigzag chains and with a potential barrier of height $U$ along the $\hat{x}$ ribbon axis. The barrier profile has a length of about 100 nm in the sample, and can be smooth or sharp, this entailing a different valley mixing degree.
When, in the barrier region, $E_{\rm F}$ lies within the region where only one electron (hole) band is active, electrons moving from left to right can only propagate at the $K$ ($K'$) valley, where the group velocity, i.e. the band slope, is positive. 
Let us consider, for example, $E_{\rm F}$ between the bottom of the first electron conduction band ($E=0$) and the bottom of the first conduction band $E\approx 200$ meV, as indicated by the green circle and dashed line in \fref{fig:valley_valve}(a). In this case, electrons are injected from the $K$ valley. As long as $U<E_{\rm F}$, they are fully transmitted along the $K$ valley in the barrier region and finally to the right contact. As shown in \fref{fig:valley_valve}(b), this entails a transmission coefficient quantized to 1. As $U>E_{\rm F}$, electrons need to pass from $K$ to $K'$ valley to propagate from left to right inside the barrier, which is not allowed as long as $U$ is smaller than the bottom of the second conduction band, as visible from window with zero transport coefficient in \fref{fig:valley_valve}(b).

A completely analogous behavior is observed when the Fermi level is between the bottoms of the first and the second conduction bands in the region outside the barrier, see the purple circle and dashed line in \fref{fig:valley_valve}(a). In this case, electrons are injected from both valleys and, for $U=0$, three conductive channels are active, see \fref{fig:valley_valve}(d). In the filtering region, electrons from the most external conduction band at $K$ are fully transmitted for $U<E_{\rm F}$, while they are largely backscattered for $U>E_{\rm F}$. However, in this case backscattering is not perfect, especially for the sharp barrier due to the fact that more than one channel is active and backscattering can occur also at the right side of the barrier. As a confirmation, the transmission coefficient shows oscillations related to Fabry-Perot interference along the length of the barrier. The residual transmission can be further suppressed by considering very long barrier with very smooth edges. 
We can thus conclude that the configuration of \fref{fig:valley_valve}(a) allows transmitting $K$-polarized electrons or to backscatter them depending on the potential barrier height $U$.

Indeed, this blocking effect is only observed when the ZGNR is composed of an even number of carbon chains, while it breaks down for an odd number of zigzag chains, as visible in \ffref{fig:valley_valve}(c,e) for $N_z=61$ and as explained in detail in \ccref{CRE_PRB77,AKH_PRB77}.
The filtering effect is triggered by the fact that the barrier potential is uniform along the transverse direction of the ribbon (it only varies along the axis direction) and thus, for an even number of zigzag chains, the Hamiltonian is invariant under mirror symmetry along the ribbon axis. As a consequence, the wave functions of the periodic ribbon have a definite parity with respect to this operation. 
In particular, the first valence and conduction bands have opposite parities.
In a $pn$ junction, this entails that electrons have to jump not only from one valley to the other, but also between wave states with opposite parity in order to be transmitted. 
If the barrier potential does not break the mirror symmetry, the matrix element is identically zero, which entails a perfect reflection as seen in \fref{fig:valley_valve}(b).  
On the contrary, when the number of zigzag chains is odd, the Hamiltonian is not invariant under mirror symmetry, see \fref{fig:valley_valve}(d), reflection is not perfect since at least one band in the barrier region and one band outside the barrier region have the same parity and then the corresponding matrix element is non-vanishing, unless the barrier is very long and smooth, as already discussed.

This proposal requires extremely well-defined and narrow (to have a large operating energy window) ZGNR, which is not easy to fabricate. A more efficient valley filter based on the same principle is proposed in \cref{NAK_PRL102}.

Gunlycke and White proposed to realize a valley filter by including a line defect in graphene \cite{GUN_PRL106}. 
The idea exploits the mirror symmetry of the considered line defects and the fact that only the symmetric component of the wave function is transmitted through the defect, while the antisymmetric component is fully reflected. 
This gives rise to a transmission through the defect that depends on the valley $\tau$ and incident angle $\alpha$ of the injected electrons (with respect to the normal of the line defect) as $T_\tau(\alpha)=(1+\sin\alpha)/2$. 
As a consequence, the line defect turns out to be semi-transparent for an unpolarized electron flux and, for $\alpha=\pm \pi/2$ the transmitted and reflected electrons are fully valley-polarized. 
The advantage of such a configuration is that it does not require fabricating high-quality ribbons with sub 10 nm width, and that the technology to realize atomically precise line defects is available \cite{CHE_PRB89}.

A different mechanism proposed by Fujita and coworkers \cite{FUJ_APL97} and by Zhai and coworkers \cite{ZHA_PRB82} is based on the combined effect of strain and magnetic field. In the Dirac Hamiltonian of graphene, strain can be included as a gauge vector potential $\vec{A}_{\rm S}$ with different orientation in the two valleys \cite{LOW_NL10}, which ensures the time-reversal symmetry invariance. The presence of a magnetic field adds a gauge vector potential $\vec{A}_{\rm M}$, which is the same for the two valleys, through the minimal substitution. This breaks the time-reversal symmetry and induces a valley anisotropy since the effective vector potential will be $\vec{A}_{\rm M}\pm \vec{A}_{\rm S}$, depending on the valley. This phenomenon can be exploited to realize valley filters, where the valley selection is performed by changing the direction of the magnetic field generated by ferromagnetic stripes \cite{FUJ_APL97,ZHA_PRB82,ZHA_JPCM23,SON_APL103,MYO_CAP14}. Other proposals are based on the valley-dependent anisotropy introduced by trigonal warping \cite{GAR_PRL100,PER_JPCM21}, slanted graphene junctions \cite{HSI_APL108}  or bilayer graphene \cite{ABE_APL95,COS_PRB92,PET_NJP14,PRA_APL104}.

\subsection{Topological valley Hall currents} \label{sec:vhe_top}
The opposite Berry curvature for electrons at different valleys suggested the possibility to generate valley-dependent transport characteristics. 
From a semi-classical point of view, the origin stems from the anomalous group velocity, which is derived from the band structure \cite{XIA_RMP82}
\begin{equation} \label{eq:anomalous_velocity}
 \vec{v}(\vec{k}) \ = \ \Frac{1}{\hbar} \ \Frac{\partial \epsilon_n(\vec{k})}{\partial \vec{k}} \ -\ \Frac{e}{\hbar} \ \vec{E}\times\vec{\Omega}_n(\vec{k}) \ ,
\end{equation} 
where $\vec{k}$ is the wave vector, $\vec{\Omega}_n(\vec{k})$ is the Berry curvature of the $n$th band and $\vec{E}$ is an external electric field. 
The last term of \eref{eq:anomalous_velocity} introduces an anomalous velocity, which is perpendicular to the electric field and the Berry curvature.
When an external perturbation breaks the sublattice (inversion) symmetry, a gap opens and the Berry curvature around the two valleys becomes finite and valley-dependent \cite{XIA_PRL99}
\begin{equation} \label{eq:vh_berry}
  \vec{\Omega}_n(\vec{k}) \ = \ 
				 \tau_z \ \Frac{3a^2 \Delta}{2(\Delta^2+4k^2\hbar^2v_F^2)^{3/2}} \ ,
\end{equation}
as can be shown from the corresponding Dirac Hamiltonian
\begin{equation} \label{eq:ham_delta}
  H \ = \ \hbar v_F \ ( \ k_x \tau_z \sigma_x \ +\ k_y \sigma_y \ ) \ + \ \Frac{\Delta}{2} \ \sigma_z \ ,  
\end{equation}
where $\Delta$ is the sublattice potential imbalance, i.e. the band gap width.
The important point is that the two Berry curvatures have opposite sign around the two valleys, and then we expect electrons to be deflected in opposite directions depending on the valley they come from.

The valley-dependent response of electrons can be better understood by looking at the velocity operator as sketched by Ando in \cref{AND_JPSJ84}.
For the $K$ valley, the Hamiltonian (\ref{eq:ham_delta}) can be written as
\begin{equation} \label{eq:ham_ando}
	H \ = \ \hbar v_F \ (\vec{k}\cdot \vec\sigma ) \ + \ \Delta\sigma_z \ = \  v_F \ (\vec{p}\cdot \vec{\sigma} ) \ + \ \Delta\sigma_z
\end{equation}
where $\vec{p}$ is the momentum operator. The velocity and the acceleration operators are
\begin{equation} \label{eq:vh_velocity}
        \vec{v} \ = \ \Frac{1}{i\hbar} [\vec{r},H] \ = \ v_F \vec{\sigma}
      \ \ \ \ \ \ {\rm and} \ \ \ \ \ \ \
	 \dot{\vec{v}} \ = \ \Frac{1}{i\hbar} [\vec{v},H] \ = \  2v_F^2\  (\vec{k}\times \vec{\sigma}) \ - \ \Frac{2\Delta}{\hbar} \ \vec{v}\times\vec{\hat{z}} \ ,
\end{equation}
where $\vec{\hat{z}}$ is the unit vector perpendicular to the graphene plane.
From this formulation, it is evident that the gap $\Delta$ induces an in-plane extra term in the acceleration operator, which is perpendicular to the velocity operator and acts as an orthogonal magnetic field with strength proportional to $\Delta$ and zero in the absence of a gap. 
Note that, very importantly, the sign of this term is opposite for the $K'$ valley.
However, the argument based on semiclassical \eref{eq:anomalous_velocity} requires finite electric field, which is difficult to reconcile with quantum transport simulations in the linear-response regime~\cite{KIR_PRB92}.

\begin{figure} [t!]
	\begin{center}
	\resizebox{12cm}{!}{\includegraphics{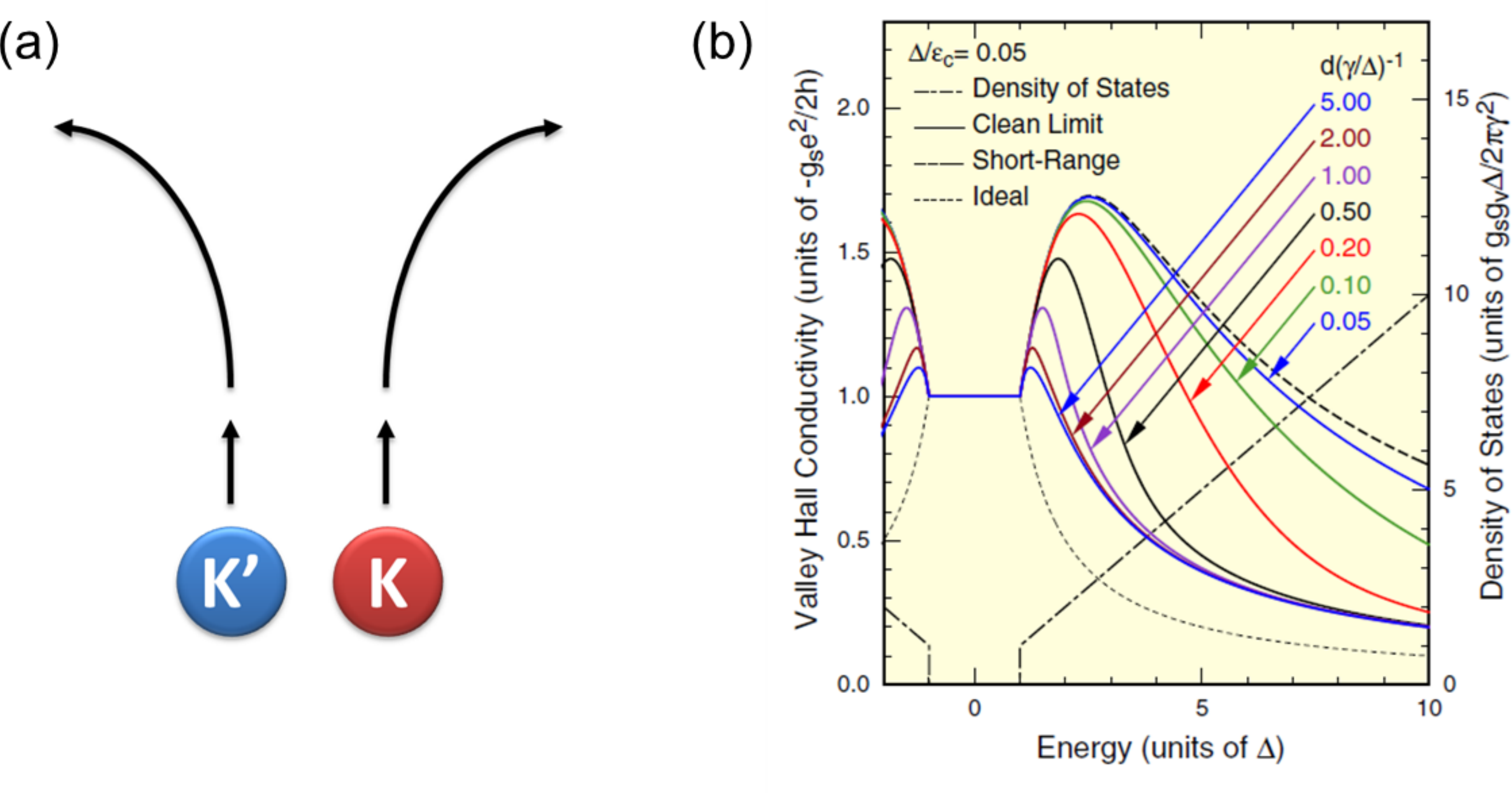}}
	\caption{(a) Separation of electrons belonging to different valleys by an acceleration term perpendicular to the velocity for graphene with a gap induced by breaking of the inversion symmetry. (b) $K$-valley Hall conductivity obtained within the Kubo formalism in the presence of disorder centers with varying potential spatial range $d$. 
	Reprinted with permission from \cref{AND_JPSJ84}, The Physical Society of Japan.}
	\label{fig:valley_ando}
	\end{center}
\end{figure}

From the Hamiltonian (\ref{eq:ham_ando}), the $K$ valley Hall conductivity at temperature $T$ and chemical potential $\mu$ can be calculated by the Kubo formula
\begin{equation} \sigma_{xy}(\mu,T)  = \Frac{2\hbar}{i\pi L^2}  \int \! dE \ f(E,\mu,T)  
                         \left\langle {\rm Tr} \left[ 
												  j_x\left(\partial_E {\rm Re} G(E+i0)\right)  j_y  {\rm Im} G(E+i0)  - (x\leftrightarrow y )
												 \right] \right\rangle \ .
												\label{eq:kubo_ando}
\end{equation}
The presence of a homogeneous distribution of disorder centers with given short-range or long-range potential profile can be included within the self-consistent Born approximation. The final result at zero temperature is \cite{AND_JPSJ84}
\begin{equation} \sigma_{xy}^K(\mu,T=0) \ = \ \left\{ 
       \begin{array}{lll} -\Frac{e^2}{h} &{\rm for} &|\mu|<\Delta \\[5mm]
			                    -\Frac{e^2}{h} \Frac{8|\delta|(\Gamma_0-\Gamma_1)[(1+\delta^2)\Gamma_0 - 2\delta^2\Gamma_1 - (1-\delta^2)\Gamma_2]}{[(1+3\delta^2)\Gamma_0-4\delta^2\Gamma_1-(1-\delta^2)\Gamma_2]^2}
                              &{\rm for} &|\mu|\geq\Delta
															\end{array}
\right.
\label{eq:sigma_ando}
\end{equation}
where $\delta=\Delta/\mu$, and $\Gamma_n$ are related to the potential profile and density of the impurity centers and vanish for clean systems.
Note that in the region of the valence and conductance bands close to the gap, the conductivity scale as $\sigma_{xy}^K=-(e^2/h)\mu/\Delta$, independently of the presence of disorder. The central result of \eref{eq:sigma_ando} is that the valley Hall conductivity is quantized to $e^2/h$ within the gap energy window, thus predicting a \QVHE\ (QVHE). While the conductivity is obtained by integrating \eref{eq:kubo_ando} over the whole Fermi sea, its quantization is determined only by the contributions just below and above the energy gap, corresponding to the regions with high Berry curvature.
An example is reported in \fref{fig:valley_ando}(b) for impurity potentials with different range. We can see that $\sigma_{xy}^K$ is quantized within the gap and enhanced at the gap edges, thus generating a double peak curve.
The sign of the $K'$ valley Hall conductivity is opposite with respect to that at $K$.
Therefore, within the gap we have a global Hall conductivity $\sigma_{xy}\equiv\sigma_{xy}^K+\sigma_{xy}^{K'}=0$ (and also $\sigma_{xx}=0$) and a valley Hall conductivity $\sigma_{xy}^v\equiv\sigma_{xy}^K-\sigma_{xy}^{K'}=2e^2/h$. This means that when applying an electric field to the insulating system, both the longitudinal and transverse net charge currents are zero. The zero transverse current can be seen as the sum of two opposite and spatially superposing valley currents. 
Note that these results neglect any intervalley scattering, which can be significant for short-range impurities and presence of edges of particular geometry. 
The QVHE, with a bulk-boundary correspondence between the $\sigma_{xy}^v$ (or equivalently the Chern numbers) and the number of edge channels, could only be observable in multilayer graphene with purely zigzag edge geometry, as discussed in \cref{REN_RPP79}.

\subsection{Valley Hall effect in tight-binding models of graphene} \label{sec:vhe_tb}

\begin{figure} [t!]
	\begin{center}
	\resizebox{13cm}{!}{\includegraphics{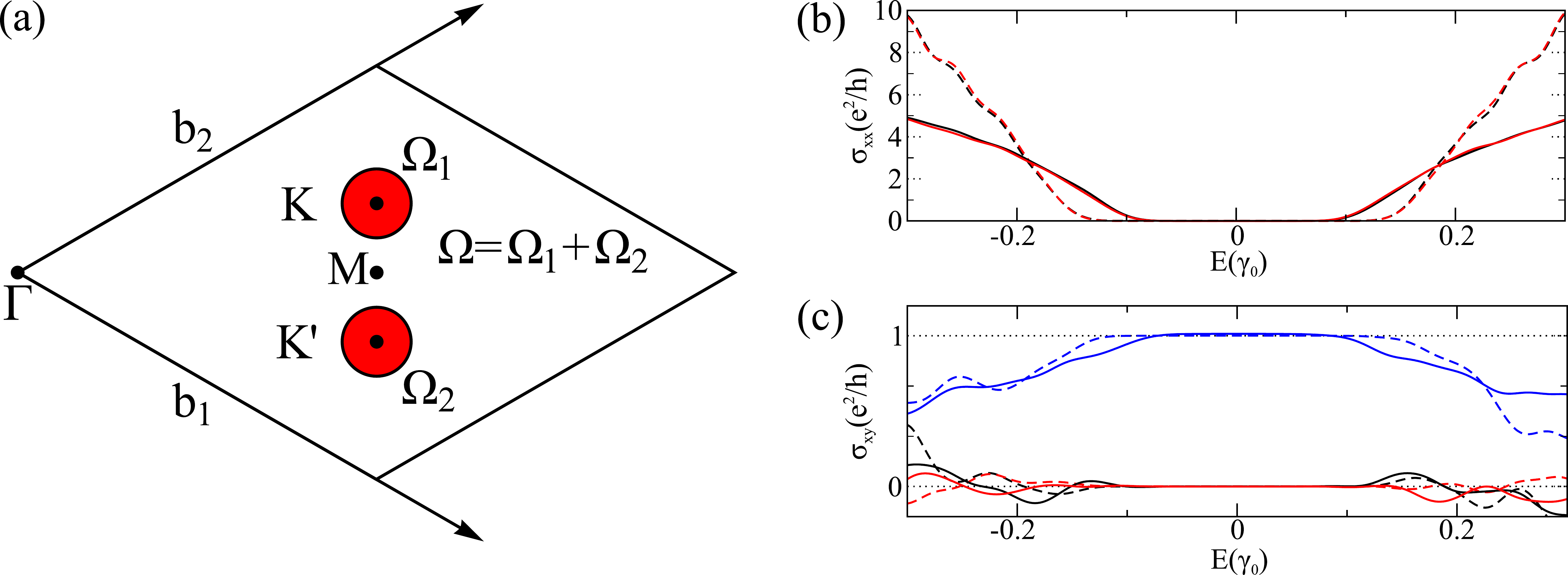}}
  \caption{(a) Regions around the K and K' Dirac points that we define as the valley region. We set the radius to $R_{K}=\frac{1}{2} |K-M|$, which is large enough to capture the physics at low energy, as shown by comparison with the full-band calculation. (b) Dissipative conductivity for $\Delta=0.2\gamma_0$ considering the full band structure $\sigma_xx$ (black) and only the valley region $\sigma_{xx}^K+\sigma_{xx}^{K'}$ for a clean system (solid line) and a disordered system with $W=0.9\gamma_0$ (dashed line). (c) Total Hall conductivity obtained considering the full band structure (black) and the valley region (red), and valley Hall conductivity (blue) for a clean system (solid line) and a disordered system with $W=0.9\gamma_0$ (dashed line).}
	\label{fig:kpm_valley_tb}
	\end{center}
\end{figure}

To analyze the VHE, the usual approach is to study the low-energy Hamiltonian assuming the absence of valley mixing (the two Dirac cones are decoupled). However, it is possible to study valley effects using the full TB Hamiltonian, through the projector
\begin{equation} \label{eq:vh_P}
		P_{\vec{K},R_K}\equiv \int_{BZ} \theta( |\vec{k}-\vec{K}|-R_K)|\vec{k}\rangle\langle\vec{k}| \ dk_x \ dk_y
\end{equation} 
where the kets $|\vec{k}\rangle$ are the eigenstates of the momentum operator and $R_K$ is the radius of the region we are going to define as our valley, the region we choose are shown in \fref{fig:kpm_valley_tb}(a). We use the KPM method elaborated in \aref{sec:kubo-formula} to compute the valley conductivity tensor $\sigma_{\alpha,\beta}^{v}\equiv \sigma_{\alpha,\beta}^{K} -\sigma_{\alpha,\beta}^{K'}$, the total conductivity tensor $\sigma_{\alpha,\beta}$, and the contribution to the conductivity tensor due to the valleys $\sigma_{\alpha,\beta}^{K+K'}\equiv \sigma_{\alpha,\beta}^{K} +\sigma_{\alpha,\beta}^{K'}$. A simple TB model of a honeycomb lattice, including a staggered potential and an Anderson disorder potential, is given by
\begin{equation} \label{eq:vh_ham}
H=\sum_{\langle i,j\rangle } a_i^\dagger b_j +{\rm h.c} +\sum_{i}\left[\left( \varepsilon_i+\frac{ \Delta}{2}  \right)a_i^\dagger a_i +\left(\varepsilon_i-\frac{ \Delta}{2}\right)b_i^\dagger b_i \right]
\end{equation}
where $(a_{i}^\dagger$ and $a_{i})\,[b_{i}^\dagger$ and $b_{i}]$ are the creation and annihilation operators for electrons in the (A)[B]  -sublattice of the honeycomb lattice, $\varepsilon_i$ is an Anderson disorder with strength $W$ and $\Delta=0.2\gamma_0$ the band gap width. The results are shown in \fref{fig:kpm_valley_tb}(b,c) for the pristine and disordered case. 
 
In \fref{fig:kpm_valley_tb}(b) we show the dissipative conductivity numerically calculated by considering the full band structure or only the valley region. These results show that, in both the pristine and the disordered case, only the electrons from the valley region contribute to transport, thus confirming that the chosen valley region is large enough to accurately capture the physics at low energy. In \fref{fig:kpm_valley_tb}(c) we show the charge and valley Hall conductivities. The charge Hall conductivity vanishes in the gap region, as required by the time-reversal symmetry, which is not broken by the staggered potential. However, we do see a quantized valley Hall conductivity, as predicted by Ando \cite{AND_JPSJ84} with analytical calculations based on the Dirac Hamiltonian. This plateau is robust against moderate Anderson disorder and persists up to a disorder strength $W=0.9\gamma_0$. Therefore, from a bulk perspective, we conclude that a staggered potential can lead to the QVHE. However, at least for the uncorrelated disorder considered here, no increase of the conductivity is seen outside the gap, which is a different result compared to what expected from the continuous model \cite{AND_JPSJ84}.
 
\begin{figure} [t!]
	\begin{center}
	\resizebox{14cm}{!}{\includegraphics{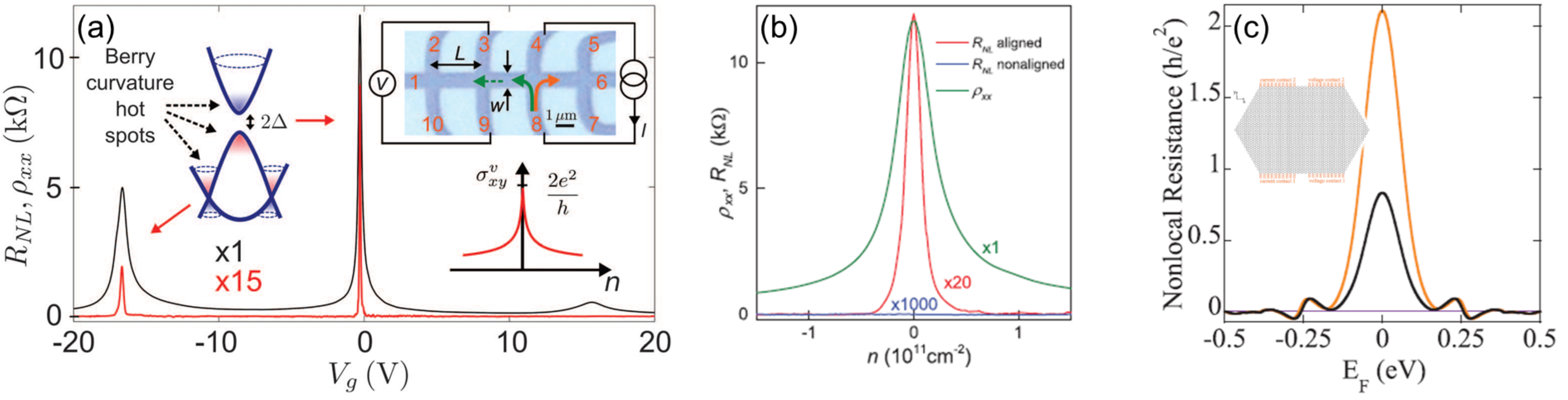}}
	\caption{(a) Experimental setup (inset) and nonlocal resistance (main panel) for a six-terminal graphene bar aligned over an h-BN substrate in the experiment by Gorbachev and coworkers \cite{GOR_SCI346}. (b) Experimental nonlocal resistance for aligned and non-aligned h-BN substrates. (c) Four terminal setup (inset) and calculated nonlocal resistance as reported in \cref{KIR_PRB92}. Adapted with permission from: (a,b) \cref{GOR_SCI346}, AAAS; (c) \cref{KIR_PRB92}, American Physical Society.}
	\label{fig:valley_exp}
	\end{center}
\end{figure}

In \cref{GOR_SCI346}, Gorbachev and coworkers adjusted the alignment between a graphene monolayer with an h-BN substrate for breaking the sublattice symmetry. As a result, a gap was formed, whereas a valley-dependent transport argument was used to interpret the large nonlocal resistance signal measured at low charge density (see \fref{fig:valley_exp}(a)).

When the layers are perfectly aligned, regions of commensurate graphene/h-BN regions form with the same A/B sublattice asymmetry sign, with surrounding strain boundaries. The size of these commensurate regions is about 10 nm, which is ten times smaller than the typical electron wave length. This indicates that a gap is locally present in the sample, even if currents can flow through the interface strained areas, thus preventing the observation of thermally activated conductance.
The result is that a non-local resistance peak around the CNP and with roughly the same width of the gap is measured, see \fref{fig:valley_exp}(a,b). 
Several aspects seem to indicate that the peak is related to a VHE. 
First of all, it disappears for misaligned layers, when the gap vanishes, see \fref{fig:valley_exp}(b). This excludes any possible SHE. 
Then, any transport along edges is excluded, and the phenomenon has a purely bulk nature.
Finally, the nonlocal resistance is observed to decay exponentially with the distance between the current and voltage terminals. The decay parameter (about 1 $\mu$m) is compatible with the intervalley scattering due to edges or disorder.

However, this experiment has raised a debate on its microscopic explanation \cite{KIR_PRB92,LEN_PRL114}.
For example, \cref{LEN_PRL114} argues that the origin of the VHE comes from the Fermi sea bulk states just beneath the gap and not from edge states, which, if present, are not topologically protected. The idea is that for $E_{\rm F}$ within the gap, two opposite valley polarized currents circulate in the system, thus generating a valley and charge-neutral current, as discussed above. Since the neutral valley current is transmitted by electrons of the Fermi sea below the gap, they are non-dissipative. Therefore, in this picture, even though the system is charge insulator and the chemical potential is within the gap, the neutral persistent valley currents arising from the Fermi sea are able to generate a non-local signal in the six-terminal experimental system, which appears counterintuitive. 

The connection between high nonlocal resistance $R_{\rm NL}$ and the topological origin of the valley Hall currents has been recently questioned by Kirczenow, who suggested that a non-local resistance can calculated by LB formalism applied to multiterminal gapped graphene described by TB Hamiltonian, see the inset of \fref{fig:valley_exp}(c) \cite{KIR_PRB92}. The four contacts are made of unidimensional carbon chains with on-site energy equal to that of the carbon of the central structure they are attached to. In the central structure, a staggered potential $\pm \Delta=60.2$ meV is applied to break the inversion symmetry. Such a signal cannot be related to the semi-classical anomalous velocity in \eref{eq:anomalous_velocity}, since the peak of the non-local resistance occurs within the gap where transport occurs by quantum-mechanical tunneling. Moreover, in the linear response regime the electric field in \eref{eq:anomalous_velocity} can be assumed to be vanishingly small. 

\begin{figure} [t!]
	\begin{center}
	\resizebox{14cm}{!}{\includegraphics{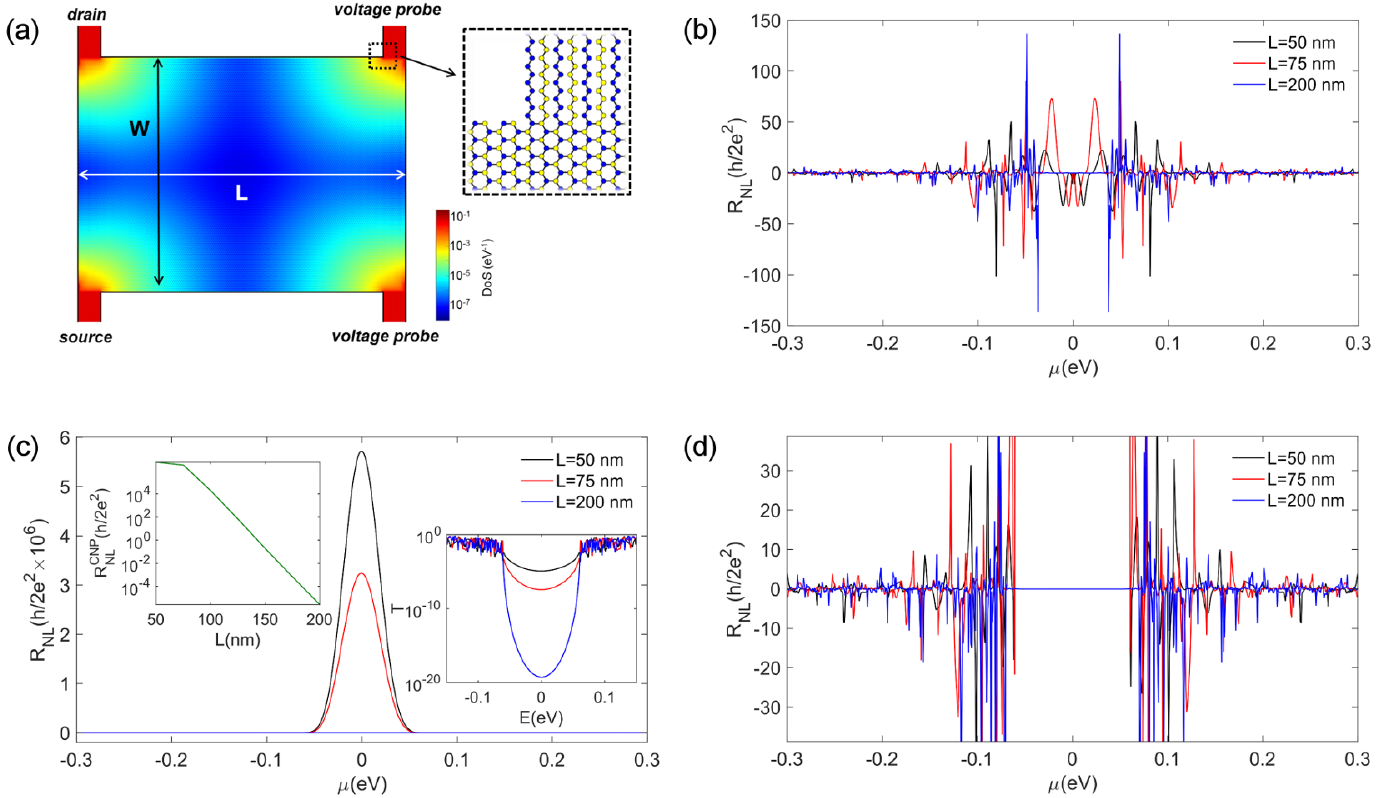}}
	\caption{(a) Structure of the simulated four-terminal graphene with and without staggered sublattice potential $\Delta=\pm60.2$ meV. The four contacts are modeled as uncoupled unidimensional chains with the same on-site energy as the carbon atoms they are attached to (see the color of the circles in the sketch). The color inside the structure represents the local DOS (on logarithmic scale) at energy $E=10$ meV and in the presence of the staggered potential. (b) Nonlocal resistance $R_\mathrm{NL}$ for the system without staggered potential as a function of the chemical potential $\mu$, at a temperature $T=4.2$ K and for different length of the system. (c) Main panel: same as (b) in the presence of the staggered potential. Left inset: Value of the nonlocal resistance at the CNP as a function of the system length $L$. Right inset: Transmission coefficient from the source contact to the bottom voltage probe as a function of the electron energy and for different system lengths. (d) Same as the main panel of (c) zoomed in the low $R_{\rm NL}$ region.}
	\label{fig:kir}
	\end{center}
\end{figure}

This phenomenon can be shown to be associated with valley currents, which however are not generated by the Berry curvature but rather by the specific geometry of the contacts, analogously to what observed for instance in \cref{RYC_NP3}, and enhanced by the staggered sublattice potential.
To further investigate this picture and to emphasize the role of tunneling transport through the gap by evanescent states, we performed four-terminal LB simulations based on a similar geometry but with larger width $W=50$ nm and for different lengths $L$ between the current and the voltage probes.
A sketch of the system is shown in \fref{fig:kir}(a).
As illustrated in \fref{fig:kir}(b), for $\Delta=0$ the nonlocal resistance is small and shows positive and negative fluctuation independently of the length. 
When $\Delta=60.2$ meV, a gap opens in the system and the nonlocal resistance shows a huge peak at the CNP and all over the width or the gap, see \fref{fig:kir}(c). However, the peak decreases when increasing $L$ and it finally disappears, see also \fref{fig:kir}(d) and the left inset of \fref{fig:kir}(c). Such a result can be easily explained by considering that within the gap transport occurs through evanescent states that rapidly vanish when penetrating into the bulk. Since the distance $W$ between source and drain is fixed, the transmission coefficient between them does not change with $L$. On the contrary, the transmission coefficient between the source or drain contacts and the voltage probes is strongly suppressed when increasing $L$, as illustrated in the right inset of \fref{fig:kir}(c). The result is that the nonlocal resistance vanishes for $L\gg W$.
The decrease of the DOS when getting away from the contacts is indicated by the color map in \fref{fig:kir}(a). 

We conclude that the observation of the VHE in gapped graphene remains debatable and a direct measure of the valley current has not yet been convincingly demonstrated. 
In particular, the connection between (not directly observable) valley Hall conductivity and $R_\mathrm{NL}$ directly observed in multi-terminal measurements calls for further theoretical and computational efforts.

%%%% ACKNOWLEDGEMENTS %%%%%%%%%%%%%%%%%%%%%%%%%%%%%%%%%%%%%%%%%%%%%%%%%%%%%%%%%%%%%%%%%%%

\acknowledgments
The authors are very grateful to M. S. Bahramy, C.-R. Chang, P.-H. Chang, C.-L. Chen, A.W. Cummings, N. Leconte, J. M. Marmolejo-Tejada, N. Nagaosa, F. Ortmann, T. Rappoport, S. O. Valenzuela, D. Van Tuan, and X. Waintal for inspiring collaboration.
This project has received funding from the European Union's Horizon 2020 research and innovation programme under grant agreement No 696656. S.R. acknowledges Funding from the Spanish Ministry of Economy and Competitiveness and the European Regional Development Fund (Project No. FIS2015-67767-P (MINECO/FEDER)), the Secretaria de Universidades e Investigaci\'{o}n del Departamento de Econom\'{i}a y Conocimiento de la Generalidad de Catalu\~{n}a, and the Severo Ochoa Program (MINECO, Grant SEV-2013-0295). We acknowledge computational resources from PRACE and the Barcelona Supercomputing Center (Mare Nostrum), under Project 2015133194.
B. K. N. was supported by NSF Grant No. ECCS 1509094. The supercomputing time was provided by in part by XSEDE, which is supported by NSF Grant No. ACI-1053575.

\appendix

%%%%% SPIN-ORBIT COUPLING IN DISORDERED GRAPHENE : PHYSICAL MECHANISM AND TIGHT-BINDING MODEL

\section{Spin-orbit coupling in disordered graphene: Physical mechanisms and tight-binding models} \label{sec:soc}

The coupling between the orbital and the spin degree of freedom of electrons is a relativistic effect described formally by the nonrelativistic expansion of the Dirac equation in external electric and magnetic fields (for which exact solutions do not exist) in powers of the inverse speed of light $c$. In the second order $v^2/c^2$, one identifies~\cite{Zawadzki2005} the SOC term
\begin{equation} \label{eq:so}
H_{\rm SO}=\frac{\hbar}{4 m^2c^2} {\bf p} \cdot [\mathbf{s} \times \nabla V({\bf r})],
\end{equation}
responsible for the entanglement of the spin and orbital degrees of freedom in the two-component nonrelativistic Pauli Hamiltonian for spin-$\frac{1}{2}$ electron. Here $m$ is the free electron mass, $\mathbf{s}=(s_x,s_y,s_z)$ is the vector of the Pauli matrices, and $V({\bf r})$ is the electric potential. 

The physical origin of \eref{eq:so} is traditionally explained as the consequence of electron magnetic dipole moment (associated with spin) interacting with magnetic field in the rest frame of an electron~\cite{Jackson1998}, which is obtained by Lorentz transforming electric field from the lab frame. However, for intuitive understanding of the effect of SOC on propagating spins it is advantageous to remain in the lab frame~\cite{Fisher1971}  where a magnetic dipole ${\bm \mu}$  moving with velocity ${\bf v}$ generates electric dipole moment ${\bf P}_{\rm lab}={\bf v} \times {\bm \mu}/c^2$. Here the right-hand side is evaluated in the electron rest frame and ${\bf P}_{\rm lab}$ is measured in the lab (both sides can be evaluated in the lab frame yielding the same result to first order in $v/c$). The potential energy of the interaction of the electric dipole with the external electric field ${\bf E}_{\rm lab}$ in the lab frame, $U_{\rm dipole}=-{\bf P}_{\rm lab} \cdot {\bf E}_{\rm lab}$, corrected for the Thomas precession (which takes into account change in rotational kinetic energy due to the precession of accelerated electron seen by the lab observer) $U_{\rm Thomas}=-U_{\rm dipole}/2$, leads to $U_{\rm SO}=U_{\rm dipole}+U_{\rm Thomas}=-{\bf P}_{\rm lab}\cdot {\bf E}_{\rm lab}/2$. Thus, replacing classical quantities in $U_{\rm SO}$ with the corresponding Hermitian operators yields \eref{eq:so}.

The nonrelativistic expansion of the Dirac equation can be viewed as a method of systematically including the effects of the negative-energy solutions on the positive energy states by starting from their nonrelativistic limit~\cite{Zawadzki2005}. The SOC effects in vacuum are small due to huge gap $2mc^2$ between positive and negative energy states. In the case of atoms, SOC is due to interaction of electron spin with the average Coulomb field of the nuclei and other electrons. In solids, $V({\bf r})$ is the sum of periodic crystalline potential $V_\mathrm{crystal}(\mathbf{r})$ and an aperiodic part containing potentials due to impurities $V_\mathrm{imp}(\mathbf{r})$, confinement, boundaries and external electric fields. Thus, in solids, large value of electric field $\mathbf{E} = -\nabla V_\mathrm{crystal}(\mathbf{r})/e$ near the nuclei competes with the huge denominator in \eref{eq:so}, so that much smaller band gap between conduction and valence band (playing the role of electron positive energy sea and positron negative energy sea, respectively) replaces $2mc^2$, thereby illustrating the origin of strong enhancement of the SOC in solids~\cite{VIG_JSNM23,Winkler2003,Engel2005}.

In addition to impurity induced SOC effects, the SOC due to $V_\mathrm{crystal}(\mathbf{r})$ in solids with bulk inversion asymmetry or an interfacial $V_\mathrm{int}(\mathbf{r})$ accompanying structural inversion asymmetry spin-splits the band structure and can give rise to intrinsic SHE~\cite{SHE2,VIG_JSNM23}. For example, the Rashba SOC~\cite{MAN_NM14,Winkler2003} in 2DEGs within heterostructures with structural inversion asymmetry is given by
\begin{equation}\label{eq:rashba}
H_\mathrm{R} = \frac{\alpha}{\hbar} (\mathbf{s} \times \mathbf{p}) \cdot \hat{z}.
\end{equation}
The quantum transport algorithms discussed in \aaref{sec:kubo-formula} and ~\ref{sec:qtalgorithms} require TB Hamiltonian as an input. In the case of 2DEGs this can be achieved by discretizing \eeref{eq:so} and \eqref{eq:rashba} on the square lattice~\cite{Nikolic2006,Nikolic2007}. 

In the case of graphene, minimal (i.e., with the smallest number of orbitals per site) effective TB model can be constructed by starting from the usual graphene Hamiltonian with single $2p_z$ orbital {\it per} site of the honeycomb lattice and by adding SOC terms permitted by the symmetries of the lattice~\cite{gmitrahydogenatedgraphene,Irmer2015,Zollner2016,Gmitra2016,KOC_arXiv2016}. This leads to the following Hamiltonian employed for transport simulations in \ssref{sec:qshe} and \ref{sec:she} 
\begin{eqnarray}\label{eq:kanemele}
H & = &  -  \gamma_0\sum_{\langle ij\rangle }c_i^\dag c_j+\frac{2i}{\sqrt{3}} V_{\rm I} \sum_{\langle\langle ij\rangle\rangle \in \mathcal{R}}c_i^\dag \mathbf{s}\cdot(\mathbf{d}_{kj} \times \mathbf{d}_{ik})c_j \ \nonumber \\
&& + i V_\mathrm{R} \sum_{\langle ij \rangle \in \mathcal{R}} c_i^\dagger \hat{z} \cdot (\mathbf{s} \times \mathbf{d}_{ij}) c_j + iV_\mathrm{PIA} \sum_{\langle\langle ij\rangle\rangle \in \mathcal{R}} c_i^\dagger \hat{z} \cdot (\mathbf{s} \times \mathbf{D}_{ij}) c_j - \mu\sum_{i\in \mathcal{R}} c_i^\dag c_i \ . 
\end{eqnarray}
Here $c_i=(c_{i\downarrow},c_{i\uparrow})$ is the pair of annihilation operators for electrons with spin down and spin up on the site $i$; $\gamma_0 = 2.7$ eV is the nearest neighbor hopping parameter; $\langle ij \rangle$ denotes sum over nearest neighbors and $\langle\langle ij \rangle\rangle$ denotes sum over next-nearest neighbors;  $\mathbf{d}_{kj}$ is the unit vector pointing from site $j$ to site $k$, with site $k$ standing in between $i$ and $j$; $\mathbf{d}_{ij}$ is the unit vector pointing from site $i$ to its nearest neighbor $j$; $\mathbf{D}_{ij}$ is the unit vector pointing from site $i$ to its next-nearest neighbor $j$; and $\mathcal{R}$ denotes a set of hexagons where respective terms are assumed to be nonzero. For quantum transport simulations of realistic graphene systems, as  performed in \ssref{sec:qshe} and ~\ref{sec:she}, the parameters $V_\mathrm{I}$, $V_\mathrm{R}$, $V_\mathrm{PIA}$, $\mu$ can be extracted by fitting  the low-energy band structure obtained from first-principles calculations on supercells of graphene with adatoms~\cite{WEE_PRX1,gmitrahydogenatedgraphene,DIN_NP10,Irmer2015,Zollner2016,Gmitra2016,KOC_arXiv2016,CHA_NL14} or graphene on different substrate materials~\cite{Gmitra2016}.

\begin{figure}[t]
	\begin{center}
		\resizebox{8cm}{!}{\includegraphics{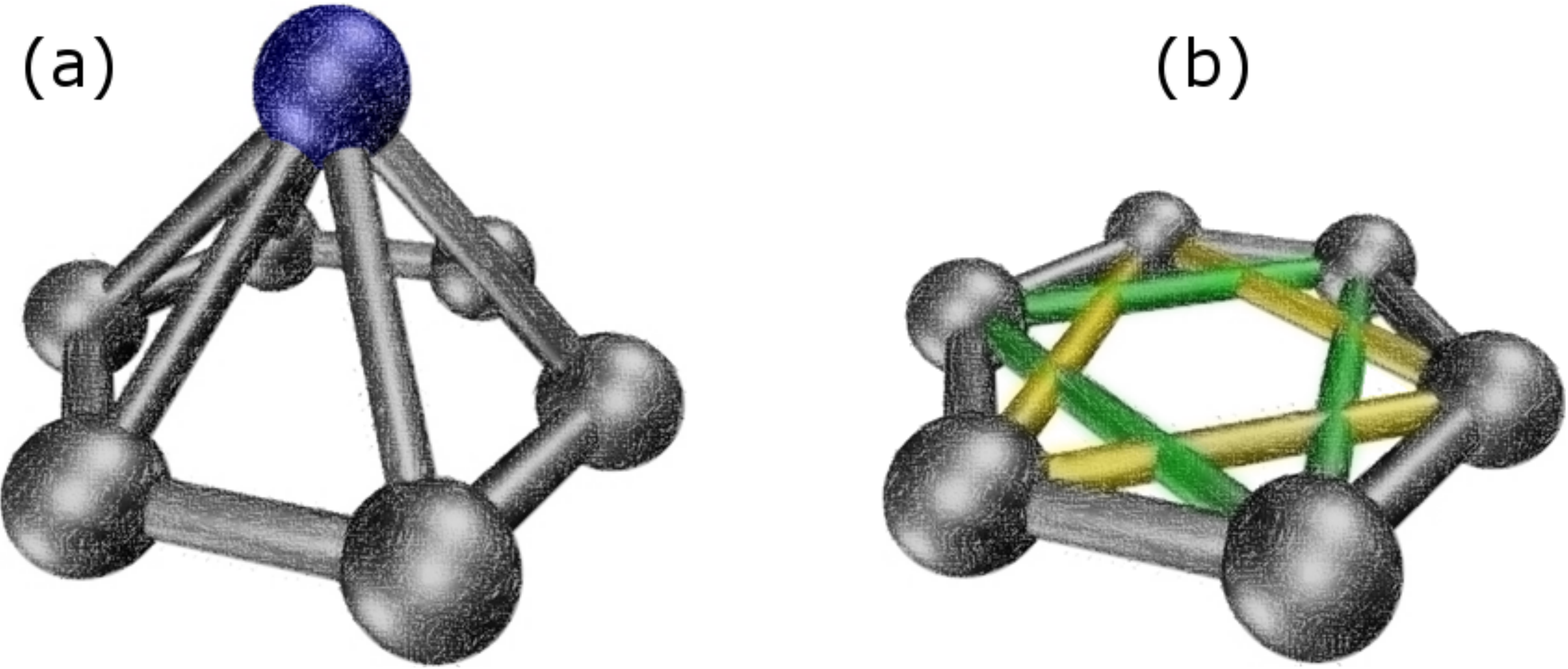}}
		\caption{(a) Graphene with a heavy In or Tl adatoms, which favor position in the center of the hexagon. (b) Visual depiction of spin-dependent hopping in the second term of \eref{eq:kanemele} whose magnitude $V_\mathrm{I}$ is enhanced by the presence of adatoms.}
		\label{fig:weeks}
	\end{center}
\end{figure}

The first and second term in \eref{eq:kanemele}, where the latter is intrinsic SOC present on all hexagons (i.e., $\mathcal{R}$ is the whole graphene lattice), comprise the so-called Kane-Mele model employed in the seminal arguments~\cite{KAN_PRL95,KAN_PRL95b} for the existence of 2D TIs. The intrinsic SOC of strength $V_\mathrm{I}$ splits Dirac cone by an energy gap $\Delta_\mathrm{I}=12 V_\mathrm{I}$, while preserving spin degeneracy due to combined time reversal and space inversion symmetry. The third term in \eref{eq:kanemele} is the Rashba SOC that appears when the inversion symmetry of graphene is broken by the substrate, external electric field or adatoms. The Rashba SOC lifts the spin degeneracy, destroys conservation of spin $s_z$ (unlike intrinsic SOC), and eventually closes~\cite{GMI_PRB80} the gap  $\Delta_\mathrm{I}$ when $V_\mathrm{R} \gtrsim V_\mathrm{I}$~\cite{Sheng2005b}. The forth term in \eref{eq:kanemele} is pseudospin inversion asymmetry (PIA) induced SOC~\cite{gmitrahydogenatedgraphene,Irmer2015,Zollner2016,Gmitra2016,KOC_arXiv2016} that arises due to the sites of the two triangular sublattices of honeycomb lattice becoming inequivalent close to the impurity site, so that matrix elements of \eref{eq:so} between $2p_z$ orbitals of carbon atom with chemisorbed adatom and next-nearest neighbor $2p_z$ orbitals (with flipped spin) become nonzero. Both Rashba and PIA SOC terms explicitly break $z \rightarrow -z$ symmetry. The fifth term can accommodate additional on-site energy $\mu$ on carbon atoms within hexagons $\mathcal{R}$ that are covered by adatoms~\cite{WEE_PRX1}.

All-electron DFT calculations~\cite{GMI_PRB80,Boettger2007} have estimated $\Delta_\mathrm{I}$ to be in the range 24--50  $\mu$eV,  which is minuscule due to the lightness of carbon atoms. The same result can be reproduced by using Slater-Koster TB Hamiltonian with a proper choice of three orbitals $2p_z$, $d_{xz}$, and $d_{yz}$ per carbon atom, where nominally unoccupied $d$ orbitals are required to fit DFT computed band structure of graphene and GNRs even in the absence of SOC~\cite{Boykin2011}. The SOC is then introduced by finding matrix elements of \eref{eq:so} in the basis of such orbitals, while spin-dependent hopping in the second, third and fourth term of \eref{eq:kanemele} is justified {\em a posteriori} through projection onto the subspace of $2p_z$ orbitals~\cite{KON_PRB82}. 

The Rashba SOC due to an external electric field is also rather small, where spin-splitting of energy levels reaches $\Delta_\mathrm{R} \simeq 5$ $\mu$eV (in Rashba spin-split Dirac cone, at each momentum there are two states with energy differing by $\Delta_\mathrm{R}$) in representative transverse electric field of strength 1 V/nm~\cite{GMI_PRB80}. Another source of Rashba SOC are ripples in graphene, but their typical curvature also leads to negligible $\Delta_\mathrm{R} \simeq 20$ $\mu$eV~\cite{Huertas-Hernando2006,Jeong2011}. Note that in contrast to Rashba SOC in \eref{eq:rashba} for 2DEG semiconductor heterostructures, in graphene it does not depend on electron momentum due to electrons having constant velocity at CNP~\cite{GMI_PRB80}.

To realize topologically protected quantum spin Hall~\cite{KAN_PRL95,KAN_PRL95b} and quantum anomalous Hall phases~\cite{Qiao2014} in graphene at room temperature, or to enable anticipated spintronic~\cite{Han2014,Mahfouzi2010} and thermoelectric applications~\cite{CHA_NL14,Xu2014a} requires to increase either $V_\mathrm{I}$ or $V_\mathrm{R}$. The exposed graphene surface makes possible new functionalities because other materials, such as ferromagnetic metals~\cite{Mahfouzi2010} and insulators~\cite{Qiao2014,YAN_PRL110}, or light~\cite{graphene_prox1} can be easily brought into direct contact with SO-coupled 2D electron system.

\begin{figure}[t]
	\begin{center}
		\resizebox{9cm}{!}{\includegraphics{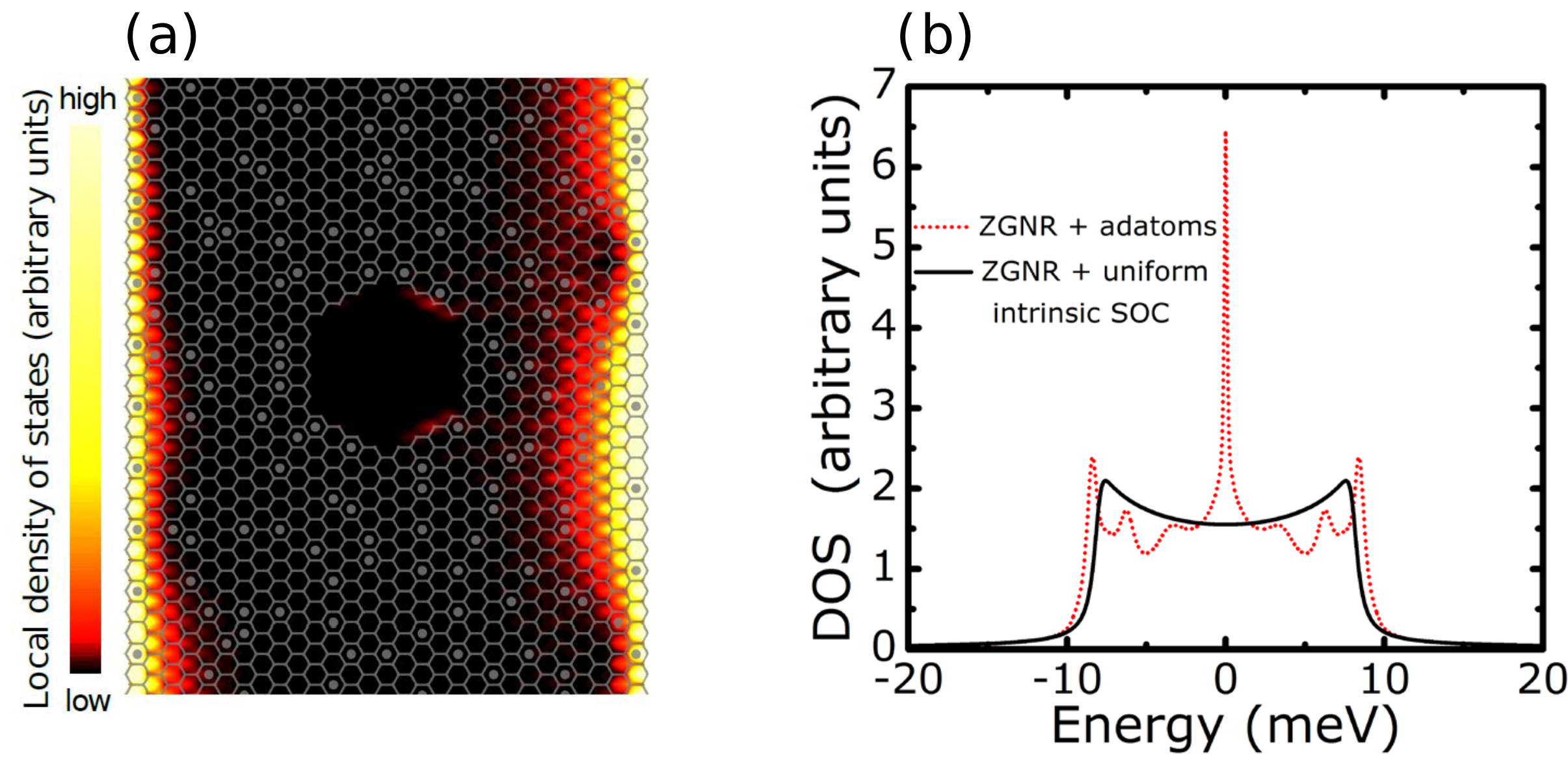}}
		\caption{(a) Local DOS at $E=1$ meV in ZGNR whose $n_\mathrm{ad} =19\%$ of hexagons are covered by In adatoms (grey dots in the center of hexagons). (b) Total DOS for ZGNR + adatoms in panel (a) 
			or ZGNR + uniform intrinsic SOC (with its magnitude tuned to open the same gap $\Delta_\mathrm{I}$ as in the case of adatoms), where the gap $\Delta_\mathrm{I} \simeq 17.3$ meV around  $E=0$ meV is filled by contributions from helical edge states and it is insensitive to the randomness of adatom configuration or spatial inhomogeneities. Adapted from \cref{CHA_NL14}.}
		\label{fig:gnrdos}
	\end{center}
\end{figure}

The DFT screening~\cite{WEE_PRX1} of heavy adatoms has predicted that $V_\mathrm{I}$ can be locally and substantially enhanced by In and Tl, which favor high-symmetry position in the center of the hexagons while being nonmagnetic and without inducing $V_\mathrm{R}$, as illustrated in \fref{fig:weeks}. The system graphene + adatoms of In or Tl is described by \eref{eq:kanemele} with $V_\mathrm{R} = V_\mathrm{PIA} = 0$ and  $V_\mathrm{I} \neq 0$ on hexagons $\mathcal{R}$ hosting the adatoms (as discussed above, uniform $V_\mathrm{I}$ on all hexagons due to carbon atoms themselves can be neglected). Remarkably, despite completely random position of heavy adatoms, such a disordered system has extremely stable 2D TI phase, which is actually stabilized by the randomness of adatom distribution~\cite{JIA_PRL109}. For example, total DOS shown in \fref{fig:gnrdos}(b) does not contain any signatures of spatial inhomogeneities, in contrast to the local DOS in \fref{fig:gnrdos}(a), which is confined around the edges and sensitive to the distribution of adatoms around the edge. This means that in transport calculations on such a system one does not need to perform disorder averaging~\cite{CHA_NL14,Shevtsov2012}. The energy gap $\Delta_\mathrm{I}=12 V_\mathrm{I} n_\mathrm{ad}$~\cite{Shevtsov2012} is controlled by the type of adatoms---$V_\mathrm{I} \approx 0.0032 \gamma_0$~\cite{CHA_NL14} for In and $V_\mathrm{I} \approx 0.017 \gamma_0$~\cite{WEE_PRX1} for Tl is extracted from DFT calculations---and their concentration $n_\mathrm{ad}$. Other choices for heavy adatoms---such as Os, Ir and Cu-Os or Cu-Ir dimers---are predicted~\cite{HU_PRL109} to generate even larger gaps $\Delta_\mathrm{I}$ using smaller adatom coverage while evading propensity of In atoms to cluster on graphene. 

The Rashba SOC can be uniformly enhanced by hybridization with $5d$-states of Au monolayer, which is intercalated at graphene/h-BN~\cite{OFarrell2016} or graphene/Ni interface~\cite{Marchenko2012}. Another experimentally demonstrated strategy~\cite{Wang2015b,Avsar2014} puts graphene onto a monolayer of transition metal dichalcogenide like WS$_2$ whose proximity effect induces Rashba and PIA SOC terms in \eref{eq:kanemele} while also enhancing the intrinsic SOC~\cite{graphene_prox1,Gmitra2016}. In fact, DFT calculations have also predicted that strong SOC proximity effect from WSe$_2$ would convert graphene into 2D TI~\cite{Gmitra2016}. Surprisingly, even light adatoms like chemisorbed hydrogen~\cite{Balakrishnan2013} or fluorine~\cite{Avsar2015} can induce nonzero $V_\mathrm{R}$ and $V_\mathrm{PIA}$, as well as  enhance $V_\mathrm{I}$, where such an effect is governed by the local lattice distortion~\cite{gmitrahydogenatedgraphene,CastroNeto2009} in the case of hydrogen or SOC of fluorine atom  itself~\cite{Irmer2015}. 

In quantum transport simulations of \sref{sec:she}, we use Au adatoms whose presence is accounted by TB model in \eref{eq:kanemele} with its parameters extracted from DFT calculations---$V_\mathrm{I} = 0.007 \gamma_{0}$, $V_\mathrm{R}=0.0165 \gamma_{0}$, $V_\mathrm{PIA}=0$ and $\mu = 0.1 \gamma_{0}$ is set on the hexagons $\mathcal{R}$ where Au adatoms are located (see \fref{NewFig1} for illustration). Hydrogen or fluorine adatoms require all terms in \eref{eq:kanemele} to be nonzero~\cite{gmitrahydogenatedgraphene,Irmer2015}, but in \sref{SHEKUBO} we selectively turn them on and off in order to understand how different SOC terms influence charge and spin Hall conductivities.

%%%%% QUANTUM TRANSPORT ALGORITHMS FOR CHARGE AND SPIN CURRENTS IN MULTITERMINAL SYSTEMS WITH AND WITHOUT DEPHASING EFFECTS 

\section{Quantum transport algorithms for charge and spin currents in multiterminal systems with and without dephasing effects}\label{sec:qtalgorithms}

The theoretical description of HE and SHE is founded on their respective conductivities, which become topological invariants in the case of QHE or QSHE~\cite{QI_RMP83}. On the other hand, the analysis of transport experiments on devices exhibiting such effects, which are embedded into circuits with many external leads~\cite{McEuen1990,Haug1993,Roth2009,Brune2010}, is typically based on the multiterminal LB formula~\cite{Buttiker1986,Datta1995}. The LB approach is actually essential to compute observable charge transport quantities, like the nonlocal resistance $R_\mathrm{NL}$ studied in \ssref{sec:she}, ~\ref{sec:zshe} and ~\ref{sec:vhe}, which are measured experimentally to confirm the existence of direct and inverse SHE and VHE. This is due to the fact that spin current and related conductance/conductivity or valley current and related conductance/conductivity are not directly observable quantities (e.g., transport of electron spin between two locations in real space is alien to Maxwell electrodynamics and no ``spin current ammeter'' exists~\cite{Adagideli2006}).

The usual multiterminal LB formula~\cite{Buttiker1986,Datta1995}
\begin{equation}\label{eq:mlbcharge}
I_p = \sum_q G_{p q}(V_p - V_q) \ ,
\end{equation}
relates total charge current $I_p$ in lead $p$ to voltages $V_q$ in all other leads {\it via} the charge conductance coefficients 
\begin{equation}\label{eq:gpq}
G_{p q} = \frac{e^2}{h} \int dE\,  \left(-\frac{\partial f}{\partial E} \right) \mathrm{Tr}[\mathbf{t}_{pq} \mathbf{t}_{pq}^\dagger] \ . 
\end{equation}
These coefficients are determined by the transmission matrix $\mathbf{t}_{p q}$ connecting transverse propagating modes within semi-infinite ideal leads $p$ and $q$. The multiterminal LB formula is valid when phase coherence is maintained in the active region of the device, while phase breaking events are assumed to be taking place only in the electron reservoirs to which the leads are attached at infinity and where electrons are equilibrated to acquire the Fermi-Dirac distribution $f_p(E)=f(E-eV_p)$. 

Using the same units for the total charge $I_p=I^\uparrow_p + I^\downarrow_p$ and total spin $I_p^{S_\alpha} = I^\uparrow_p - I^\downarrow_p$ currents flowing through lead $p$, which are constructed from spin-resolved charge currents $I_p^s$ with the spin quantization axis for $s=\uparrow,\downarrow$ chosen along the $\alpha$-direction ($\alpha=x,y,z$), Eq.~\eqref{eq:mlbcharge} can be extended to describe how total spin current $I_p^{S_\alpha}$ in lead $p$ 
\begin{equation}\label{eq:mlbspin}
I_p^{S_\alpha} = \sum_q G_{p q}^{S_\alpha} (V_p - V_q) \ .
\end{equation}
The total spin current $I_p^{S_\alpha}$ is conserved quantity within lead $p$ (i.e., $I_p^{S_\alpha}$ is the same as at each cross section of the lead) on the proviso that the lead does not contain any spin-dependent interactions~\cite{Nikolic2006}. The spin conductance coefficients in Eq.~\eqref{eq:mlbspin} are given by 
\begin{equation}\label{eq:gspinpq}
G_{p q}^{S_\alpha} = \frac{e^2}{h} \int dE\,  \left(-\frac{\partial f}{\partial E} \right) \mathrm{Tr}[s_\alpha \mathbf{t}_{pq} \mathbf{t}_{pq}^\dagger] \ ,
\end{equation}
where $s_\alpha$ is the Pauli matrix. 

We note that there has been a lively debate~\cite{Nikolic2005,Scheid2007} in the literature on the proper derivation of Eq.~\eqref{eq:mlbspin}. The debate was spurred by one of the early derivations~\cite{Pareek2004} using the traditional scattering matrix approach~\cite{Datta1995}, which predicted unphysical $I_p^{S_\alpha} \neq 0$ in equilibrium, i.e for constant $V_p$. The pitfalls~\cite{Scheid2007} in such a derivation can be evaded by starting from NEGF~\cite{Stefanucci2013} based expression for spin current in lead $p$ (the so-called Meir-Wingreen formula~\cite{Meir1992})
\begin{equation} \label{eq:mw}
I_p^{S_\alpha} =  \frac{e}{h} \int \!\! dE \, \mathrm{Tr} \, \{ s_\alpha [{\bm \Sigma}_p^<(E) {\bf G}^>(E) - {\bm \Sigma}^>_p(E) {\bf G}^<(E)] \} \ ,
\end{equation}
which assumes that any inelastic scattering events are localized within the active region of the device (i.e., they do not occur in the attached ideal leads). The two fundamental objects of NEGF formalism for steady state transport that enter into this formula are the retarded $\mathbf{G}(E)$ and the lesser $\mathbf{G}^<(E)$ Green's functions (GFs), which describe the density of available quantum states and how electrons occupy those states, respectively~\cite{Stefanucci2013}. They are given by
\begin{eqnarray}\label{eq:negf}
{\bf G}(E) & = & \left[E - {\bf H} - \sum_p {\bm \Sigma}_p(E) - {\bm \Sigma}_{\rm int}(E) \right]^{-1}, \label{eq:negf1} \\
{\bf G}^<(E) & = & {\bf G}(E) \left[\sum_p {\bm \Sigma}^<_p(E) + {\bm \Sigma}^<_{\rm int}(E)\right] {\bf G}^\dagger(E) \ . \label{eq:negf2}
\end{eqnarray}
In the elastic transport regime, the self-energies due to interactions with other electrons or bosonic quasiparticles are zero, ${\bm \Sigma}_\mathrm{int}(E) = 0 = {\bm \Sigma}^<_\mathrm{int}(E)$, so that retarded ${\bm \Sigma}_p(E)$ and lesser  ${\bm \Sigma}^<_p(E) = i f_p(E) {\bm \Gamma}_p(E)$ self-energies are generated only by attached leads, where ${\bm \Gamma}_p=i [{\bm \Sigma}_p(E) - {\bm \Sigma}^\dagger_p(E)]$ is the level broadening matrix determining the escape rates for electrons to exit into the attached leads. This makes it possible to rewrite \eref{eq:mw} for the total spin current in lead $p$ in the elastic transport regime as
\begin{eqnarray} \label{eq:lbspin}
I_{p}^{S_\alpha}  =   \frac{e}{h} \sum_{q} \int \!\! dE \, \mathrm{Tr} \, [s_{\alpha} {\bm \Gamma}_{q}(E)\mathbf{G}(E)\mathbf{\bm \Gamma}_{p}(E)\mathbf{G}^{\dagger}(E)] \left\{ f_{p}(E)-f_{q}(E) \right\} \ .
\end{eqnarray}
By expanding \mbox{$f_p(E)-f_q(E)$} to linear order in $V_{p}-V_{q}$, we obtain the desired multiterminal LB formula for spin currents 
\begin{eqnarray} \label{eq:lbtempspin}
I_{p}^{S_\alpha}  = \frac{e^2}{h} \sum_{q}\int \!\! dE \,  \left(-\frac{\partial f}{\partial E} \right) \mathrm{Tr}\, [s_\alpha \mathbf{\Gamma}_{q}(E)\mathbf{G}(E)\mathbf{\Gamma}_{p}(E)\mathbf{G}^{\dagger}(E)]  (V_p - V_q) \ ,
\end{eqnarray}
where $\mathrm{Tr}\, [s_\alpha\mathbf{\Gamma}_{q}(E)\mathbf{G}(E)\mathbf{\Gamma}_{p}(E)\mathbf{G}^{\dagger}(E)]$ expression in terms of NEGF quantities is equivalent to $\mathrm{Tr}[s_\alpha \mathbf{t}_{pq} \mathbf{t}_{pq}^\dagger]$ expressions in terms of the transmission submatrix component~\cite{Groth2014} of the full scattering matrix. Note that charge current $I_p$ is obtained also from \eref{eq:mw} by replacing $s_\alpha$ with a unit $2 \times 2$ matrix.

The calculation of spin and charge conductance coefficients in \eeref{eq:gpq} and \eqref{eq:gspinpq} was performed in \sref{sec:vhe} by means of home-made codes and in \sref{sec:she} by using the {\tt KWANT} package, which employs highly efficient and robust algorithms to calculate scattering matrix, while being able to significantly outperform commonly used recursive GF methods~\cite{Kazymyrenko2008} for multiterminal systems containing a large number of atoms~\cite{Groth2014}. The {\tt KWANT} packages also avoids the usual instabilities that occur with many commonly used algorithms (such as in dealing with the evanescent modes of complex leads)~\cite{Groth2014}. In the computation of nonlocal resistance $R_{\rm NL}$ measured in recent SHE, ZSHE and VHE experiments on multiterminal graphene, we inject charge current $I_1$ through lead 1 and current $-I_1$ flows through lead $2$ while $I_p \equiv 0$ in all other leads of six-terminal graphene devices illustrated in \ffref{NewFig1} and ~\ref{fig:fig1_zshe}. We then compute voltages that develop in the leads $p=3,4,5,6$ (labeled in \ffref{NewFig1} and ~\ref{fig:fig1_zshe}) in response to injected current $I_1$. The nonlocal resistance is obtained as $R_{\rm NL} = (V_3-V_4)/I_{1}$. The spin Hall angle for graphene devices in \ffref{NewFig1} and ~\ref{fig:fig1_zshe} is obtained from $\theta_\mathrm{sH} = I_5^{S_z}/I_1$.

The equivalence between NEGF and scattering matrix approaches holds only in the elastic or phase-coherent transport regime. When electron-electron~\cite{Thygesen2008,Profumo2015}, electron-phonon~\cite{Frederiksen2007} and electron-magnon~\cite{Mahfouzi2014} scattering events occur in the active region of the device, LB formulas become inapplicable. On the other hand, NEGF formalism~\cite{Stefanucci2013} offers a rigorous prescription for including such a process by starting from a microscopic Hamiltonian and by constructing the interacting self-energies ${\bm \Sigma}_{\rm int}(E)$ and ${\bm \Sigma}^<_\mathrm{int}(E)$ in some approximation to yield spin and charge current through \eref{eq:mw}. 

Although the NEGF formalism is also capable of scaling to systems with large number of atoms~\cite{Areshkin2010,Sanvito2011}, the self-consistent evaluation of Feynman diagrams for interacting self-energies (which yields coupled system of nonlinear integral equations) is at present prohibitively expensive for devices containing realistic number of atoms~\cite{Mahfouzi2014}.  Thus, to include dephasing processes due to many-body interactions in devices containing few thousands of carbon atoms, in Sec.~\ref{sec:zshe} we employ the phenomenological model of \cref{Golizadeh-Mojarad2007a}, which is conceptually and numerical simple while making it possible to adjust the degree of phase and momentum relaxation independently. In the ``momentum-conserving'' model of dephasing, the interacting self-energies are given by~\cite{Golizadeh-Mojarad2007a}
\begin{eqnarray} \label{eq:mc}
{\bm \Sigma}_{\rm int}(E) & = & d_p {\bf G}(E), \\
{\bm \Sigma}^<_{\rm int}(E) & = & d_p {\bf G}^<(E) \ ,
\end{eqnarray}
while in the ``momentum-relaxing'' model
\begin{eqnarray}\label{eq:mr}
{\bm \Sigma}_{\rm int}(E) & = & \mathcal{D}[d_m {\bf G}(E)], \\
{\bm \Sigma}^<_{\rm int}(E) & = & \mathcal{D}[d_m {\bf G}^<(E)] \ .
\end{eqnarray}
The operator $\mathcal{D}[\ldots]$ selects the diagonal elements of the matrix on which it acts while setting to zero all the off-diagonal elements. Any linear combination of these two choices can be used to adjust the phase and momentum relaxation lengths independently. When computed self-consistently together with ${\bf G}(E)$ and ${\bf G}^<(E)$, both of these choices for ${\bm \Sigma}_{\rm int}(E)$ and ${\bm \Sigma}^<_{\rm int}(E)$  ensure the conservation of charge current, $\sum_p I_p=0$.

For both momentum-conserving and momentum-relaxing dephasing (or their linear combination) one has to solve for ${\bf G}(E)$ and ${\bm \Sigma}_{\rm int}(E)$ using a self-consistent loop where the initial guess is
\begin{equation}
{\bf G}_{\rm in}(E) = \left[E- {\bf H} - \sum_p {\bm \Sigma}^{r}_p(E) \right]^{-1} \ .
\end{equation}
Then
\begin{equation}
{\bf G}_{\rm out}(E) = \left[E- {\bf H} - \sum_p {\bm \Sigma}_p(E) - d_p {\bf G}_{\rm in}(E) \right]^{-1} \ ,
\end{equation}
in the case of ``momentum-conserving'' dephasing or
\begin{equation}
{\bf G}_{\rm out}(E) = \left\{ E- {\bf H} - \sum_p {\bm \Sigma}_p(E) - \mathcal{D}[d_m {\bf G}_{\rm in}(E)] \right \}^{-1},
\end{equation}
in the case of momentum-relaxing dephasing is used as the input ${\bf G}_{\rm in}(E)$ of next iteration. We assume that the self-consistent loop has converged when $||{\bf G}_{\rm out}(E) -  {\bf G}_{\rm in}(E)||<10^{-4}$.

Using the converged ${\bf G}(E)$ matrix, the next step is to compute ${\bf G}^<(E)$, which proceeds differently for momentum-conserving and momentum-relaxing dephasing while yielding the same generalization of Eq.~\eqref{eq:mlbcharge} 
\begin{equation}\label{eq:mld}
I_p  = \sum_q (G_{pq}^{\rm coh}+G_{pq}^{\rm incoh})(V_p-V_q) \ .
\end{equation}
Here the ``coherent'' contribution to charge conductance coefficients is given by
\begin{equation}\label{eq:tcoh}
G_{pq}^{\rm coh}(E) = \frac{e^2}{h} \int dE\,  \left(-\frac{\partial f}{\partial E} \right) \mathrm{Tr}\, [ {\bf \Gamma}_p(E) {\bf G}(E) {\bf \Gamma}_q(E) {\bf G}^\dagger(E) ] \ ,
\end{equation}
while the ``incoherent'' contribution is given by
\begin{equation}\label{eq:tincoh}
G_{pq}^{\rm incoh}(E) = \frac{e^2}{h} \int dE\,  \left(-\frac{\partial f}{\partial E} \right) \mathrm{Tr}\, [ {\bf \Gamma}_p(E) {\bf G}(E) {\bf \Gamma}_q^{d}(E){\bf G}(E)].
\end{equation}
Although $G_{pq}^\mathrm{coh}$ in \eref{eq:tcoh} resembles \eref{eq:gpq} for phase-coherent transport of single electron exhibiting elastic scattering only, it actually takes into account the many-body interaction effects through  ${\bf G}(E)$ in \eref{eq:negf1}, which includes ${\bm \Sigma}_{\rm int}(E)$.

In the case of momentum-conserving dephasing, the matrix ${\bm \Gamma}_{\alpha}^{d}$ in \eref{eq:tincoh} is obtained from
\begin{equation}\label{eq:sylvester}
[{\bf G}^{r}_0]^{-1} {\bm \Gamma}_{\alpha}^{d} - d_p {\bm \Gamma}_{\alpha}^{d} {\bf G}^{a}_0 - d_p {\bm \Gamma}_{\alpha} {\bf G}^{a}_0 = {\bf 0} \ .
\end{equation}
This is recognized as the Sylvester equation~\cite{Golub1996} of matrix algebra, ${\bf A} {\bf X} + {\bf X}{\bf B} + {\bf C}={\bf 0}$, where we identify unknown matrix as ${\bf X} = {\bm \Gamma}_{\alpha}^{d}$ while the known coefficients are \mbox{${\bf A} = [{\bf G}^{r}_0]^{-1}$}, \mbox{${\bf B} = - d_p {\bf G}^{a}_0$}, and ${\bf C}= - d_p {\bm \Gamma}_{\alpha} {\bf G}^{a}_0$ \ .

In the case of momentum-relaxing dephasing, the diagonal elements of the matrix ${\bm \Gamma}_{\alpha}^{d}$ in \eref{eq:tincoh} are obtained from
\begin{equation}
[{\bf \Gamma}_{\beta}^{d}]_{jj} = d_m \sum_v [{\bf Q}]_{jv} [{\bf G}^{r}_0{\bf \Gamma}_{\beta}{\bf G}_0^{a}]_{vv} \ ,
\end{equation}
using~\cite{Cresti2008a} ${\bf Q}=[1-d_m {\bf P}]^{-1}$ and $[{\bf P}]_{jv}=[{\bf G}_0^{r}]_{jv} [{\bf G}_0^{a}]_{vj}$. Here the notation $[{\bf M}]_{jv}$ denotes the matrix element of ${\bf M}$.

We note that phenomenological dephasing effects are often introduced~\cite{Kilgour2016} into quantum transport simulations {\it via} the B\"{u}ttiker voltage probe scheme~\cite{Buttiker1985}. Such probes are attached to the active region as leads with no net charge current flowing through them, so that for every electron that enters the probe and is absorbed by its macroscopic reservoir at infinity another one has to come out, which is not coherent with the one going in. For example, one possible way to apply this method to multiterminal graphene devices in \fref{NewFig1} or \fref{fig:fig1_zshe} is to attach one-dimensional leads to each site~\cite{Metalidis2006} of the honeycomb lattice. This is equivalent to adding a complex energy $-i\eta$ to on-site potential of graphene Hamiltonian (parameter $\eta$ is related to the dephasing time $\eta=\hbar/2\tau_\phi$). In addition, one has to solve the ensuing multiterminal LB formula by imposing that current through additional 1D leads is zero~\cite{Metalidis2006}. However, besides blurring phase-coherence-generated oscillations in $G_{pq}$, B\"uttiker voltage probes can also introduce additional scattering that reduces the average value of $G_{pq}$ in an uncontrolled fashion~\cite{Golizadeh-Mojarad2007a}. 

Since we find that momentum-conserving model of dephasing discussed cannot reproduce experimental results~\cite{Abanin2011}, in \sref{sec:zshe} we employ  the momentum-relaxing model. An interested reader can find detailed comparison of momentum-conserving, momentum-relaxing and traditional B\"uttiker voltage probes~\cite{Buttiker1985,Metalidis2006} phenomenological methods to introduce dephasing in quantum transport in \cref{Golizadeh-Mojarad2007a} for a simple example of disordered wire attached to two ideal semi-infinite leads.

The GFs also provide other interesting quantities \cite{CRE_PRB68,CRE_EPJB46} as the local density-of-states $\rho_i(E)$ on the carbon atom with index $i$
\begin{equation}\label{eq:ldos}
			\mathbf{\rho}_i(E) \ = \ -\Frac{1}{\pi} \ \Im \left[ \mathbf{G}_{ii}(E) \right] \ ,
\end{equation}
the local density-of-occupied-states
\begin{equation}\label{eq:lodos}
			\rho_i^{\rm occ}(E) \ = \ \Frac{1}{2\pi} \  \mathbf{G^<}_{ii}(E)  \ ,
\end{equation}
and the local spectral current distribution $I_{i\rightarrow j}$ flowing between the atom with index $i$ and the atom with index $j$
\begin{equation}\label{eq:lcur}
			I_{i\rightarrow j}(E) \ = \ \Frac{e}{h} \ \Im \left[ \mathbf{H}_{ij} ~ \mathbf{G}^<_{ji}(E) \right] \ ,
\end{equation}
where $H_{ij}$ is the Hamiltonian matrix element.
These local quantities allow us to understand where charges flow and accumulate within the system.

%%%%% THE KUBO FORMULA FOR DIFFERENT CONDUCTIVITY TENSORS %%%%%%%%%%%%%%%%%%%%%%%%%%%%%%%

\section{The Kubo formula for different conductivity tensors}\label{sec:kubo-formula}

The Kubo formula allows the investigation of the linear response of a system to an external perturbation. When this perturbation is an electric field, a possible result is the generation of a current density, which, in the context of this review, is associated with transport of charge or different types of spins (pseudospin, isopin, intrinsic spin). In this case, the linear coefficient is defined through the following equation
\begin{equation} \label{def:cond}
   \Gamma_\alpha=\sigma_{\alpha,\beta} E_\beta \ , 
\end{equation}
where $\Gamma_\alpha$ is the generated macroscopic current in the system and the tensor $\sigma_{\alpha,\beta}$ is  the conductivity. 
Here, we will derivate the formalism along the same lines as in \cref{Mahan}
The starting point is a Hamiltonian of the form
\begin{equation} \label{eq:kubo_ham}
    H=H_0 + \lim_{s\rightarrow 0 }\text{e}^{st}H' \ ,
\end{equation}
where $H_0$ is the many-body Hamiltonian of the system when the perturbation is absent, and $H'$ is the coupling to the perturbation. 
The prefactor $\text{e}^{st}$ is explicitly placed so that the perturbation vanish at $t\rightarrow-\infty$, and the limit $s\rightarrow 0$ is placed to ensure the field is turned on adiabatically. One fundamental assumption to construct the Kubo formula is the existence of a unique many-body ground state at temperature $T$ and chemical potential $\mu$, which is the  result of a previous thermalization process at $t=-\infty$.  This mean there is no quantum coherent states at the initial time, excluding as a consequence initial currents or polarized states. 

The macroscopic current in \eref{def:cond} is connected to the microscopic \emph{many-body operator} $\gamma^{\text{MB}}_\alpha$, through its thermal average
\begin{equation} \label{def:mean}
   \Gamma_\alpha= \text{Tr}\left [\, \rho(\mu,T) \, \gamma^{\text{MB}} _\alpha\,\right]
\end{equation}
where $\rho$  is the many-body density matrix in a grand canonical ensemble at given temperature $T$ and chemical potential $\mu$. In order to find the conductivity, one must connect \eref{def:mean} with \eref{def:cond} within the linear regime by calculating the perturbed density matrix in \eref{def:mean}. 
In the linear response regime, this quantity can be written as
\begin{equation}
    \rho(t,\mu,T)=\rho_0(\mu,T)+\lim_{s\rightarrow 0} \text{e}^{st} \delta\rho(\bm{E}) \ ,
\end{equation}
where $\rho_0$ is the equilibrium density matrix in the absence of the electric field and  $\delta\rho(\bm{E})$ is a small modification due to the presence of the field, which is assumed to vanish for $t\rightarrow-\infty$, in the same way as the electric field. After the electric field is turned on, the equilibrium density will evolve adiabatically to a new density. This process is described by the Liouville-Von Neumann equation
\begin{equation}
   i\hbar \frac {\partial\rho(t) }{\partial t} = [H(t),\rho(t) ] \ .
\end{equation}
By replacing both the Hamiltonian and the density operator with those in \eeref{eq:kubo_ham} and \eqref{def:mean}, and excluding all terms that are nonlinear in the electric field, the final result is
\begin{equation} \label{eq:rho1}
   i\hbar \frac {\partial \left[\text{e}^{st} \right ]}{\partial t}\delta\rho - \text{e}^{st}[H_0,\delta\rho] =\text{e}^{st} [H',\rho_0] \ .
\end{equation}
Before proceeding any further, let us introduce the definition for the evolution operator of a time-independent Hamiltonian $H_0$
\begin{equation}
   U(t,t_0)\equiv \text{e}^{-i\frac{(t-t_0)}{\hbar} H_0 } \ .
\end{equation}
In the Heisenberg picture, this operator is responsible for the time evolution of observables 
\begin{equation}
    A(t-t_0)\equiv U^\dagger(t,t_0) A U(t,t_0) \ .
\end{equation}
where $A$ is a time-independent operator. Using these definitions, one can prove the following identity:
\begin{equation}
  i\hbar  U(t,0) \frac{d\, f(t)A(t)}{dt} U^\dagger(t,0)=i\hbar  U(t,0) \frac{d}{dt}\Big( f(t)U^\dagger(t,0) \,A \, U(t,0)\Big) U^\dagger(t,0)=   - i\hbar \frac{\partial f(t)}{\partial t} A -f(t)[H_0, A] 
\end{equation}
with $f(t)$ a time-dependent scalar function. Setting $f(t)=e^{st}$ and $A=\delta\rho$, we can use this identity to express equation \eref{eq:rho1} as:
\begin{equation}
   i\hbar U(t,0)\frac{d\text{e}^{st}\delta\rho(t)}{dt}U^\dagger(t,0)= \text{e}^{st} [H',\rho_0] \ .
\end{equation}
It is straightforward to prove that the time evolution of $A(t)$ is determined by
\begin{equation}
   i\hbar\frac{\partial A(t_0)}{\partial t} -[H_0,A(t_0)] =    i\hbar\frac{dA(t)}{dt}  \ .
\end{equation}
By setting $A(t)=\text{e}^{st}U(t,0)^\dagger\delta\rho U(t,0)=\text{e}^{st}\delta\rho(t)$, we have a relation that seems to be very similar to left side of equation \eref{eq:rho1}
\begin{equation}
   i\hbar\frac{\partial \text{e}^{st}\delta\rho(t) }{\partial t} -[H_0,\text{e}^{st}\delta\rho(t)] =    i\hbar\frac{d\text{e}^{st}\delta\rho(t)}{dt}  \ ,
\end{equation}
in terms of a total derivative, and then
\begin{equation}
     U(t,0)\frac{d\text{e}^{st}\delta\rho(t)}{dt}U^\dagger(t,0)= -\frac{i}{\hbar}\text{e}^{st} [H',\rho_0] \ ,
\end{equation}
which can be immediately integrated as
\begin{equation}
    \delta\rho= -\frac{i}{\hbar}\int_{-\infty}^t dt'\text{e}^{s(t'-t)} [H'(t'-t),\rho_0] \ .\label{eqconm}
\end{equation}
In order to eliminate the commutator, we first demonstrate the following identity
\begin{equation}
    [H'(-t),\rho_0] = -i\rho_0 \hbar \int_0^\lambda d\lambda' \frac{\partial H'(-t-i\hbar\lambda')}{\partial t} \ ,
\end{equation}
where $\lambda=1/T$ with $T$ the temperature. To prove it, we first notice that
\begin{equation}
 \int_0^\lambda  \frac{\partial H'(-t-i\hbar\lambda')}{\partial \lambda} d\lambda'=i\hbar \int_0^\lambda  \frac{\partial H'(-t-i\hbar\lambda')}{\partial t} d\lambda' = H'(-t)-H'(-t-i\hbar\lambda) ,
\end{equation}
then, multiplying by the equilibrium density matrix $\rho_0(\mu,T)=\text{e}^{-(H-\mu N)/T}$, we obtain
\begin{equation}
 i\hbar\rho_0 \int_0^\lambda  \frac{\partial H'(-t-i\hbar\lambda')}{\partial t} d\lambda' = \rho_0(\mu,T)[ \, H'(-t)-\text{e}^{H/T}H'(-t) \text{e}^{-H/T} \,]=[\rho_0,H(-t)] \ ,
\end{equation}
which can be replaced in \eref{eqconm} after performing the following change of variable $t'-t\rightarrow -t$
\begin{equation} \label{non-equilibriuim-density}
     \delta\rho= -\rho_0\int_{0}^{\infty}dt\text{e}^{-s t}   \int_0^\lambda d\lambda' \frac{\partial H'(-t-i\hbar\lambda')}{\partial t} \ . 
\end{equation}
This last relation is a very important one because it relates the nonequilibrium density matrix with an arbitrary external perturbation. 
We specify now the perturbation Hamiltonian by considering the minimal coupling interaction term
\begin{equation}
     H'(t)= \int_{V} d{r}^3 \phi(\bm{r}) \rho_e(\bm{r},t) \ ,
\end{equation}
where $V$ is the volume of the sample, $\phi(\bm{r})$ is a time-independent scalar potential, which is related to the electric field through the relation $\bm{E}=-\bm{\nabla} \phi$  and $\rho_e^{\text{MB}}(\bm{r},t)$ is the many-body charge density. By taking the time derivative of the interaction term, one can use the continuity equation for charge to obtain
\begin{equation}
    \frac{\partial H'(t)}{\partial t}= \int_V d{r}^3 \phi(\bm{r}) \frac{\partial\rho_e^{\text{MB}}(\bm{r},t)}{\partial t} =  \int d{r}^3 \phi\cdot \bm{\nabla}\cdot\bm{j}^{\text{MB}}(\bm{r},t) \ .
\end{equation}
where $\bm{j}^{\text{MB}}$ is the many body charge current density.
After integrating by parts, the previous equation can be rewritten as
\begin{equation}
    \frac{\partial H'(t)}{\partial t}=  -\int_V d{r}^3 \bm{E} \cdot\bm{\bm{j}^{\text{MB}}}(\bm{r},t) \ . 
\end{equation}
The last step consists in using the previous relation in \eref{non-equilibriuim-density}
to obtain
\begin{equation}
     \delta\rho= \lim_{s\rightarrow 0 }\rho_0\int_{0}^{\infty}dt\text{e}^{-s t}  \int_0^\lambda d\lambda'  \sum_\beta\int_V  dr^3 E_\beta j^{\text{MB}}_\beta(\bm{r},-t-i\hbar\lambda') \ .
\end{equation}
Finally, from \eref{def:mean} we obtain the conductivity tensor
\begin{equation}
    \sigma_{\alpha\beta}(\bm{r})=\lim_{s\rightarrow 0 }\int_{0}^{\infty}dt\text{e}^{-s t}  \int_0^\lambda d\lambda'  \int_V  dr^3 \left[\rho_0{j}^{\text{MB}}_\beta(\bm{r},-t-i\hbar\lambda')\gamma^{\text{MB}}_\alpha \right]) \ . 
\end{equation}
This last equation is the Kubo formula.

\subsection*{Different representations of the Kubo formula}

The previous derivation of the Kubo formula was obtained for a general system, provided that the initial assumptions are satisfied. In the following, we will focus on the non-interacting electron approximation and derive the different versions of the formula that are used in the main text. 

In general, the many-body Hamiltonian can be written as
\begin{equation}
    H=\sum_{n} c_{n}^\dagger c_n \varepsilon_{n} \ ,
\end{equation}
where $c_{n}^\dagger$ and $c_n$ are the creation and annihilation operators of an electron at a given energy $\varepsilon_n$, which is an eigenvalue of $H$. Additionally, the many-body current operator can represented as
\begin{equation}
    \gamma^{\text{MB}}_\alpha=\sum_{n}c_{n}^\dagger c_m \langle m|\gamma_\alpha|n\rangle
\end{equation}
where $|n\rangle$ are single-particle eigenstates and $\gamma_\alpha$  is the single-particle current operator. 

In the independent electron approximation the trace in \eref{def:mean} can be calculated by using the following result
\begin{equation}
    \text{Tr}\left[\, \rho_0 \,c_m^\dagger c_n c_p^\dagger c_q  \, \right]=  \delta_{np} \delta_{mq}f(\mu,T,\varepsilon_m )[\,1 -f(\mu,T,\varepsilon_n )\,] \ .
\end{equation} 
By using this property, one obtains
\begin{equation}
    \sigma_{\alpha,\beta}=  \lim_{s\rightarrow 0 }\int_{0}^{\infty}dt\text{e}^{-s t}  \sum_{m,n}\int_0^\lambda d\lambda' f(\mu,T,\varepsilon_m ) \langle m|\gamma_\alpha|n\rangle \,[\,1 -f(\mu,T,\varepsilon_n ) ]\,\langle n|j_\beta(-t-i\hbar\lambda')|m\rangle \ .
\end{equation}
In order to perform the integration in $\lambda'$, we will cast the evolution operator for $t=-i\hbar\lambda$, such that we can write
\begin{align}
    \sigma_{\alpha,\beta}&= \lim_{s\rightarrow 0 } \int_{0}^{\infty}dt\text{e}^{-s t}  \sum_{m,n} 
                            f(\varepsilon_m )(1 -f(\varepsilon_n ) ) 
                            \int_0^\lambda d\lambda' \text{e}^{-\lambda(\varepsilon_n-\varepsilon_m)} 
                            \langle m|\gamma_\alpha|n\rangle
                            \,\,\langle n|j_\beta(-t)|m\rangle \ . \nonumber\\
\sigma_{\alpha,\beta}&=  \lim_{s\rightarrow 0 }\int_{0}^{\infty}dt\text{e}^{-s t}  \sum_{m,n}  \left[
                            f(\varepsilon_m )(1 -f(\varepsilon_n ) )
                            \frac{1-\text{e}^{-\lambda(\varepsilon_n-\varepsilon_m)} }{\varepsilon_n-\varepsilon_m}\right]
                            \langle m|\gamma_\alpha|n\rangle
                            \,\,\langle n|j_\beta(-t)|m\rangle \ .
\end{align}
The expression in brackets can be replaced by the following identity
\begin{equation}
    \frac{1-\text{e}^{-\lambda(\varepsilon_n-\varepsilon_m)}}{\varepsilon_n-\varepsilon_m} f(\varepsilon_m)[1-f(\varepsilon_n)]=\frac{f(\varepsilon_n)-f(\varepsilon_m)}{\varepsilon_n-\varepsilon_m} \ ,
\end{equation}
so that the conductivity can rewritten as
\begin{equation}
    \sigma_{\alpha,\beta}=  \lim_{s\rightarrow 0 }\int_{0}^{\infty}dt\text{e}^{-s t}  \sum_{m,n}\int_0^\lambda d\lambda' 
    \frac{f(\varepsilon_n)-f(\varepsilon_m)}{\varepsilon_n-\varepsilon_m}
    \langle m|\gamma_\alpha|n\rangle
    \,\,\langle n|j_\beta(-t)|m\rangle \ .
\end{equation}
When the time integration is performed and we replace the current operator by the velocity operator, this expression is the same as the initial expression for the conductivity presented in \sref{sec:intro_hall}.

By developing this equation further, one can insert the identity $\int_{-\infty}^\infty d\varepsilon \delta(\varepsilon - \varepsilon_n) =1$ in order to rewrite the Kubo formula as
\begin{equation}
    \sigma_{\alpha,\beta}= \lim_{s\rightarrow 0 } \int_{-\infty}^\infty d\varepsilon\int_{0}^{\infty}dt\text{e}^{-s t}  \sum_{m,n}\left(\frac{ \delta(\varepsilon -\varepsilon_n)f(\varepsilon)}{    \varepsilon-\varepsilon_m}- \frac{\delta(\varepsilon -\varepsilon_m)f(\varepsilon)}{\varepsilon_n-\varepsilon}\right)
    \langle m|\gamma_\alpha|n\rangle
    \,\langle n|j_\beta(-t)|m\rangle \ ,
\end{equation}
which can be then represented as a trace
\begin{eqnarray}
    \sigma_{\alpha,\beta} &=& \lim_{s\rightarrow 0 }\lim_{\eta\rightarrow0} \int_{-\infty}^\infty d\varepsilon\int_{0}^{\infty }dt\text{e}^{-s t}  f(\varepsilon)\text{Tr}\left[
    \delta(\varepsilon -H)j_\beta(-t)\frac{1}{H-\varepsilon+i\eta}\gamma_\alpha \right. \\[3mm]
                          & & \ \ \ \ \ \ \ \left. - \delta(\varepsilon -H)\gamma_\alpha\frac{1}{H-\varepsilon-i\eta}j_\beta(-t)\right] \ , \nonumber
\end{eqnarray}
where the $\eta$ parameters is introduced to ensure the convergence of the integrals. This second expression is the starting point of the real-space method developed by Ortmann and coworkers \cite{Ortmann2015}, which was presented in \sref{sec:qshe}. After integrating in time, one obtains the following expression 
\begin{eqnarray}
    \sigma_{\alpha,\beta}&=&\lim_{s\rightarrow 0 } \lim_{\eta\rightarrow0}\int_{-\infty}^\infty d\varepsilon f(\varepsilon)\text{Tr}\left[
    \delta(\varepsilon -H)j_\beta\frac{1}{H-\varepsilon+i\eta}\frac{1}{H-\varepsilon+is}\gamma_\alpha \right. \\[3mm]
                          & & \ \ \ \ \ \ \ \left.   - \delta(\varepsilon -H)\gamma_\alpha\frac{1}{H-\varepsilon-i\eta}\frac{1}{H-\varepsilon-is}j_\beta\right]
													\ . \nonumber
\end{eqnarray}
By using the definition of the retarded $G^{-}(\varepsilon,H)$ and advanced $G^{+}(\varepsilon,H)$ GFs for non-interacting electrons
\begin{equation}
    G^\pm(\varepsilon,H)=\lim_{\eta\rightarrow 0}\frac{1}{H-\varepsilon\pm i\eta}
\end{equation}
and the following identity
\begin{equation}
    \frac{dG^\pm(\varepsilon,H)}{d\varepsilon}=-\frac{1}{(H-\varepsilon\pm i\eta)^2}
\end{equation}
one can find the following representation of the Kubo formula
\begin{equation}
    \sigma_{\alpha,\beta}=  \int_{-\infty}^\infty d\varepsilon f(\varepsilon) \text{Tr}\left[
    \delta(\varepsilon -H)j_\beta\frac{dG^+(\varepsilon,H)}{d\varepsilon}\gamma_\alpha
  - \delta(\varepsilon -H)\gamma_\alpha\frac{dG^-(\varepsilon,H)}{d\varepsilon}j_\beta\right] \ ,
\end{equation}
which is known as the Kubo-Bastin formula \cite{bastin1971}, and it is used in \sref{sec:qshe} to compute the spin conductivity tensor. 

As a final simplification, we consider the case where only the diagonal elements of the GFs are relevant. In this case, one can use the property
\begin{equation}
    G^+(\varepsilon,H)-G(\varepsilon,H)^-= \frac{i}{\pi}\delta(H-\varepsilon)
\end{equation}
to simplify the Kubo-Bastin formula as
\begin{equation}
    \sigma_{\alpha,\alpha}=  \frac{1}{\pi^2} \int_{-\infty}^\infty d\varepsilon f(\varepsilon)\text{Tr}\left[
    \delta(\varepsilon -H)j_\alpha\frac{d\delta(\varepsilon-H)}{d\varepsilon}j_\alpha
    \right] \ ,
\end{equation}
which now can be written as
\begin{equation}
    \sigma_{\alpha,\alpha}=   \int_{-\infty}^\infty d\varepsilon\frac{\partial f(\varepsilon)}{\partial\varepsilon}\text{Tr}\left[
    \delta(\varepsilon -H)j_\alpha\delta(\varepsilon-H)j_\alpha
    \right] \ .
\end{equation}
This last expression is known as the Kubo-Greenwood formula, which is widely used for transport calculation and is the starting point of the wave package evolution technique described briefly in \sref{sec:intro_hall}.

\subsection*{Numerical implementation of the Kubo formula in real-space: Kernel polynomial method}

As can be seen from all the previous equations, to calculate the conductivity it is essential to obtain the GFs of the system. To this aim, a very convenient approach is the KPM \cite{SilverKPM,WeisseKPM}. 
The first step, is to rescale all the energies of the system between the $[-\alpha,\alpha]$ interval with $ \alpha\rightarrow1$, which can be done by the following transformations 
\begin{equation} \label{eq:sh_hamil}
  \tilde{H}=\frac{2\alpha}{\Delta E}\left(H -\frac{E^{+}+E^{-}}{2}\right), \quad
  \tilde{\varepsilon}=\frac{2\alpha}{\Delta E}\left(\varepsilon -\frac{E^{+}+E^{-}}{2}\right)
\end{equation}
where $\Delta E$ is the bandwidth  and $E^-$ and $E^+$ the lower and upper band edge respectively. Then, one can proceed to approximate the GFs of the system by using an orthogonal basis of Chebyshev polynomials. For the Kubo-Bastin formula, for example, this leads to the following expression for the conductivity \cite{Garcia2016}
\begin{equation}
  \sigma_{{\alpha\beta}}^z(\mu, T)= \frac{4\hbar}{V}\frac{4}{\Delta E^2}
  \int_{-1}^{1}d\tilde{\varepsilon}
  \frac{f(\tilde{\varepsilon})}{(1-\tilde{\varepsilon}^2)^2}\sum_{m,n}^M\Gamma_{{nm}}
  (\tilde { \varepsilon } )\mu^{\alpha\beta}_{{nm}}\label{ChebyshevKubo}
\end{equation} 
where 
\begin{equation} \label{eq:sh_gamma}
  {\Gamma_{mn}(\tilde{\varepsilon})\equiv[(\tilde{\varepsilon}-i
  n\sqrt{1-\tilde{\varepsilon}^2})\text{e}^{in\,\text{acos}(\tilde{\varepsilon})}T_m(\tilde{\varepsilon})}
  +(\tilde{\varepsilon}+im\sqrt{1-\tilde{\varepsilon}^2})\text{e}^{-im\,\text{acos}(\tilde{\varepsilon})}T_n(\tilde{\varepsilon})]
\end{equation}
is an energy dependent function that does not depend on the details of the Hamiltonian and  
\begin{align}
  \mu^{\alpha\beta,z}_{{mn}}\equiv \frac{g_m g_n}{(1+\delta_{{n0}})(1+\delta_{{m0}})}\text{Tr}\left[\gamma_\alpha^z T_m(\tilde{H})j_\beta T_n(\tilde{H})\right] 
\end{align}
is the so-called Chebyshev expansion moments, with $T_n(H)$ the Chebyshev polynomials defined recursively as
\begin{equation} \label{eq:sh_T}
    T_n(H)=2HT_{n-1}(H)-T_{n-2}(H),\quad T_1(H)=H ,\quad T_0(H)=1, 
\end{equation}
and $g_n$ as the Jackson's $g$-factor used to reduce Gibbs oscillations \cite{WeisseKPM} due to truncation at order $M$ of the sum in \eref{ChebyshevKubo}. Calculating the Chebyshev moments is numerically very time-consuming, because it requires $M^2$ matrix-vector multiplication. However, thanks to the sparse nature of the usual TB Hamiltonian, all these operations are of order $N$.

%%%%% BIBLOGRAPHY %%%%%%%%%%%%%%%%%%%%%%%%%%%%%%%%%%%%%%%%%%%%%%%%%%%%%%%%%%%%%%%%%%%%%%%

%

%%%%%%%%%%%%%%%%%%%%%%%%%%%%%%%%%%%%%%%%%%%%%%%%%%%%%%%%%%%%%%%%%%%%%%%%%%%%%%%%%%%%%%%%%

\end{document}